\documentclass[12pt]{article}

\usepackage{amssymb,amsmath,amsfonts,eurosym,geometry,ulem,graphicx,caption,color,setspace,sectsty,comment,footmisc,caption,natbib,pdflscape,array,hyperref,subcaption,threeparttable, amsthm, rotating,chngcntr,lscape,subfloat, bbm, breqn, comment, booktabs}

\newcolumntype{L}[1]{>{\raggedright\let\newline\\arraybackslash\hspace{0pt}}m{#1}}
\newcolumntype{C}[1]{>{\centering\let\newline\\arraybackslash\hspace{0pt}}m{#1}}
\newcolumntype{R}[1]{>{\raggedleft\let\newline\\arraybackslash\hspace{0pt}}m{#1}}

\usepackage[skip=6pt]{caption}

\def\sym#1{\ifmmode^{#1}\else\(^{#1}\)\fi}

\begin{document}

\begin{titlepage}
\title{Effect of State and Local Sexual Orientation Anti-Discrimination Laws on Labor Market Differentials\thanks{We would like to thank Sandra Black, Rich Murphy, Steve Trejo, Gerald Oettinger, Dan Hamermesh, Kitt Carpenter, Dario Sansone, Chris Jepsen, Luis Faundez, and all the seminar participants of the University of Texas at Austin. No financial support was given to support this research. Correspondence: d.vamossy@pitt.edu.}}
\author{Scott Delhommer\thanks{University of Texas at Austin; sdelhommer@utexas.edu} \and Domonkos F. Vamossy\thanks{University of Pittsburgh, d.vamossy@pitt.edu}}
\date{\today}
\maketitle
\begin{abstract}

This paper presents quasi-experimental research examining the effect of both local and state anti-discrimination laws on sexual orientation on the labor supply and wages of lesbian, gay, and bisexual (LGB) workers. To do so, we use the American Community Survey data on household composition to infer sexual orientation and combine this with a unique panel dataset on state and local anti-discrimination laws. Leveraging variation in law implementation across localities over time and between same-sex and different-sex couples, we find that anti-discrimination laws significantly narrow gaps in labor force participation and employment for men in same-sex couples relative to men in different-sex couples, and also increase their percentile rank in the wage distribution. Our analysis reveals mostly null effects for female same-sex couples; however, in metropolitan areas these laws significantly reduce their employment compared to women in different-sex couples. One explanation for the reduced labor supply is that female same-sex couples begin to have more children in response to the laws. Finally, we present evidence that state anti-discrimination laws significantly and persistently increased support for same-sex marriage. This research shows that anti-discrimination laws can be an effective policy tool for reducing labor market inequalities across sexual orientation and improving sentiment toward LGB Americans.


\bigskip
\end{abstract}
\setcounter{page}{0}
\thispagestyle{empty}
\end{titlepage}
\pagebreak \newpage

\doublespacing

\section{Introduction} \label{sec:introduction}

On June 15, 2020, the U.S. Supreme Court held in Bostock v. Clayton County, 590 U.S. 644 (2020), that Title VII’s prohibition of sex discrimination in employment includes discrimination based on sexual orientation and gender identity. Prior to this ruling, more than half of U.S. states lacked explicit statewide employment protections based on sexual orientation, though many localities had adopted their own ordinances. As a result, protections varied substantially across and within states throughout the period studied. Wisconsin became the first state to enact a statewide sexual orientation employment non-discrimination law in 1982. There has been continued legislative interest in expanding federal protections, most notably the introduction of the ``Equality Act'' in early 2019, which would add sexual orientation and gender identity to federal civil rights statutes.

This variation in state and local protections provides the basis for our empirical analysis. In this paper, we examine how sexual orientation employment non-discrimination laws affect labor market differentials for lesbian and gay workers.

We exploit the differential roll-out of state and local laws from 2005--2019 in a modern difference-in-differences-in-differences (DDD) framework to analyze how these anti-discrimination laws impact the wages and labor supply of LGB workers. The economics literature consistently finds that gay men (or men in male same-sex couples) face earnings and labor-supply penalties, while lesbian women (or women in female same-sex couples) exhibit earnings and labor-supply premia relative to comparable heterosexual counterparts.\footnote{For a review of the impacts of legal access to same-sex marriage, see \cite{badgett2024review}. For a broader review of the economics related to sexual orientation and gender identity, see \cite{badgett2023review}. In addition, see \cite{badgett1995wage}; \cite{klawitter1998effects}; \cite{allegretto2001empirical}; \cite{black2003earnings}; \cite{carpenter2005self}; \cite{antecol2008sexual}; \cite{klawitter2011multilevel}; \cite{klawitter2015meta}; \cite{jepsen2017self}; \cite{carpenter2017does}; \cite{Burn2018}.} The differences in pay are attributed to a host of factors including discrimination and intra-household labor allocation and specialization.\footnote{Occupational sorting and concealment may also mediate gaps, with sexual minorities selecting into jobs with lower customer discrimination or greater autonomy (\cite{tilcsik2015concealable,black2007economics}).} Consistent with administrative complaints, sexual-orientation discrimination occurs at rates comparable to sex-based discrimination (\cite{ramos2008evidence}). We test the role of discrimination by examining how the passage of anti-discrimination laws affects the labor supply and wage differences between LGB and heterosexual workers.

This paper provides quasi-experimental evidence on both local and state sexual orientation anti-discrimination laws. It is also, to our knowledge, the first study to evaluate the effects of local anti-discrimination ordinances, whether based on sex, race, or sexual orientation, within a modern difference-in-differences framework.\footnote{Legal changes and enforcement have been used to study race- and sex-based discrimination (e.g., \cite{neumark2006labor}; see also \cite{donohue1991continuous}). These analyses do not incorporate local ordinances, which can bias estimates that rely on state-only policy variation.}

We use the 2005-2019 American Community Survey (ACS) and household composition to infer sexual orientation and create a unique and novel panel dataset on the passage of local and state anti-discrimination laws. We collected information on local laws from a host of sources including media reports, FOIA requests, and an advocacy group. We find a significant reduction in differences between LGB workers and heterosexual workers across labor supply and wage measures, due to anti-discrimination laws. Anti-discrimination laws reduce the labor force participation and employment gaps for men in male same-sex couples by 3.8 percentage points (78\% of the baseline gap) and 3.7 percentage points (77\%), respectively. The laws improve men in male same-sex couples’ relative position in the wage distribution: their percentile ranks in the hourly wage and annual earnings distributions rise by about 2.6 and 2.9 percentile points, respectively. By contrast, the wage distribution for women in same-sex couples remains largely unaffected by these laws. Event-study estimates show no pre-trends, supporting the parallel-trends assumption and the comparability of treated and control regions.

We explore potential mechanisms for the differing effects on men and women in the Discussion section, without imposing a single unified explanation; as in the broader literature, some differences remain an empirical puzzle. Reduced discrimination improves job security for sexual minorities (\cite{badgett2006discrimination}), which may allow the secondary earner to reduce market participation in favor of increased household production, including child rearing, a pattern consistent with household bargaining models (\cite{chiappori1992collective, lundberg1993separate}). These differences in household specialization are less pronounced in same-sex couples, but they may become more similar for women in female same-sex couples following the passage of an anti-discrimination law if it gives greater protection to the higher wage earner. We show empirically that female same-sex households have significantly more children than male same-sex households after the passage of an anti-discrimination law. 

Finally, our last contribution is the use of polling data on the support for same-sex marriage to examine the relationship between anti-discrimination laws and state sentiment toward LGB workers. One may expect that sentiment toward LGB workers would increase right before the passage of the laws, creating a selection issue. Alternatively, the anti-discrimination laws may normalize being a sexual minority, and improve sentiments toward LGB workers. We compiled state–year polls on same-sex marriage from the Pew Research Center. We show that the passage of statewide sexual orientation anti-discrimination laws persistently increases favorability toward LGB people through increased support for same-sex marriage. The states that passed anti-discrimination laws had parallel pre-trends in support of same-sex marriage before the passage of the laws. Afterward, those states had a significant and persistent increase in their support for same-sex marriage. The observed increase in support following the enactment of the law, rather than before, implies that policy changes can indeed influence public opinion. This matches the findings of \citet{sansone2019pink}, suggesting that legislative actions can lead to shifts in societal attitudes. This is in contrast to the thermostatic model proposed by \citet{wlezien1995public}, which posits that public sentiment drives policy changes rather than the other way around. Increased favorability toward LGB people may act in conjunction with greater job security to improve labor market outcomes for LGB Americans. 

Our study complements and extends prior state-focused analyses (e.g., \cite{klawitter2011multilevel, martell2013endas, Burn2018}) by introducing a novel city–county ENDA panel and applying modern staggered-adoption estimators. Unlike work centered on marriage laws (e.g., \cite{hansen2020labor,hansen2022tolerance}), we isolate employment protections and link them to labor supply, wage ranks, fertility, and opinion dynamics.

\subsection{Related Literature}

Research on sexual-orientation anti-discrimination laws (ENDAs) spans cross-sectional, causal state-level, and related policy literatures. Early cross-sectional work linked protections to narrower wage gaps but could not establish causality: \cite{klawitter1998effects} use the 1990 Census and find no significant wage differences in protected places, while \cite{gates2009impact} use the 2000 Census and report small premia. \cite{tilcsik2011pride} found that resumes signaling LGB status received significantly fewer callbacks in localities without sexual orientation anti-discrimination laws. \cite{klawitter2011multilevel} estimate multi-level earnings models with 2000 Census data, documenting systematic associations between state/local policies and earnings but not exploiting within-place policy changes over time. These past studies on anti-discrimination laws, while informative, fail to fully account for changes over time. 

Quasi-experimental state-level studies strengthen identification. \cite{martell2013endas} implement a state DiD design and find sizable wage convergence for gay men post-ENDAs, and \cite{Burn2018} adopt a triple-difference framework, still at the state level. A limitation of these state-only designs is that local ordinances, the likely first binding margin for many workers, are omitted; moreover, conventional TWFE estimators can be biased under staggered adoption and heterogeneous effects. To highlight these issues, Appendix B replicates a state-only specification and shows how excluding local ordinances can mask meaningful within-state variation, underscoring the value of incorporating local laws into the analysis.

Two adjacent strands inform mechanisms. First, policy shocks around marriage legalization affect household specialization and fertility: \cite{hansen2020labor} show changes in labor supply following marriage legalization; \cite{hansen2022tolerance} link tolerance to paid work among cohabiting gays and lesbians. These patterns match our household-reallocation results for women and the fertility margin we study. Second, \cite{mann2025antidiscrimination} document mental-health benefits from anti-discrimination laws, suggesting channels beyond pecuniary outcomes and complementing our evidence on opinion dynamics.

Broader anti-discrimination evidence underscores that legal protections and enforcement can shift hiring and employment margins across protected classes: disability (audit evidence in \cite{ArmourButtonHollands2018}; policy evidence in \cite{Button2018Expanding}) and age (field-experiment evidence in \cite{NeumarkBurnButtonChehras2019}). While these are not sexual-orientation ENDAs, they corroborate that statutory protections and their salience can meaningfully change labor-market opportunities.

\section{Data} \label{sec:data}

A common challenge in analyzing pay disparities between different types of same-sex and different-sex couples is the limited availability of high-quality data on sexual orientation, wages, and employment. We follow the literature in inferring sexual orientation by examining household relationship structure in the ACS (\citet{klawitter1998effects}; \citet{gates2009impact}; \citet{jepsen2017self}). Specifically, we classify a person as being in a same-sex couple if they report a spouse or ``unmarried partner'' who is the same sex as themselves.\footnote{The ACS measures self-reported ``sex'' rather than gender identity. Following the existing literature, we interpret this variable as a proxy for gender in couple-type inference, but this may misclassify some transgender or gender-nonconforming respondents. See, for example, \cite{gates2009impact} and \cite{jepsen2017self}. We exclude individuals whose sex or relationship status was allocated following \cite{gates2009same}. } To identify ``unmarried partners,'' the ACS asks each household member to report their relationship to the householder from a fixed list of categories. The category ``unmarried partner'' is defined by the Census Bureau as a person who has a close personal relationship with the householder. Importantly, the questionnaire provides a separate category for ``housemate/roommate,'' which reduces the likelihood that unrelated adults, such as different-sex roommates, are erroneously classified as unmarried partners.

We use the 2005-2019 yearly ACS from \cite{ipumsusa}. We use wages, defined as salaried wages from an employer. The ACS reports wage earnings in annual terms. We use the annual figures and convert them to hourly wage figures using variables on average weekly hours worked and weeks worked.\footnote{``Weeks worked" in the ACS is a categorical variable giving a range of weeks worked. We take the median value given in the range to compute weeks worked.} Finally, we limit our sample to prime-age working adults and only examine those who are ages 25-65. We use 25 as the lower age cutoff to allow workers to complete college and have more fully developed human capital, which is typically thought of as an important factor for the differences in wages between LGB and straight workers (\cite{black2007economics}). 

We obtained data on the passage of state anti-discrimination laws from LGBTMap.org, an LGBT advocacy group. The website gives information on which states passed anti-discrimination laws and when. We focus solely on sexual orientation anti-discrimination laws that give protection in employment. LGBTMap.org also provided incomplete data detailing the passage of local city and county laws with many cities missing years for the start of their anti-discrimination laws. We completed the dataset for the relevant years using old media reports, correspondence with local officials, and FOIA requests. Building on earlier city-level analyses such as \cite{klawitter2011multilevel}, this dataset provides the first complete city-level dataset covering 15 years of sexual orientation anti-discrimination laws in the U.S. We focus on cities reported in the ACS and match those city laws with their corresponding counties to merge with the ACS. The ACS only reports county of residence for those in metro areas, so any rural counties are lumped together within a state.

\begin{figure}[!htbp]
\centering

\begin{subfigure}{\textwidth}
    \centering
    \includegraphics[width=0.525\textwidth]{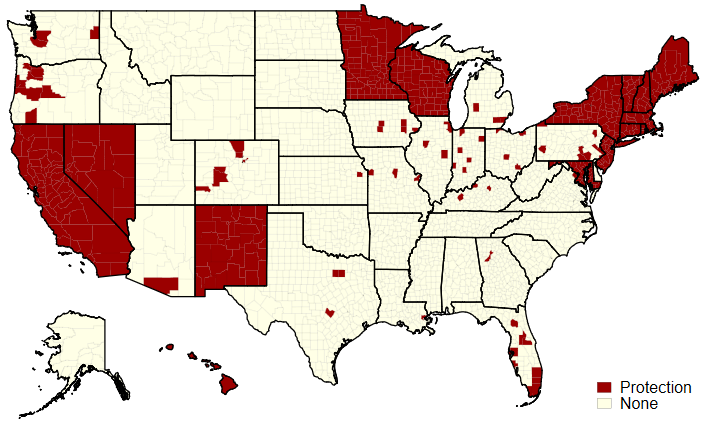}
    \caption{2005}
\end{subfigure}

\hfill

\begin{subfigure}{\textwidth}
    \centering
    \includegraphics[width=0.525\textwidth]{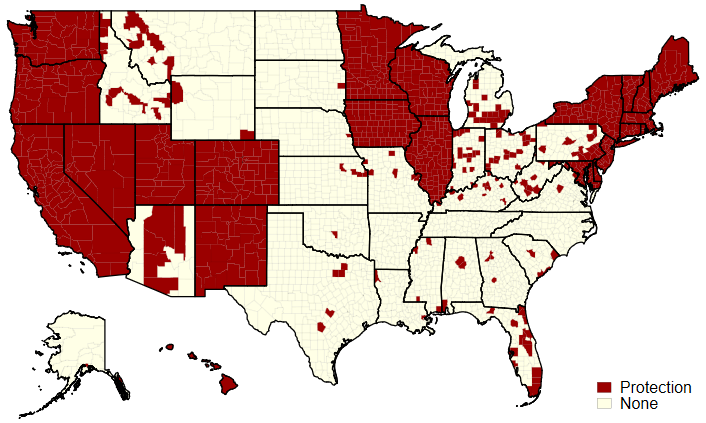}
    \caption{2019}
\end{subfigure}

\caption{Sexual Orientation and Anti-Discrimination Laws}
\label{fig:county}

\begin{flushleft}
\scriptsize
Notes: State and local sexual orientation employment anti-discrimination laws in 2005 (a) and 2019 (b). Data on laws obtained from LGBTMap.org and authors' own investigation using media reports and FOIA requests.
\end{flushleft}

\centering
\includegraphics[width=0.525\textwidth]{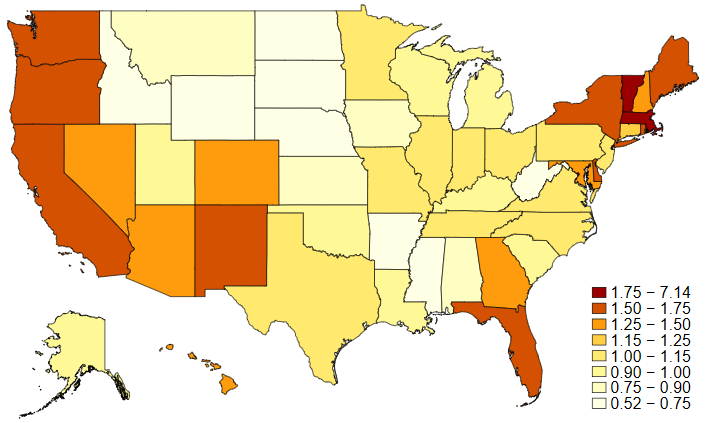}
\captionof{figure}{Percentage of Same Sex Couples: 2005--2019}
\label{fig:per_ssp}

\begin{flushleft}
\scriptsize
Notes: Percentage of couples that are same-sex for each state and DC over 2005--2019 using the American Community Survey. Author's calculations. District of Columbia’s share (7.14\%) is a significant outlier relative to the other states.
\end{flushleft}

\end{figure}

We report the state, city, and county laws and the year they were enacted in Table \ref{tab:laws}.\footnote{Since August 2018, Pennsylvania has interpreted discrimination based on sexual orientation and gender identity as covered under the ``sex'' category of the Pennsylvania Human Relations Act, allowing LGBT individuals to file complaints for discrimination in employment, housing, education, and public accommodations.}  In Figure \ref{fig:county}, we show which counties had sexual orientation anti-discrimination laws in 2005 and 2019, respectively. The enactment of anti-discrimination legislation is not random. The states that have these protections are generally considered friendlier to LGB workers than those without these laws and have a larger share of same-sex couples than those states without these laws. Also, many counties with anti-discrimination laws have large cities with a larger concentration of LGB workers than rural counties. However, this is not universal. For instance, Utah extended protection to LGB workers despite being a relatively conservative state, and certain liberal cities like Houston are noticeably absent from extending protection to LGB workers.\footnote{Houston passed a sexual orientation and gender identity anti-discrimination ordinance, but it was only in effect for 3 months before being challenged. The law was put up to a public vote and lost, repealing the law.} In Figure \ref{fig:per_ssp}, we present the distribution of same-sex couples by state from 2005 to 2019, and in Table \ref{tab:pop}, we present the 10 counties with the largest share of same-sex couples. Unsurprisingly, the distribution of same-sex couples by state and by county is skewed toward more progressive states and counties with large cities in them that are known for having a large LGB population like San Francisco, the District of Columbia, New York City, and Boston. 

We present descriptive statistics showing the differences in the same-sex couple population compared to the different-sex couple population broken down by education level and sex in Table \ref{tab:sum}. Descriptively, there is a clear difference in labor market outcomes and characteristics between men and women in same-sex couples and those in different-sex couples when controlling for education.

\begin{table}[!ht]
\centering
\scriptsize
\begin{threeparttable}
\caption{Timing of Sexual Orientation Anti-discrimination Laws}\label{tab:laws}
\begin{tabular}{r|r|l}
\multicolumn{1}{l|}{Year} & \multicolumn{1}{l|}{State} & City or County \\
\midrule
\multicolumn{1}{l|}{2005 and Before} & \multicolumn{1}{l|}{CA, CT, } & Boulder, CO; Denver, CO; Fort Collins, CO; Gainsville, FL; Hialeah, FL; Hollywood, FL;  \\
      & \multicolumn{1}{l|}{DC, HI, } & Key West, FL; Miami, FL; Orlando, FL; Pembroke, FL; Saint Petersburg, FL;  \\
      & \multicolumn{1}{l|}{ME, MD, } & Tampa, FL; West Palm, FL; Atlanta, GA; Ames, IA; Cedar Rapids, IA; Davenport, IA;  \\
      & \multicolumn{1}{l|}{MA, MN, } & Des Moines, IA; Iowa City, IA; Campaigne, IL; Chicago, IL; Peoria, IL; Urbana, IL;  \\
      & \multicolumn{1}{l|}{NV, NH, } & Bloomington, IN; Fort Wayne, IN; Michigan City, IN; Terre Haute, IN; Lawerence, KS;  \\
      & \multicolumn{1}{l|}{NJ, NM, } & Covington, KY; Lexington, KY; Louisville, KY; New Orleans, LA; Ann Arbor, MI;  \\
      & \multicolumn{1}{l|}{NY, RI, } & Detroit, MI; Grand Rapids, MI; Ypsilanti, MI; Columbia, MO; Kansas City, MO;  \\
      & \multicolumn{1}{l|}{VT, WI} & Saint Louis, MO; Cleveland, OH; Columbus, OH; Toledo, OH; Eugene, OR;  \\
      &       & Portland, OR; Benton County, OR; Salem, OR; Allentown, PA; Erie, PA; Harrisburg, PA; \\
      &       & Lancaster, PA; Philadelphia, PA; Pittsburgh; Scranton, PA; Austin, TX; Dallas, TX;  \\
      &       & Fort Worth, TX; Alexandria, VA; Arlington, VA; Seattle, WA; Spokane, WA; Tacoma, WA \\
\midrule
2006  & \multicolumn{1}{l|}{IL, WA} & Dubuque, IA; Indianapolis, IN, Ferndale, MI; Lansing, MI; \\
      &       &  Cincinnati, OH; Easton, PA; West Chester, PA; Charleston, SC \\
\midrule
2007  & \multicolumn{1}{l|}{CO, IA, OR} & Waterloo, IA; Coshocton, OH; Dayton, OH; Newark, OH; Charleston, WV \\
\midrule
2008  &       & Columbia, SC \\
\midrule
2009  & \multicolumn{1}{l|}{DE} & Allegheny, PA; Reading, PA; Salt Lake City, UT \\
\midrule
2010  &       & Tallahassee, FL; Traverse City, MI; Missoula, MT;  \\
      &       & Lower Merion, PA; Grand County, UT; Summit County, UT \\
\midrule
2011  &       & Volusia County, FL; Evansville, IN; University City, MO; East Cleveland, OH,  \\
      &       & Bethlehem, PA; Conshohocken, PA; Haverford, PA; Ogden, UT \\
\midrule
2012  &       & St. Augustine, FL; Boise, ID; New Albany, IN; South Bend, IN; Flint, MI;  \\
      &       & Muskegon, MI; Maplewood, MO; Helena, MT; Omaha, NE; Canton, OH;  \\
      &       & Abington, PA; Cheltenham, PA; Morgantown, WV \\
\midrule
2013  &       & Phoenix, AZ; Pocatello, ID; Frankfort, KY; Shreveport, LA; Battle Creek, MI;  \\
      &       & Bristol, PA; Pittston, PA; San Antonio, TX; Charlottesville, VA; Huntington, WV \\
\midrule
2014  &       & Tempe, AZ; Adrian, MI; Macomb County, MI; Butte, MT; Atlantic Beach, FL \\
\midrule
2015  & \multicolumn{1}{l|}{UT} & Anchorage, AK; Osceola County, FL; Anderson, IN; Clinton, IN; Hammond, IN; Muncie, IN \\
\midrule
2016  &       & Kokomo, IN; Manahattan, KS; St. Charles, MO; Jackson, MS; Lakewood, OH;  \\
      &       & Carlisle, PA; Dickson City, PA; Wilkes-Barre, PA; Martinsburg, WV; Wheeling, WV \\
2017  &     &  Sitka, AK; Glendale, AZ; \\
2018  &     &  \\
2019  &      &  Fort Lauderdale, FL; Fernandina Beach, FL; Decatur, GA\\
2020  & VA     &  \\
     \bottomrule
\end{tabular}%
\begin{tablenotes}
\scriptsize
\item Notes: List of states, cities, and counties with sexual orientation anti-discrimination laws pulled from LGBTMap.org, an advocacy group, as well as through media reports and local FOIA requests. We only list city or county laws if there is no state law. 
\end{tablenotes}
\end{threeparttable}
\end{table}%

\begin{table}[!ht]
\scriptsize
\centering
\begin{threeparttable}
\caption{Counties with Largest LGB Populations}\label{tab:pop}%
\begin{tabular}{cc|c}
County & State & Percent of SSPs \\
\midrule
San Francisco County & CA    & 7.37 \\
District of Columbia & DC    & 7.20 \\
New York County & NY    & 5.43 \\
Suffolk County & MA    & 4.37 \\
St. Louis city & MO    & 4.01 \\
Multnomah County & OR    & 3.99 \\
Alexandria city & VA    & 3.70 \\
Santa Fe County & NM    & 3.43 \\
Baltimore city & MD    & 3.37 \\
\bottomrule
\end{tabular}%
\begin{tablenotes}
\scriptsize
\item Notes: Using ACS and person weights to recover the percentage of partnerships that are same-sex partnerships by county over 2005-2019.
\end{tablenotes}
\end{threeparttable}
\end{table}%

\begin{table}[!ht]
\centering
\scriptsize 
\begin{threeparttable}
\caption{Summary Statistics}\label{tab:sum}%
\begin{tabular}{l|c|c|c|c|c|c}
\toprule
\multicolumn{7}{c}{Panel A: Men} \\
\midrule
      & \multicolumn{3}{c|}{High School Grad or Lower} & \multicolumn{3}{c}{Some College or Higher} \\
      & \multicolumn{1}{l|}{SSP} & \multicolumn{1}{l}{DSP} &       & \multicolumn{1}{l|}{SSP} & \multicolumn{1}{l}{DSP} &  \\
      & \multicolumn{1}{l|}{n = 21,737} & \multicolumn{1}{l|}{n = 2,805,327} &       & \multicolumn{1}{l|}{n = 80,456} & \multicolumn{1}{l|}{n = 4,943,301} &  \\
\midrule
Variable & \multicolumn{1}{l|}{Mean} & \multicolumn{1}{l|}{Mean} & \multicolumn{1}{l|}{Difference} & \multicolumn{1}{l|}{Mean} & \multicolumn{1}{l|}{Mean} & \multicolumn{1}{l}{Difference} \\
\midrule
In Labor Force       &       0.738&       0.829&       -0.091\sym{***}&       0.863&       0.898&       -0.035\sym{***}\\
                     &     (0.440)&     (0.377)&                     &     (0.344)&     (0.303)&                     \\
Employed             &       0.687&       0.783&       -0.096\sym{***}&       0.831&       0.871&       -0.041\sym{***}\\
                     &     (0.464)&     (0.412)&                     &     (0.375)&     (0.335)&                     \\
Real Wage            &   28,991.2&   36,602.8&    -7,611.6\sym{***}&   71,867.9&   79,338.7&    -7470.8\sym{***}\\
                     & (40,363.5)& (39,963.0)&                     & (89,316.6)& (90,257.4)&                     \\
Real Hourly Wage     &      16.735&      18.661&       -1.925\sym{***}&      35.718&      37.691&       -1.974\sym{***}\\
                     &    (58.837)&    (44.692)&                     &    (76.987)&   (115.923)&                     \\
Age                  &      46.074&      47.440&       -1.367\sym{***}&      45.609&      46.786&       -1.177\sym{***}\\
                     &    (10.932)&    (10.912)&                     &    (10.645)&    (11.071)&                     \\
Number of Children   &       0.484&       1.187&       -0.702\sym{***}&       0.234&       1.124&       -0.890\sym{***}\\
                     &     (1.014)&     (1.280)&                     &     (0.692)&     (1.187)&                     \\
Asian                &       0.029&       0.029&       -0.000         &       0.046&       0.062&       -0.017\sym{***}\\
                     &     (0.169)&     (0.169)&                     &     (0.209)&     (0.242)&                     \\
Black                &       0.079&       0.080&       -0.001         &       0.041&       0.057&       -0.016\sym{***}\\
                     &     (0.269)&     (0.271)&                     &     (0.198)&     (0.232)&                     \\
Hispanic             &       0.115&       0.118&       -0.003         &       0.069&       0.050&      0.020\sym{***}\\
                     &     (0.319)&     (0.323)&                     &     (0.254)&     (0.218)&                     \\
White                &       0.789&       0.798&       -0.009\sym{**} &       0.863&       0.842&      0.021\sym{***}\\
                     &     (0.408)&     (0.401)&                     &     (0.344)&     (0.365)&                     \\
\midrule
\multicolumn{7}{c}{Panel B: Women} \\
\midrule
      & \multicolumn{3}{c|}{High School Grad or Lower} & \multicolumn{3}{c}{Some College or Higher} \\
      & \multicolumn{1}{l|}{SSP} & \multicolumn{1}{l}{DSP} &       & \multicolumn{1}{l|}{SSP} & \multicolumn{1}{l}{DSP} &  \\
      & \multicolumn{1}{l|}{n = 22,238} & \multicolumn{1}{l|}{n = 2,654,209} &       & \multicolumn{1}{l|}{n = 79,414} & \multicolumn{1}{l|}{n = 5,464,738} &  \\
\midrule
Variable & \multicolumn{1}{l|}{Mean} & \multicolumn{1}{l|}{Mean} & \multicolumn{1}{l|}{Difference} & \multicolumn{1}{l|}{Mean} & \multicolumn{1}{l|}{Mean} & \multicolumn{1}{l}{Difference} \\
\midrule
In Labor Force       &       0.724&       0.596&      0.128\sym{***}&       0.862&       0.749&      0.114\sym{***}\\
                     &     (0.447)&     (0.491)&                     &     (0.345)&     (0.434)&                     \\
Employed             &       0.673&       0.558&      0.115\sym{***}&       0.833&       0.722&      0.111\sym{***}\\
                     &     (0.469)&     (0.497)&                     &     (0.373)&     (0.448)&                     \\
Real Wage            &   25,032.0&   16,555.5&   8,476.5\sym{***}&   56,374.5&   39,254.0&  17,120.6\sym{***}\\
                     & (33,315.4)& (24,328.6)&                     & (64,852.8)& (50,889.8)&                     \\
Real Hourly Wage     &      14.331&      10.497&      3.834\sym{***}&      29.083&      22.721&      6.363\sym{***}\\
                     &    (37.390)&    (39.943)&                     &    (62.691)&    (86.710)&                     \\
Age                  &      45.437&      48.151&       -2.714\sym{***}&      44.860&      45.382&       -0.522\sym{***}\\
                     &    (11.356)&    (10.973)&                     &    (11.001)&    (11.134)&                     \\
Number of Children   &       0.716&       1.088&       -0.371\sym{***}&       0.497&       1.103&       -0.606\sym{***}\\
                     &     (1.093)&     (1.287)&                     &     (0.902)&     (1.179)&                     \\
Asian                &       0.024&       0.045&       -0.021\sym{***}&       0.029&       0.067&       -0.038\sym{***}\\
                     &     (0.152)&     (0.206)&                     &     (0.168)&     (0.250)&                     \\
Black                &       0.107&       0.065&      0.042\sym{***}&       0.057&       0.056&      0.001         \\
                     &     (0.309)&     (0.247)&                     &     (0.231)&     (0.230)&                     \\
Hispanic             &       0.099&       0.120&       -0.021\sym{***}&       0.058&       0.054&      0.004\sym{***}\\
                     &     (0.299)&     (0.325)&                     &     (0.234)&     (0.226)&                     \\
White                &       0.778&       0.799&       -0.022\sym{***}&       0.861&       0.836&      0.025\sym{***}\\
                     &     (0.416)&     (0.400)&                     &     (0.346)&     (0.370)&                     \\
\bottomrule
\end{tabular}%
\begin{tablenotes}
\scriptsize
\item Notes: Data is derived from the American Community Survey (ACS) conducted annually between 2005 and 2019. Summary statistics are presented by education level. T-tests were conducted to determine significant differences between those in same-sex partnerships and different-sex partnerships.  \sym{*} \(p<0.10\), \sym{**} \(p<0.05\), \sym{\sym{***}} \(p<0.01\).
\end{tablenotes}
\end{threeparttable}
\end{table}%

\clearpage

\section{Identification Strategy \& Estimation} \label{sec:identification}

The differential roll-out of anti-discrimination laws by state and locality over time, and between same-sex and different-sex couples, naturally lends itself to a saturated triple-differences framework, where the primary outcomes of interest are changes in labor supply and pay. This strategy uses the variation presented in Figure \ref{fig:county}, which shows how the laws changed over time by state and county. Formally, we estimate this equation on labor supply: 

\begin{equation}\label{eq_1}
\begin{aligned}
Y_{ipct}
&=
\alpha_{g}
\;+\;
\beta_{1g}
\bigl[\mathrm{SSP}_{ipct} \times \mathrm{ENDA}_{ct}\bigr]
\;+\;
\lambda_{s(c), g}
\bigl[\mathrm{SSP}_{ipct} \times \mathrm{SSM}_{s(c), t}\bigr]
\\
&\quad
+ \theta_{gct}
+ \mu_{gpc}
+ \kappa_{gpt}
+ X_{ipct}' \, \delta_g
+ \epsilon_{ipct}.
\end{aligned}
\end{equation}

In our regression, equation (1) defines the labor market outcomes \(Y_{ipct}\) for individual \(i\) of sex \(g\), in partnership type \(p\), residing in county \(c\) in state \(s\), at time \(t\). These outcomes include labor force participation, employment status, and the percentile rank of real wages (in 2019 dollars). The binary variable \(\mathrm{SSP}_{ipct}\) indicates whether an individual is in a same‐sex partnership (1) or a different‐sex partnership (0). \(\mathrm{SSM}_{s(c), t}\) indicates whether the state \(s(c)\) recognizes same‐sex marriage in year \(t\), where \(s(c)\) denotes the state corresponding to county \(c\). Our key coefficient of interest is \(\beta_{1g}\), which measures the sex-specific effect of the interaction of same‐sex partnership status with the presence of a county‐level Employment Non‐Discrimination Act, \(\mathrm{ENDA}_{ct}\). Treatment status is computed using the earliest year of ENDA implementation at the county or state level, with individuals remaining classified as ``treated'' thereafter. All coefficients ($\alpha_g, \beta_{1g}, \lambda_{s(c), g}, \delta_g$) and fixed effects ($\theta_{gct}, \mu_{gpc}, \kappa_{gpt}$) are fully interacted with sex $g$, so that the model is estimated jointly but allows each sex to have its own coefficients. The error term \( \epsilon_{ipct} \) is clustered at the county level to account for within-county correlation.\footnote{The ACS only identifies metro counties, so the county fixed effect is close to a pseudo-county fixed effect. Rural counties within a state are compared together with a true county fixed effect for metro area counties.}

We restrict the sample to individuals aged 25--65 and estimate all models jointly by sex using ACS person weights. Following \citet{chen2024logs}, we compute weighted within-year percentile ranks for real annual earnings and real hourly wages (expressed in 2019 dollars) by assigning individuals to weighted percentiles each year. We then construct three distributional indicators, at the 25th, 50th, and 75th percentiles, set to one when an individual’s percentile rank meets or exceeds each threshold. Because ACS earnings reflect a 23-month reference period (\cite{hansen2020labor}), measured earnings may not align exactly with ENDA adoption dates, likely attenuating treatment effects.

We include three sets of high-dimensional fixed effects, each fully interacted with sex to reflect the joint estimation. First, the county-by-year fixed effects $\theta_{gct}$ absorb all time-varying shocks or policy changes common to individuals of sex $g$ within county $c$ and year $t$. Second, the partnership-by-county fixed effects $\mu_{gpc}$ capture time-invariant characteristics specific to partnership type $p$ within county $c$ for each sex $g$. Third, the partnership-by-year fixed effects $\kappa_{gpt}$ absorb year-specific shocks that differentially affect partnership types for each sex. The vector $X_{ipct}'\delta_g$ includes demographic controls (race, age, and education), also allowed to vary by sex. 

The coefficient of interest, $\beta_{1g}$, identifies how ENDA coverage $\mathrm{ENDA}_{ct}$ differentially affects same-sex partners relative to different-sex partners for each sex $g$, net of any confounding from same-sex marriage legalization. The interacted term $\lambda_{s(c),g}\bigl[\mathrm{SSP}_{ipct} \times \mathrm{SSM}_{s(c),t}\bigr]$ partials out state-specific labor-market impacts of same-sex marriage legalization on same-sex partners for each sex.

The identifying variation for $\beta_{1g}$ comes from differences in the timing of county-level ENDA adoption, combined with how these policy changes differentially affect individuals in same-sex versus different-sex couples within counties, holding constant sex-specific fixed effects. This allows us to isolate how employment protections change labor supply and wage outcomes for same-sex partners relative to different-sex partners of the same sex. Thus, $\beta_{1g}$ captures the sex-specific differential impact of ENDA on labor market outcomes for individuals in same-sex couples.\footnote{We recognize that the effects of anti-discrimination laws may vary across regions (e.g., \citet{Burn2018}), but our objective is to estimate an average effect for those directly subject to the laws.} To interpret $\beta_{1g}$ as the causal effect of enacting employment protections for sexual orientation, we assume that no other factors differentially shift outcomes for same-sex versus different-sex partners in counties that do and do not adopt ENDAs. Below, we outline the threats to this assumption and describe how our empirical design addresses each one.

\subsection{Threats to Identification}

We group the main identification concerns into three categories: (i) threats that can be addressed directly through sample restrictions, (ii) threats mitigated through design choices and high-dimensional fixed effects, and (iii) residual concerns that cannot be fully ruled out but primarily generate attenuation rather than spurious effects.

\subsubsection{Threats Addressed}
\paragraph{Selective Migration.}
Individuals in same-sex couples may relocate to jurisdictions with stronger LGBT protections, altering the composition of treated counties. To assess this possibility, we augment the specification with an additional layer of interactions distinguishing non-mover females, non-mover males, mover females, and mover males. This four-way structure lets ENDA effects differ flexibly by migration status. Comparing coefficients across these groups provides a direct test for migration-induced bias, and the results show no meaningful differences.\footnote{Long-run unobserved mobility may still shift the composition of same-sex couples in treated areas, though our mover–non-mover comparisons indicate limited bias.}

\paragraph{Household Formation and Sample Composition.}
A potential concern is that ENDAs may affect the likelihood that individuals enter or remain in a same-sex partnership, thereby altering who appears in our coupled-household sample. If anti-discrimination protections increase relationship stability or the willingness to publicly identify a partner, then treatment could shift the composition of the same-sex population in ways correlated with labor market outcomes. To assess this concern, we estimate whether the share of adults living in same-sex couples changes following ENDA adoption at the county level. We find no evidence of differential changes for men or women. This suggests that ENDAs do not meaningfully affect partnership formation or household structure, and thus do not induce endogenous entry into the coupled sample. As a result, the identifying variation in our main specification is unlikely to reflect compositional changes in who is observed as part of a same-sex household.

\paragraph{Residence–Work Geography Mismatch.} 
ENDAs apply to the county of employment, whereas ACS geography reflects the county of residence. As with migration, this creates the possibility of misclassifying some individuals with respect to treatment if they commute across county (or state) lines. We evaluate this concern using the same strategy as our mover–non-mover analysis: we allow ENDA effects to differ flexibly across groups likely to experience geographic mismatch. The estimates are stable across these specifications, indicating that any residence–work misalignment does not materially influence the results. As with other sources of geographic measurement error, this type of misclassification primarily generates attenuation rather than spurious treatment effects.

\subsubsection{Threats Mitigated}

\paragraph{Staggered Adoption.}
To address potential biases from staggered policy adoption, we employ the methodology of \citet{borusyak2021revisiting}. Unlike standard TWFE models, which may misattribute post-treatment effects from early-treated units to later-treated ones, their estimator uses only untreated observations to form counterfactuals for treated units. Because we estimate the model jointly by sex, this approach is applied separately within each sex group, yielding the sex-specific treatment effects $\beta_{1g}$ without contamination from negative weighting. Our identification hinges on the parallel trends assumption: in the absence of ENDAs, labor market trends for same-sex and different-sex partners would have evolved similarly within each sex group across treated and untreated counties.\footnote{This requires no anticipatory responses and equal pre-treatment trends within sex-by-partnership cells across treatment timing groups.} We assess this assumption with event-study analyses (Figures~2--3), which show no evidence of differential pre-trends.\footnote{We conduct a sensitivity analysis following \cite{RambachanRoth2023}. As detailed in Appendix Table \ref{tab:sensitivity} and discussed in Appendix \ref{app:sensitivity}, we find that our extensive margin estimates for men are robust to post-treatment trend violations up to 40\% of the maximal pre-trend deviation, while the null results for women persist even under relaxed assumptions.}

\paragraph{Same-Sex Marriage Legalization.}
Same-sex marriage (SSM) legalization may confound the effects of ENDAs by shifting household specialization or labor supply among same-sex partners. We directly partial out these effects by including sex-specific interactions of same-sex partnership status with SSM legalization status. This absorbs any differential impacts of state-level marriage reforms on same-sex couples.\footnote{While these controls mitigate potential bias, overlapping policies may still exert residual confounding influences on the estimated effects.} As an additional check, Section~\ref{subsec:marriage} restricts the sample to pre-SSM years within each state, again yielding similar estimates.

\paragraph{Differential Local Trends and Endogenous Adoption.}
Jurisdictions adopting ENDAs tend to be more supportive toward sexual minorities. If positive sentiment drives both ENDA passage and improvements in labor market outcomes, estimates may reflect omitted variables. Our saturated triple-difference specification addresses this by including county-year, partnership-by-year, and partnership-by-county fixed effects, all interacted with sex. These fixed effects ensure comparisons are made only within county-year cells and within sex-by-partnership cells, substantially reducing concerns about correlated local shocks or endogenous policy timing.

\subsubsection{Residual Threats}

\paragraph{Misclassification of Sexual Orientation.}
Our comparison relies on household composition rather than self-identified sexual orientation. Bisexual individuals in different-sex relationships enter the control group, and some gay or lesbian workers may also be misclassified into that group. Because lesbian women tend to earn more than heterosexual women, gay men tend to earn less than heterosexual men, and bisexual workers earn less than both (\cite{mize2016sexual}), such misclassification does not shift group earnings in a uniform direction. In addition, bisexual workers may respond differently to ENDAs than gay or lesbian workers, so the sign of any resulting bias in the estimated ENDA coefficients is theoretically ambiguous. At the same time, misclassification that places gay or lesbian individuals into the control group reduces the contrast between partnership types and therefore operates through the usual attenuation mechanism. Overall, these forms of misclassification complicate interpretation but are unlikely to generate spurious positive treatment effects.

\paragraph{Overlap and Heterogeneity in Local and State ENDA Laws.}
Local and state ENDAs vary in scope and enforcement. Our coding focuses on the first year in which a county receives broad protections in private employment and housing. For example, although Delaware extended protections to state employees in 1999, we code treatment beginning in 2009 when comprehensive protections were enacted. Similarly, some local ordinances initially covered only public employees before broader laws passed. These earlier, narrower protections may still have influenced social attitudes or employer behavior prior to full coverage. In addition, the strength of protections differs across jurisdictions, some laws exempt small firms or specific sectors, and many large employers adopt nondiscrimination policies independent of legal requirements (e.g., firms scoring highly on the Corporate Equality Index). Collapsing this heterogeneity into a single binary ENDA indicator necessarily compresses meaningful variation. This generates classical attenuation bias, pushing estimated effects toward zero and making our findings conservative.

\paragraph{Geographic Misclassification From PUMA-to-County Assignment.}
ACS microdata identify geography at the PUMA level, which sometimes maps imperfectly to counties. A small share of individuals may thus be assigned to counties with classification error. We do not exclude ambiguous PUMAs; however, such misclassification behaves like classical measurement error in the treatment indicator and therefore attenuates estimated treatment effects rather than creating false positives.

To summarize, our empirical strategy exploits a fully saturated triple-differences design estimated jointly by sex. We control for confounding policy changes such as same-sex marriage legalization using sex-specific marriage interactions, and we implement the estimator of \citet{borusyak2021revisiting} to mitigate biases arising from staggered ENDA adoption. Together, these features allow us to identify how local ENDAs affect the labor market outcomes of individuals in same-sex couples relative to their different-sex counterparts within each sex group. As we show in Appendix \ref{sec:app_alternative}, the estimates for males and females are similar in sign and magnitude across alternative estimators, supporting the robustness of our findings.

\section{Primary Findings} \label{sec:results}

\subsection{Extensive \& Intensive Margins}

In Table \ref{tab:ext_int}, we present regression results that examine the extensive and intensive margins of labor supply for same-sex couples. The coefficient \(\beta_{1g}\) captures the effect of anti-discrimination laws on the labor supply outcomes for men in male same-sex couples and for women in female same-sex couples.

\begin{table}[htbp]
\scriptsize
\centering
\begin{threeparttable}
\caption{Effect of Anti-Discrimination Laws: Extensive \& Intensive Margin of Labor Supply}\label{tab:ext_int}
\begin{tabular}{l|cc|cccc}
\toprule
& (1) & (2) & (3) & (4) & (5) & (6) \\
& \multicolumn{2}{c}{\underline{Extensive Margin}} 
& \multicolumn{4}{c}{\underline{Intensive Margin}} \\
& Labor Force & Employed 
& Weekly Hours & Weeks Worked 
& Weekly Hours & Weeks Worked \\
\midrule

ENDA $\times$ SSP: Female  
& -0.0037 & -0.0050 & 0.9519\sym{*} & -0.4143 & 0.7017\sym{**} & -0.8915\sym{***} \\
& (0.0116) & (0.0135) & (0.5332) & (0.6235) & (0.3080) & (0.3349) \\[4pt]

ENDA $\times$ SSP: Male  
& 0.0379\sym{***} & 0.0370\sym{***} & 1.1289\sym{**} & 1.5089\sym{**} & -0.0920 & -0.0719 \\
& (0.0130) & (0.0135) & (0.4984) & (0.6276) & (0.3727) & (0.2821) \\

\midrule

$p$-value: M $-$ F  
& 0.0032 & 0.0177 & 0.7908 & 0.0082 & 0.1160 & 0.0465 \\

\midrule 

Employed Only  
&  &  &  &  & X & X \\

Observations  
& 7,781,358 & 7,781,358 & 7,781,358 & 7,781,358 & 5,765,580 & 5,765,580 \\

Mean (F)  
  &       0.818         &       0.782         &       34.42         &       39.51         &       41.25         &       47.27         \\

Mean (M)  
    &       0.825         &       0.790         &       35.53         &       39.96         &       42.41         &       47.69         \\

\bottomrule
\end{tabular}

\begin{tablenotes}
\scriptsize
\item Notes: Data is derived from the American Community Survey (ACS) conducted annually between 2005 and 2019. We compare individuals in same-sex partnerships with those in different-sex partnerships within four years of ENDA enactment. The regressions are estimated jointly, allowing the treatment effect to differ for men and women through interaction terms. Columns (1–2) focus on the extensive margin, examining changes in labor force participation and employment. Columns (3–6) explore the intensive margin using variation in weeks worked per year and usual weekly hours; Columns (5–6) restrict the sample to employed individuals. The ``Mean (F)" and ``Mean (M)" rows report pre-treatment averages of the outcome variables (including never-treated counties) for women and men in same-sex couples, respectively. Reported \(p\)-values test whether the ENDA effect differs significantly between men and women. Standard errors are clustered at the county level. \sym{*} \(p<0.10\), \sym{**} \(p<0.05\), \sym{***} \(p<0.01\).
\end{tablenotes}

\end{threeparttable}
\end{table}

The joint estimation allows the effects of anti-discrimination laws to differ for men and women through interaction terms. Columns (1) and (2) show that, for men in male same-sex couples, anti-discrimination laws reduce both the labor force participation and employment gaps by approximately 3.8 and 3.7 percentage points, which translates into reductions of about 78\% and 77\%, respectively. These coefficients are statistically significant at the 1\% level. Columns (3) through (6) address the intensive margin, indicating modest increases in weekly hours and weeks worked among all men. However, when the sample is restricted to employed men (Columns (5) and (6)), these effects become statistically insignificant. This pattern suggests that the primary channel through which these laws operate is by increasing the rate of employment rather than by altering hours worked among those already employed.

The estimates for women in same-sex couples follow a different pattern. Along the extensive margin (Columns (1) and (2)), the coefficients are small, negative, and statistically insignificant, on the order of 0.3–0.5 percentage points, which are not only imprecise but also economically small relative to mean participation rates. On the intensive margin, Column (3) shows a modest increase of 0.95 weekly hours (significant at the 10\% level), while Column (4) shows a small, statistically insignificant reduction in weeks worked. When focusing on employed women (Columns (5) and (6)), we observe a statistically significant increase of 0.70 weekly hours and a significant decrease of 0.89 weeks worked. These mixed but generally modest effects underscore that, for women, the evidence for labor-market responses is weaker and less systematic than for men, consistent with the possibility that adjustments for women occur outside the labor-market margins examined here.

Finally, the joint estimation allows for explicit comparison across sexes. The $p$-values reported in the table indicate that the extensive margin effects differ significantly between men and women, with men experiencing large, positive, and statistically significant gains while women show no measurable changes. Additionally, the intensive margin effect on weeks worked is also significantly larger for men than for women. These differences underscore meaningful heterogeneity in how anti-discrimination laws affect male and female same-sex couples in the labor market.

\subsection{Wages}

As shown in Table \ref{tab:ext_int}, anti-discrimination laws significantly affect labor force participation and employment outcomes, thereby potentially altering the composition of the employed sample and raising endogeneity concerns for wage analyses. Table \ref{tab:wage} shows regression results on the wage gap, using real hourly wages and annual earnings in 2019 dollars, measured by percentile ranks and computed jointly for men and women with heterogeneous ENDA effects by sex.\footnote{We first compute weighted within-year percentile ranks for real annual earnings and real hourly wages by assigning individuals to weighted percentiles each survey year; we then construct three distributional indicators, set to one when an individual’s percentile rank meets or exceeds the 25th, 50th, and 75th percentiles, respectively.} This robust, scale-invariant metric reduces the impact of outliers and reflects an individual's relative wage position, allowing us to evaluate how anti-discrimination laws shift wage standings for individuals in same-sex couples.\footnote{Following \cite{chen2024logs}, including non-employed individuals in the wage-rank distribution improves transparency by combining extensive- and intensive-margin responses into a single metric. However, this approach does not recover the causal effect of ENDAs on potential wages for all individuals, since ranks for non-employed individuals are mechanically assigned rather than observed wage offers.}

\begin{table}[htbp]
\scriptsize
\centering
\begin{threeparttable}
\caption{Effect of Anti-Discrimination Laws: Wages and Earnings}\label{tab:wage}
\begin{tabular}{l|cccc|cccc}
\toprule
& \multicolumn{4}{c}{Hourly Wage} & \multicolumn{4}{c}{Annual Earnings} \\
\midrule
& (1) & (2) & (3) & (4) & (5) & (6) & (7) & (8) \\
& Percentile & $\ge$25th & $\ge$50th & $\ge$75th 
& Percentile & $\ge$25th & $\ge$50th & $\ge$75th \\
\midrule

ENDA $\times$ SSP: Female  
& -0.3333 & 0.0045 & -0.0117 & -0.0061 
& -0.5994 & 0.0051 & -0.0204 & -0.0016 \\
& (0.8450) & (0.0118) & (0.0143) & (0.0154) 
& (0.8315) & (0.0118) & (0.0134) & (0.0161) \\[4pt]

ENDA $\times$ SSP: Male  
& 2.6310\sym{**} & 0.0366\sym{***} & 0.0326\sym{*} & 0.0232 
& 2.9315\sym{***} & 0.0365\sym{***} & 0.0467\sym{**} & 0.0334\sym{**} \\
& (1.1771) & (0.0139) & (0.0182) & (0.0178)
& (1.1019) & (0.0139) & (0.0185) & (0.0154) \\

\midrule
$p$-value: M $-$ F  
& 0.0379 & 0.0340 & 0.0776 & 0.2093 
& 0.0066 & 0.0382 & 0.0049 & 0.0713 \\
\midrule 
Observations  
& 7,781,358 & 7,781,358 & 7,781,358 & 7,781,358 
& 7,781,358 & 7,781,358 & 7,781,358 & 7,781,358 \\

Mean (F)              &       48.73         &       0.799         &       0.513         &       0.247         &       48.68         &       0.798         &       0.515         &       0.241         \\
Mean (M)              &       51.51         &       0.791         &       0.557         &       0.305         &       51.69         &       0.791         &       0.561         &       0.305         \\

\bottomrule
\end{tabular}

\begin{tablenotes}
\scriptsize
\item Notes: Data are derived from the American Community Survey (ACS) conducted annually between 2005 and 2019. We compare individuals in same-sex partnerships with those in different-sex partnerships within four years of ENDA enactment. The regressions are estimated jointly, allowing the treatment effect to differ for men and women through interaction terms. Columns (1–4) report effects on hourly real wage ranks, while Columns (5–8) report effects on annual real earnings ranks. Columns (2–4) and (6–8) estimate the probability of being above the 25th, 50th, and 75th percentiles of the respective income distributions. All earnings measures are expressed in 2019 dollars.  The ``Mean (F)" and ``Mean (M)" rows report pre-treatment averages of the outcome variables (including never-treated counties) for women and men in same-sex couples, respectively. Reported \(p\)-values test whether the ENDA effect differs significantly between men and women. Standard errors are clustered at the county level. 
\item \sym{*} \(p<0.10\), \sym{**} \(p<0.05\), \sym{***} \(p<0.01\).
\end{tablenotes}

\end{threeparttable}
\end{table}

For men in same-sex couples, the increased employment rates reported in Table \ref{tab:ext_int} are accompanied by significant wage gains. Specifically, anti-discrimination laws boost hourly wages by roughly 2.6 percentiles and annual earnings by 2.9 percentiles on average. Furthermore, these laws raise the probability of ranking above the 25th and 50th percentiles in the hourly wage distribution by 3.7 and 3.3 percentage points, respectively, although the effect at the 75th percentile remains statistically insignificant. In the annual earnings distribution, anti-discrimination laws improve the likelihood of being in the upper tiers by 3.6, 4.7, and 3.3 percentage points at the 25th, 50th, and 75th percentiles, respectively. For women in same-sex couples, the wage distribution remains largely unaffected by anti-discrimination laws. The estimates are statistically insignificant and also small in magnitude—typically well under one percentile point, suggesting limited scope for economically meaningful wage effects. This pattern is consistent with the muted employment responses for women reported in Table \ref{tab:ext_int}.

Joint estimation allows us to formally compare effects across sexes. The $p$-values reported in Table \ref{tab:wage} show that the wage effects for men are significantly larger than those for women at most points in the wage and earnings distributions, reinforcing that anti-discrimination laws generate substantial improvements in labor market standing for men in same-sex couples but have comparatively limited effects for women.

These findings suggest that the wage gains for men in same-sex couples operate primarily through increased employment on the extensive margin. New entrants to the labor force, individuals who were previously non-employed or concentrated at the bottom of the distribution, enter at various points in the wage distribution, shifting the overall distribution upward, particularly around the lower and middle percentiles where such workers are most likely to sort. Specialization models among couples imply that primary earners are typically already located high in the earnings distribution; thus, if ENDAs primarily induce secondary earners to enter work or increase their labor supply, we should expect relatively muted effects at the upper end of the hourly wage distribution. This is precisely what we observe: the 75th percentile of the hourly wage distribution shows no significant response to ENDAs. By contrast, the 75th percentile of the annual earnings distribution rises significantly. While this may appear surprising at first glance, it likely reflects a combination of wage gains, increased hours or weeks worked, and the fact that some new entrants can move into higher annual earnings ranks through expanded labor supply. Overall, the evidence points to selection through secondary earners entering employment and adjusting labor supply, rather than upward movements among already high-earning primary earners.

\subsection{Impact Dynamics}

\begin{figure}[!ht]
\footnotesize
\begin{center}
	\begin{subfigure}{0.9\textwidth}  
		\centering
		\includegraphics[width=0.8\textwidth]{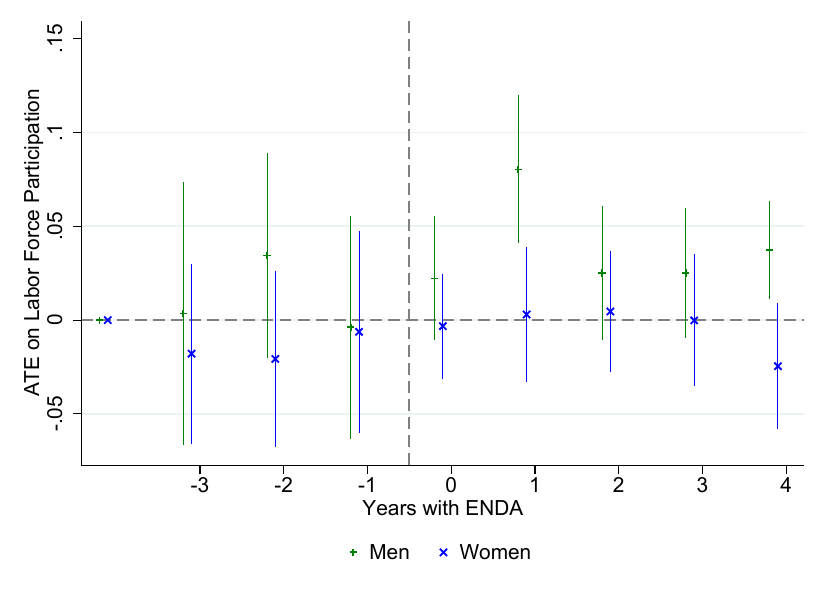}
		\caption{Labor Force Participation}
	\end{subfigure}
 
    \hfill  
    
	\begin{subfigure}{0.9\textwidth}  
		\centering
		\includegraphics[width=0.8\textwidth]{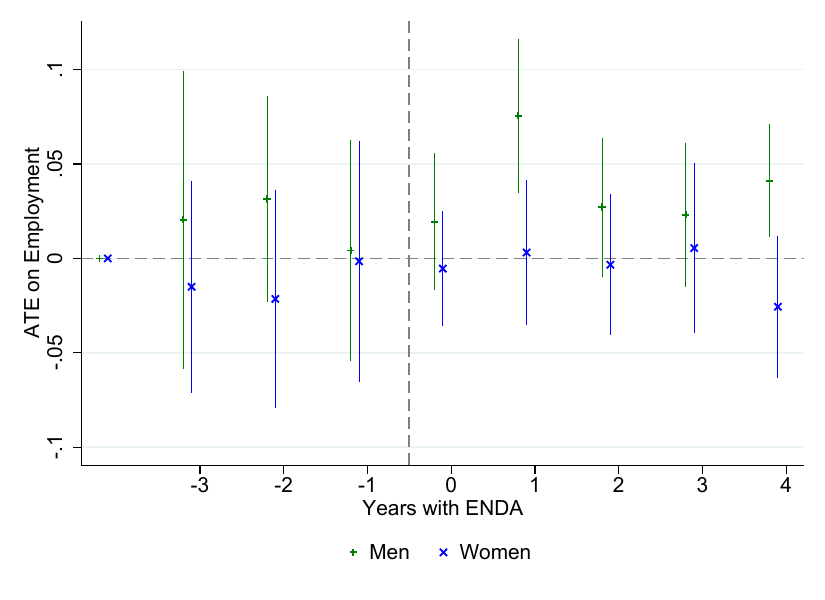}
		\caption{Employment}
	\end{subfigure} \\
\end{center}
	\caption{Labor Supply Impacts on the Extensive Margin}\label{fig:event_ls}   
\begin{flushleft}
\scriptsize{Notes: Event study plots on the difference in labor supply between people in same-sex couples and different-sex couples broken down by sex following the county-level and state-level anti-discrimination laws. Coefficients with 95\% confidence intervals. Standard errors are clustered at the county level. Reference year is -4.}
\end{flushleft} 
\end{figure}

We estimate Equation (1) using event study regressions following \citet{borusyak2021revisiting} to trace the evolution of the impact estimates. Figure \ref{fig:event_ls} displays the results for extensive labor supply outcomes: Panel (a) shows labor force participation rates, while Panel (b) presents employment levels. For men in male same-sex couples, both panels show an immediate rise in labor force participation and employment following ENDA adoption, with the largest effects appearing in the early post-treatment years. While the point estimates decline somewhat over time, our data do not extend far enough to determine whether these effects are temporary or simply leveling off. In contrast, the estimates for women in same-sex couples are small and statistically insignificant, indicating little evidence of a labor-market response. The pre-treatment trends are largely parallel, supporting the plausibility of the identifying assumption, and the post-treatment dynamics again highlight pronounced gender differences in responses to ENDAs.

\begin{figure}[!ht]
\footnotesize
\begin{center}
	\begin{subfigure}{0.9\textwidth}  
		\centering
		\includegraphics[width=0.8\textwidth]{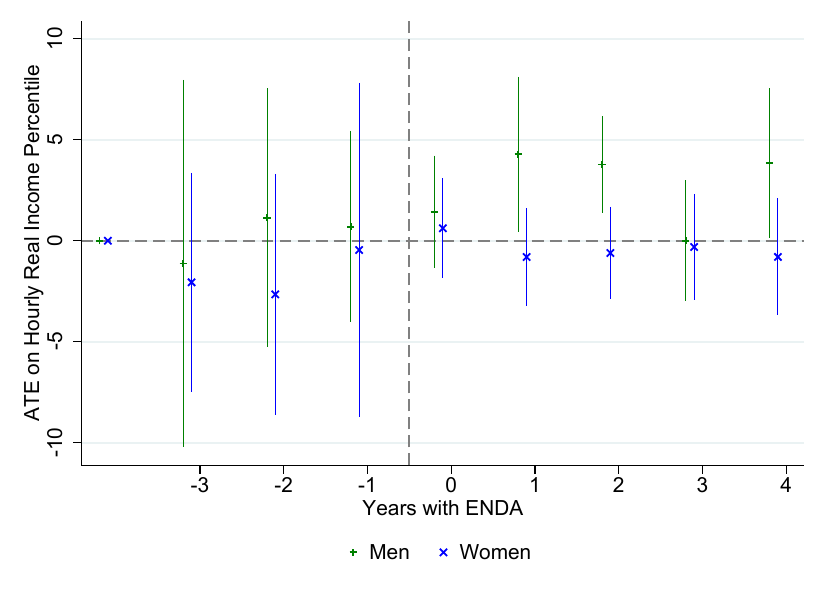}
		\caption{Percentile Rank of Real Hourly Wage}
	\end{subfigure}
 
    \hfill  
    
	\begin{subfigure}{0.9\textwidth}  
		\centering
		\includegraphics[width=0.8\textwidth]{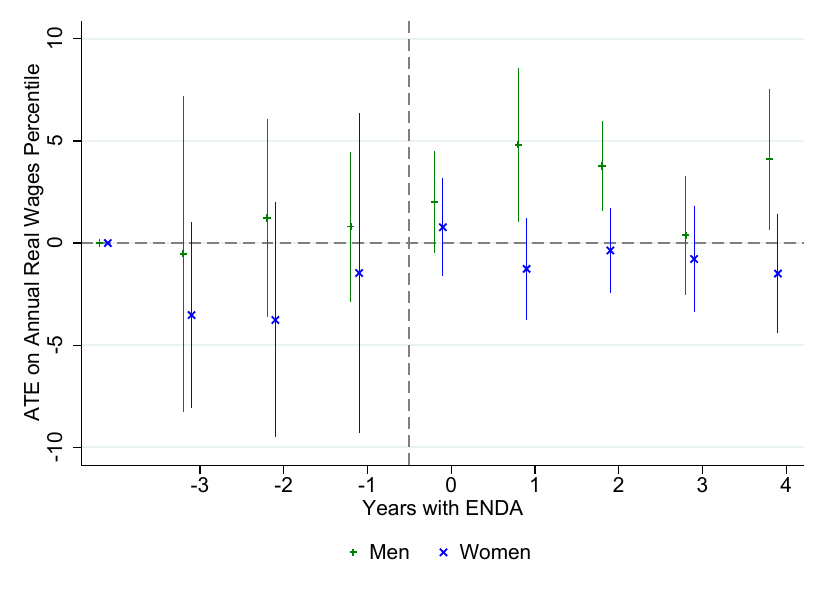}
		\caption{Percentile Rank of Annual Wage}
	\end{subfigure} \\
\end{center}
	\caption{Impacts on Wages}\label{fig:event_wage}   
\begin{flushleft}
\scriptsize{Notes: Event study plot on the difference in pay between people in same-sex couples and different-sex couples broken down by sex following the county-level and state-level anti-discrimination laws. Coefficients with 95\% confidence intervals. Standard errors are clustered at the county level. Reference year is -4.}
\end{flushleft} 
\end{figure}

Figure \ref{fig:event_wage} presents the event study regressions for male and female pay disparities. The results mirror the trends observed in Figure \ref{fig:event_ls}, with a significant wage jump occurring one year after the law's enactment for men in same-sex couples. However, the difference between the one-year and four-year post‑enactment estimates is relatively modest, suggesting that the initial wage improvement for men stabilizes and remains persistent over time. We find no statistically significant effects on wage disparities for women in any post‑ENDA period. These findings indicate that while anti‑discrimination laws yield immediate and lasting wage benefits for men in same‑sex couples, the impact on wage outcomes for women in same-sex couples is insignificant.

\section{Robustness Checks} \label{sec:robustness}

To assess the robustness of our findings and address key threats to identification, we conduct several sensitivity analyses. These tests examine (i) the endogeneity of law adoption using state-level support for same-sex marriage, (ii) potential confounding from same-sex marriage legalization, (iii) migration responses, (iv) metro–non-metro heterogeneity, (v) sensitivity to commuting, and (vi) reporting changes from the ACS redesign.

For exercises (ii)–(vi), we estimate a fully interacted framework that splits men and women into subsample and complementary groups, allowing us to recover ENDA effects for each subgroup and test differences across sex and across samples.

\subsection{Endogenous Adoption of Anti-Discrimination Laws}

Anti-discrimination laws are not randomly distributed. Locally, anti-discrimination laws are concentrated in larger cities, and state laws are to be concentrated in more liberal states that presumably are more accepting of and more favorable to LGB workers. The main concern is that there is an unobservable factor like general sentiment toward LGB workers that affects both the passage of laws as well as the labor market outcomes for LGB workers. In our main regressions, we alleviate this concern by using county-year fixed effects.

Nonetheless, we investigate the endogenous adoption of anti-discrimination laws at the state level. Specifically, we use state-level polling information on support for same-sex marriage as a proxy for general sentiment toward LGB workers. It is possible that controlling for state-level polling is not the best way to capture sentiment toward LGB workers since it is possible to discriminate against people based on sexual orientation and still support their right to marry. However, it seems plausible that the changes in state-level support for same-sex marriage are highly correlated with changes in sentiment toward LGB workers such that it will suffice for a suitable proxy. To better get at the question of endogenous adoption of laws, we create an event-study plot showing how state laws change support for same-sex marriages. Specifically, we estimate this equation:

\begin{equation}
\begin{aligned}
\text{Support}_{st} 
&=
\sum_{\substack{j=-4 \\ j \neq -4}}^{4} \psi_j \, \mathbbm{1}\Bigl(\text{ENDA}_{st}=j\Bigr)
+ \phi_s
+ \delta_t
+ \epsilon_{ts}.
\end{aligned}
\end{equation}

\begin{figure}[!ht]
\centering
\includegraphics[width=0.7\textwidth]{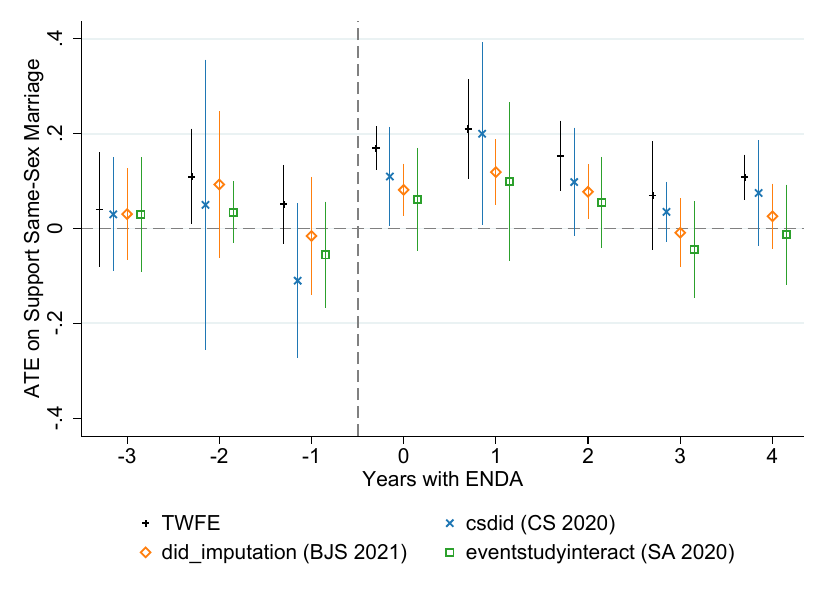}
\caption{Impacts on Same-Sex Marriage Support}
\begin{flushleft}
\scriptsize{Notes: Event study plot showing how support for same-sex marriage changes following state-level anti-discrimination laws. Polling data comes from Pew Polling and encompasses 2005-2017 for every state except Hawaii and Alaska, which are missing for 2005-2008. Standard errors are clustered at the state level.}
\label{fig:event_support}
\end{flushleft} 
\end{figure}

Support$_{st}$ denotes the level of support for same-sex marriage in state $s$ during year $t$, while $\mathbbm{1}(\text{ENDA}_{st}=j)$ is an indicator variable equal to 1 if state $s$ has had a sexual orientation anti-discrimination law for $j$ years by time $t$. We estimate the coefficients $\psi_j$, which capture the effect of anti-discrimination laws on support for same-sex marriage, using an event study framework within a difference-in-differences design. This strategy relies on the identifying assumption that, in the absence of anti-discrimination laws, support for same-sex marriage in states that passed these laws would have followed parallel trends to those in states that did not. Figure \ref{fig:event_support} presents the event study plot.

The event study indicates that state-level anti-discrimination laws significantly boost support for same-sex marriage. Prior to the legislation, trends in support moved in parallel across states regardless of their legal stance, suggesting that, absent the laws, these trajectories would have remained similar. Although one might argue that the passage of such laws was driven by a pre-existing increase in support for same-sex marriage, an endogenous explanation would predict a gradual rise in support leading up to the law's enactment, with little change at the moment of passage. Instead, we observe a distinct, significant jump in support coinciding with the implementation of the legislation. This pattern implies that the legislative changes themselves are likely driving the increase in support for same-sex marriage, and possibly improving overall attitudes toward LGB individuals.

Moreover, the observed jump in state-level sentiment toward LGB workers following the passage of anti-discrimination laws may reflect an unobserved corresponding increase in local sentiment. This local-level change could act as a mechanism through which these laws impact the labor supply decisions of LGB workers.

\subsection{Marriage Laws}\label{subsec:marriage}

Same-sex marriage legalization may confound our identification strategy. For example, Utah legalized same-sex marriage in 2015, the same year it passed its statewide sexual orientation anti-discrimination bill, while Iowa and Arizona experienced similar overlaps (with Iowa enacting its anti-discrimination law in 2007 and legalizing same-sex marriage in 2009, and Arizona in 2014). Prior studies have shown that same-sex marriage legalization can differentially affect employment outcomes for same-sex couples. To address this issue, we included fixed effects that interact same-sex couple status with state-level same-sex marriage legalization. Nevertheless, to further isolate the impact of anti-discrimination laws, we restrict our sample to periods before same-sex marriage became legal in each state (e.g., excluding observations from 2008 onward for states that legalized in 2008). 

\begin{table}[htbp]
\scriptsize
\centering
\begin{threeparttable}
\caption{Effects of Anti-Discrimination Laws on Labor Supply: Pre–Same-Sex Marriage Legalization}\label{tab:ext_int_ssm}
\begin{tabular}{l|cc|cccc}
\toprule
& (1) & (2) & (3) & (4) & (5) & (6) \\
& \multicolumn{2}{c}{\underline{Extensive Margin}} &
\multicolumn{4}{c}{\underline{Intensive Margin}} \\
& Labor Force & Employed &
Weekly Hours & Weeks Worked &
Weekly Hours & Weeks Worked \\
\midrule

\multicolumn{7}{l}{\underline{Pre–SSM Legalization}} \\[2pt]

ENDA $\times$ SSP: Female  
& -0.0157 & -0.0215 & 1.1714\sym{*} & -0.6654 & 1.0376\sym{***} & -1.2330\sym{***} \\
& (0.0153) & (0.0158) & (0.6726) & (0.7928) & (0.3682) & (0.3997) \\[4pt]

ENDA $\times$ SSP: Male  
& 0.0283 & 0.0311 & 0.8892 & 0.4945 & -0.1655 & -0.7909\sym{**} \\
& (0.0186) & (0.0193) & (0.6512) & (0.8727) & (0.4310) & (0.3950) \\[4pt]
\midrule 
$p$: M -- F (Pre–SSM)  
& 0.0186 & 0.0202 & 0.7304 & 0.2214 & 0.0192 & 0.4193 \\

$p$: M (Pre–SSM -- Post-SSM)  
& 0.1196 & 0.7317 & 0.0293 & 0.0048 & 0.0000 & 0.0004 \\

$p$: F (Pre–SSM -- Post-SSM)  
& 0.0702 & 0.0039 & 0.0024 & 0.0270 & 0.0299 & 0.0125 \\
\midrule 
Observations  
& 7,779,108 & 7,779,108 & 7,779,108 & 7,779,108 & 5,763,735 & 5,763,735 \\

Mean (F)  
&       0.817         &       0.774         &       34.44         &       39.13         &       41.28         &       46.85         \\

Mean (M)  
&       0.818         &       0.778         &       35.39         &       39.44         &       42.40         &       47.27         \\

\bottomrule
\end{tabular}

\begin{tablenotes}
\scriptsize
\item Notes: Data are derived from the American Community Survey (ACS) conducted annually between 2005 and 2019. We compare individuals in same-sex partnerships with those in different-sex partnerships within four years of ENDA enactment. All regressions in this table are estimated jointly with a fully interacted model that allows ENDA effects to vary across four groups: men and women, each in the pre–same-sex-marriage and post–same-sex-marriage periods. This framework nests the separate-by-sex estimations and enables formal tests of equality across groups. We report the ENDA effects estimated on the Pre–SSM Legalization sample and include $p$-values testing whether these coefficients differ from those obtained in the complementary Post–SSM sample. Columns (1–2) focus on the extensive margin, examining changes in labor force participation and employment. Columns (3–6) analyze intensive-margin outcomes, usual weekly hours and weeks worked, where Columns (5–6) restrict the sample to employed individuals. The ``Mean (F)" and ``Mean (M)" rows report pre-treatment averages of the outcome variables (including never-treated counties) for women and men in same-sex couples in the Pre-SSM sample, respectively. Reported $p$-values test differences across sex and across robustness conditions. Standard errors are clustered at the county level. \sym{*} \(p<0.10\), \sym{**} \(p<0.05\), \sym{***} \(p<0.01\).
\end{tablenotes}

\end{threeparttable}
\end{table}

Table \ref{tab:ext_int_ssm} presents results from the jointly estimated model, reporting ENDA effects separately for the pre–same-sex-marriage subsample while also providing $p$-values testing whether these effects differ significantly from those in the complementary post-legalization period. Although this approach removes nearly half of the same-sex partnership observations and reduces statistical power, the qualitative patterns remain robust. For men in same-sex couples, the point estimates for both labor force participation and employment remain comparable to the baseline, exceeding 3 percentage points, with labor force participation retaining statistical significance. The employment coefficient declines from 0.037 to 0.0311 and becomes borderline insignificant, consistent with the loss of power, and similar attenuations appear on the intensive margin. For women, the intensive-margin estimates are slightly larger in the restricted sample, though still small in magnitude and statistically weak.

The pre–post $p$-value tests reported in Table \ref{tab:ext_int_ssm} indicate that several coefficients differ across the two periods, particularly on the intensive margin. Still, the qualitative pattern of the ENDA effects is preserved: the signs of the coefficients remain stable, and the largest impacts continue to appear on the extensive margin for men and on the intensive margin for women. Moreover, even within the pre–SSM subsample, the male–female comparison on the extensive margin remains statistically significant and positive, reinforcing that men in same-sex couples experience systematically larger gains in labor force participation and employment. Thus, while same-sex marriage legalization may shift the magnitudes of some estimates, it does not overturn the core conclusion that anti-discrimination laws improve labor market outcomes for individuals in same-sex couples. This confirms that our identification strategy is not driven solely by concurrent same-sex marriage policies.

\subsection{Migration}\label{subsec:migration}

Next, we account for the possibility that migration patterns could influence our results. If LGB individuals were disproportionately moving to areas with recently enacted anti-discrimination laws, it could bias the treatment effects. To address this concern, we restrict our sample to individuals who had not moved in the past year and re-estimate the models presented in Table \ref{tab:ext_int}. By focusing on a more stable population, we can better isolate the effects of anti-discrimination laws on labor outcomes. The corresponding estimates, reported in Table \ref{tab:ext_int_migration}, remain consistent with the full sample results, providing further robustness and suggesting that migration does not significantly alter our conclusions. Interestingly, the estimates for men in same-sex couples are slightly larger for this subset, which may indicate that the positive employment effects are even more pronounced among those who are less mobile. 

\begin{table}[htbp]
\scriptsize
\centering
\begin{threeparttable}
\caption{Effects of Anti-Discrimination Laws on Labor Supply: Analysis of Non-Movers}\label{tab:ext_int_migration}
\begin{tabular}{l|cc|cccc}
\toprule
& (1) & (2) & (3) & (4) & (5) & (6) \\
& \multicolumn{2}{c}{\underline{Extensive Margin}} &
\multicolumn{4}{c}{\underline{Intensive Margin}} \\
& Labor Force & Employed &
Weekly Hours & Weeks Worked &
Weekly Hours & Weeks Worked \\
\midrule

\multicolumn{7}{l}{\underline{Non-Movers}} \\[2pt]

ENDA $\times$ SSP: Female  
& 0.0050 & 0.0048 & 1.1290\sym{*} & 0.1140 & 0.7276\sym{**} & -0.5031 \\
& (0.0142) & (0.0147) & (0.6189) & (0.7471) & (0.3084) & (0.3944) \\[4pt]

ENDA $\times$ SSP: Male  
& 0.0421\sym{***} & 0.0424\sym{***} & 1.2112\sym{**} & 1.7705\sym{**} & -0.0583 & 0.2447 \\
& (0.0145) & (0.0153) & (0.5690) & (0.7291) & (0.4024) & (0.2941) \\[4pt]

\midrule 
$p$: M -- F (Non-Movers) &      0.0209         &      0.0412         &      0.9143         &      0.0503         &      0.1284         &      0.1084         \\ 
$p$: M (Non-Movers -- Movers)&      0.3954         &      0.2811         &      0.6663         &      0.5295         &      0.9676         &      0.4559         \\
$p$: F (Non-Movers -- Movers)&      0.0504         &      0.0763         &      0.2704         &      0.0013         &      0.2691         &      0.0146         \\

\midrule

Observations  
& 7,781,169 & 7,781,169 & 7,781,169 & 7,781,169 & 5,765,391 & 5,765,391 \\

Mean (F)          &       0.815         &       0.783         &       34.34         &       39.57         &       41.32         &       47.55         \\
Mean (M)         &       0.823         &       0.791         &       35.44         &       39.99         &       42.45         &       47.89         \\
\bottomrule
\end{tabular}

\begin{tablenotes}
\scriptsize
\item Notes: Data are derived from the American Community Survey (ACS) conducted annually between 2005 and 2019. We compare individuals in same-sex partnerships with those in different-sex partnerships within four years of ENDA enactment. All regressions in this table are estimated jointly with a fully interacted model that allows ENDA effects to vary across four groups: men and women, each in the mover subsample and in the non-mover subsample. This framework nests the separate-by-sex estimations and enables formal tests of equality across groups. We report the ENDA effects estimated for non-movers and include $p$-values testing whether these coefficients differ from those obtained in the complementary sample of recent movers. Columns (1–2) focus on the extensive margin, examining changes in labor force participation and employment. Columns (3–6) analyze intensive-margin outcomes, usual weekly hours and weeks worked, where Columns (5–6) restrict the sample to employed individuals. The ``Mean (F)" and ``Mean (M)" rows report pre-treatment averages of the outcome variables (including never-treated counties) for women and men in same-sex couples in the non-mover sample, respectively. Reported $p$-values test differences across sex and across the mover versus non-mover samples. Standard errors are clustered at the county level. \sym{*} \(p<0.10\), \sym{**} \(p<0.05\), \sym{***} \(p<0.01\).
\end{tablenotes}

\end{threeparttable}
\end{table}

The joint-estimation framework, which interacts ENDA exposure with both gender and migration status (male–mover, male–non-mover, female–mover, female–non-mover), also allows us to test whether the ENDA effects differ between movers and non-movers. For men, the $p$-value tests show no statistically significant differences across the two samples, reinforcing that migration does not meaningfully drive the male results. Among women, several coefficients do differ between movers and non-movers, but the overall qualitative pattern of effects is preserved and most estimates remain small or statistically insignificant. Importantly, the male–female comparison on the extensive margin remains statistically significant within the non-mover subsample, indicating that the gender differences highlighted in the main specification are not driven by migration behavior.

\subsection{Metropolitan Areas}\label{subsec:metro}

We re-estimate our model specifically for residents in metropolitan areas to assess whether the effects of anti-discrimination laws vary by geographic context. Table \ref{tab:ext_int_metro} shows that, within metro areas, the qualitative patterns for men in same-sex couples remain consistent with the main results: extensive-margin effects are positive and statistically significant, and the joint-estimation framework again shows that these effects are significantly larger for men than for women.

\begin{table}[htbp]
\scriptsize
\centering
\begin{threeparttable}
\caption{Effects of Anti-Discrimination Laws on Labor Supply: Analysis of Metro Areas}\label{tab:ext_int_metro}
\begin{tabular}{l|cc|cccc}
\toprule
& (1) & (2) & (3) & (4) & (5) & (6) \\
& \multicolumn{2}{c}{\underline{Extensive Margin}} &
\multicolumn{4}{c}{\underline{Intensive Margin}} \\
& Labor Force & Employed &
Weekly Hours & Weeks Worked &
Weekly Hours & Weeks Worked \\
\midrule

\multicolumn{7}{l}{\underline{Metro Areas}} \\[2pt]

ENDA $\times$ SSP: Female  
& -0.0262\sym{**} & -0.0284\sym{**} & -0.4601 & -1.3804\sym{**} & -0.4479 & -1.3355\sym{***} \\
& (0.0115) & (0.0129) & (0.7244) & (0.6511) & (0.6693) & (0.3471) \\[4pt]

ENDA $\times$ SSP: Male  
& 0.0352\sym{**} & 0.0398\sym{**} & 0.8139 & 1.4503\sym{*} & -0.4661 & -0.2430 \\
& (0.0159) & (0.0162) & (0.5418) & (0.7403) & (0.3735) & (0.3455) \\[4pt]

\midrule 
$p$: M -- F (Metro)  
& 0.0005 & 0.0007 & 0.1692 & 0.0013 & 0.9819 & 0.0248 \\ 

$p$: M (Metro -- Non-Metro)  
& 0.6882 & 0.5923 & 0.6851 & 0.2116 & 0.6688 & 0.0463 \\

$p$: F (Metro -- Non-Metro)  
& 0.0263 & 0.0621 & 0.0065 & 0.1878 & 0.0019 & 0.1576 \\

\midrule

Observations  
& 7,781,334 & 7,781,334 & 7,781,334 & 7,781,334 & 5,765,559 & 5,765,559 \\

Mean (F)           &       0.834         &       0.801         &       35.20         &       40.42         &       41.43         &       47.52         \\
Mean (M)           &       0.847         &       0.812         &       36.50         &       41.09         &       42.62         &       47.96         \\

\bottomrule
\end{tabular}

\begin{tablenotes}
\scriptsize
\item Notes: Data are derived from the American Community Survey (ACS) conducted annually between 2005 and 2019. We compare individuals in same-sex partnerships with those in different-sex partnerships within four years of ENDA enactment. All regressions in this table are estimated jointly with a fully interacted model that allows ENDA effects to vary across four groups: men and women, each in the metro subsample and in the non-metro subsample. This framework nests the separate-by-sex estimations and enables formal tests of equality across groups. We report the ENDA effects estimated for the metro subsample and include $p$-values testing whether these coefficients differ from those in the non-metro population. Columns (1–2) focus on the extensive margin, examining labor force participation and employment. Columns (3–6) analyze intensive-margin outcomes, usual weekly hours and weeks worked, with Columns (5–6) restricted to employed individuals.  The ``Mean (F)" and ``Mean (M)" rows report pre-treatment averages of the outcome variables (including never-treated counties) for women and men in same-sex couples in the metro-area sample, respectively. Standard errors are clustered at the county level. \sym{*} \(p<0.10\), \sym{**} \(p<0.05\), \sym{***} \(p<0.01\).
\end{tablenotes}

\end{threeparttable}
\end{table}

For women in same-sex couples, the metro-area estimates become negative across several outcomes, including labor force participation, employment, and weeks worked. The joint model further indicates that the coefficients for women differ significantly between metro and non-metro samples, whereas for men no such differences are detected. One interpretation is that urban labor markets, combined with higher childcare costs, may amplify intra-household specialization, leading some women to reduce labor supply when partners experience greater employment stability. The metro-area analysis reinforces the gender differences observed in the main results: male extensive-margin gains remain robust, while female responses vary substantially by geography.

\subsection{Residence - Work Mismatch}\label{subsec:commuting}

We next examine whether commuting patterns moderate the effects of anti-discrimination laws. In particular, we distinguish individuals who both live and work in the same county (“non-commuters”) from those who commute across county lines. Table~\ref{tab:ext_int_commute} reports the estimates for non-commuters based on the fully interacted joint-estimation framework.

\begin{table}[htbp]
\scriptsize
\centering
\begin{threeparttable}
\caption{Effects of Anti-Discrimination Laws on Labor Supply: Analysis of Work \& Home Residence}\label{tab:ext_int_commute}
\begin{tabular}{l|cc|cccc}
\toprule
& (1) & (2) & (3) & (4) & (5) & (6) \\
& \multicolumn{2}{c}{\underline{Extensive Margin}} &
\multicolumn{4}{c}{\underline{Intensive Margin}} \\
& Labor Force & Employed &
Weekly Hours & Weeks Worked &
Weekly Hours & Weeks Worked \\
\midrule

\multicolumn{7}{l}{\underline{Work–Residence County Matched (Same County)}} \\[2pt]

ENDA $\times$ SSP: Female  
& -0.0020 & -0.0041 & 1.3169\sym{**} & -0.2497 & 1.1835\sym{***} & -0.6671 \\
& (0.0133) & (0.0158) & (0.5689) & (0.7315) & (0.3187) & (0.4222) \\[4pt]

ENDA $\times$ SSP: Male  
& 0.0262\sym{**} & 0.0231\sym{*} & 0.5135 & 1.0258\sym{*} & -0.4456 & -0.1132 \\
& (0.0133) & (0.0139) & (0.5081) & (0.6102) & (0.3420) & (0.3172) \\[4pt]

\midrule 

$p$: M -- F (Matched)  
& 0.0858 & 0.2049 & 0.2584 & 0.1563 & 0.0011 & 0.2927 \\

$p$: M (Matched -- Mismatched)  
& 0.0490 & 0.0960 & 0.0809 & 0.0271 & 0.2188 & 0.3734 \\

$p$: F (Matched -- Mismatched)  
& 0.8780 & 0.7938 & 0.0155 & 0.3999 & 0.0291 & 0.5921 \\

\midrule

Observations  
& 7,781,313 & 7,781,313 & 7,781,313 & 7,781,313 & 5,765,525 & 5,765,525 \\

Mean (F)         &       0.779         &       0.736         &       32.70         &       37.46         &       40.87         &       46.77         \\
Mean (M)         &       0.785         &       0.741         &       33.69         &       37.83         &       42.07         &       47.23         \\

\bottomrule
\end{tabular}

\begin{tablenotes}
\scriptsize
\item Notes: Data are derived from the American Community Survey (ACS) conducted annually between 2005 and 2019. We compare individuals in same-sex partnerships with those in different-sex partnerships within four years of ENDA enactment. All regressions in this table are estimated jointly with a fully interacted model that allows ENDA effects to vary across four groups: men and women, each in the commuter subsample and in the non-commuter subsample. This framework nests the separate-by-sex estimations and enables formal tests of equality across groups. We report ENDA effects for the matched sample (work county = residence county) and include p-values comparing these estimates to those obtained for individuals with a work–home county mismatch. Columns (1–2) focus on the extensive margin, examining labor force participation and employment. Columns (3–6) analyze intensive-margin outcomes, usual weekly hours and weeks worked, where Columns (5–6) restrict the sample to employed individuals.  The ``Mean (F)" and ``Mean (M)" rows report pre-treatment averages of the outcome variables (including never-treated counties) for women and men in same-sex couples in the matched sample, respectively. Standard errors are clustered at the county level. \sym{*} \(p<0.10\), \sym{**} \(p<0.05\), \sym{***} \(p<0.01\).
\end{tablenotes}

\end{threeparttable}
\end{table}

Among non–commuters, the qualitative patterns for men in same–sex couples are broadly consistent with the main results: ENDA generates positive and statistically significant effects on labor force participation and employment (Columns~1–2), while intensive–margin effects are more modest. For women, extensive–margin coefficients remain close to zero, but we find sizable and statistically significant increases in usual weekly hours. The joint model shows that, within the non–commuter subsample, gender differences on the extensive margin are at most marginally significant, whereas women exhibit significantly larger ENDA effects than men on weekly hours among the employed. These marginal or insignificant gender differences likely reflect reduced precision due to the substantial sample restriction rather than substantive shifts in underlying patterns.

The $p$–values comparing non–commuters to the commuter sample indicate that men’s ENDA effects change somewhat across commuting regimes on both the extensive and intensive margins, but the signs and overall pattern remain stable. For women, by contrast, the intensive–margin responses in hours differ more sharply between commuters and non–commuters. 

\subsection{Reporting Changes}\label{subsec:reporting}

Prior research has highlighted higher rates of misclassification errors before 2008, where different-sex couples were sometimes incorrectly identified as same-sex couples due to typos in the sex variable. In response, and considering the 2008 ACS redesign aimed at reducing such errors \citep{o2009changes}, we restrict our sample to 2008–2019 to test the robustness of our findings reported in Table~\ref{tab:ext_int}. Table~\ref{tab:ext_int_2008} presents the estimates based on this restricted sample, using the joint estimation framework.

\begin{table}[htbp]
\scriptsize
\centering
\begin{threeparttable}
\caption{Effects of Anti-Discrimination Laws on Labor Supply: ACS Redesign (2008–2019)}\label{tab:ext_int_2008}
\begin{tabular}{l|cc|cccc}
\toprule
& (1) & (2) & (3) & (4) & (5) & (6) \\
& \multicolumn{2}{c}{\underline{Extensive Margin}} &
\multicolumn{4}{c}{\underline{Intensive Margin}} \\
& Labor Force & Employed &
Weekly Hours & Weeks Worked &
Weekly Hours & Weeks Worked \\
\midrule

\multicolumn{7}{l}{\underline{ACS Redesign Sample (2008–2019)}} \\[2pt]

ENDA $\times$ SSP: Female  
& 0.0217\sym{*} & 0.0272 & 1.3482\sym{**} & 0.6313 & 0.5888 & -0.1525 \\
& (0.0129) & (0.0195) & (0.6583) & (0.7446) & (0.4430) & (0.4733) \\[4pt]

ENDA $\times$ SSP: Male  
& 0.0459\sym{***} & 0.0492\sym{***} & 1.5759\sym{**} & 3.1075\sym{***} & -0.2548 & 0.8835\sym{***} \\
& (0.0140) & (0.0156) & (0.7054) & (0.8309) & (0.7122) & (0.3363) \\[4pt]

\midrule 
$p$: M -- F (ACS Redesign)  
& 0.1635 & 0.3301 & 0.8272 & 0.0116 & 0.3877 & 0.0347 \\

$p$: M (ACS Redesign -- Rest)  
& 0.7712 & 0.6221 & 0.5605 & 0.0837 & 0.8041 & 0.0165 \\

$p$: F (ACS Redesign -- Rest)  
& 0.1277 & 0.0633 & 0.7380 & 0.1901 & 0.7515 & 0.0070 \\

\midrule

Observations  
& 7,778,320 & 7,778,320 & 7,778,320 & 7,778,320 & 5,763,114 & 5,763,114 \\

Mean (F)        &       0.824         &       0.787         &       34.27         &       39.72         &       41.09         &       47.51         \\
Mean (M)        &       0.837         &       0.802         &       35.66         &       40.48         &       42.41         &       48.11         \\

\bottomrule
\end{tabular}

\begin{tablenotes}
\scriptsize
\item Notes: Data are derived from the American Community Survey (ACS) conducted between 2005 and 2019, following the major ACS redesign that altered survey implementation and imputation procedures. All regressions are estimated jointly using a fully interacted model that allows ENDA effects to vary across men and women within the redesign subsample. We report the ENDA effects estimated for the 2008–2019 sample and include $p$-values testing whether (i) male and female coefficients differ within the redesign sample, and (ii) redesign coefficients differ from those estimated in the 2005–2007 sample. Columns (1–2) report extensive-margin outcomes (labor force participation and employment). Columns (3–6) report intensive-margin outcomes (usual weekly hours and weeks worked), with Columns (5–6) restricted to employed individuals. The ``Mean (F)" and ``Mean (M)" rows report pre-treatment averages of the outcome variables (including never-treated counties) for women and men in same-sex couples in the ACS redesign sample (2008-2019), respectively. Standard errors are clustered at the county level. \sym{*} \(p<0.10\), \sym{**} \(p<0.05\), \sym{***} \(p<0.01\).
\end{tablenotes}

\end{threeparttable}
\end{table}

For men in same-sex couples, the results remain qualitatively similar to our full-sample analysis: ENDA continues to generate positive and statistically significant effects on labor force participation and employment, with magnitudes close to those in Table~\ref{tab:ext_int}. Among women, extensive-margin coefficients remain small, with the labor-force estimate marginally significant, and the employment estimate statistically insignificant. The intensive-margin results are broadly consistent with the full sample, though with some variation in magnitude.

The joint model also allows us to test whether the 2008–2019 coefficients differ from those estimated in the 2005–2007 sample. For men, most outcomes do not differ significantly across samples, though the effect on weeks worked does. For women, we detect significant differences in employment and in weeks worked among the employed. The male–female comparison on the extensive margin is not statistically significant in the ACS Redesign subsample, although the differences remain qualitatively similar to those in the full sample.

Taken together, these results indicate that potential misclassification errors in earlier ACS waves do not drive our main conclusions. The consistency between the full and redesign-restricted samples reinforces the robustness of our empirical strategy and the validity of our conclusions regarding the impact of anti-discrimination laws on labor market outcomes for same-sex couples.

\section{Discussion} \label{sec:discussion}

\subsection{Differences in Response by Sex}

The estimates reveal meaningful heterogeneity by sex within same-sex partnerships. For men in male same-sex couples, ENDA adoption significantly reduces the labor supply differential relative to men in different-sex partnerships. By contrast, we do not detect statistically significant average effects for women in female same-sex partnerships. Importantly, the absence of an average effect for women does not imply that ENDAs only affect men or that a single mechanism must operate symmetrically across groups. Our heterogeneity tests show that the extensive-margin responses for men and women differ significantly: men exhibit clear increases in labor supply, whereas the corresponding female estimates are near zero (and negative in metro areas). The appropriate interpretation is therefore one of heterogeneous adjustment, not opposing effects from a unified mechanism. Men respond on labor-market margins, while women appear to adjust through channels outside those margins, such as fertility decisions, with subsequent evidence also showing increased specialization for both male and female households.

A natural explanation draws on household models of specialization. If ENDAs reduce discrimination and increase job security for sexual minority workers, households may respond by reallocating labor between market and non-market activities. Prior work documents that reduced discrimination can strengthen the position of the primary earner within sexual minority households (\cite{badgett2006discrimination}). In collective and bargaining models of the household, an improvement in one partner’s labor market opportunities increases the returns to specialization: the partner with stronger market prospects supplies more labor, while the other reallocates time toward household production (\cite{chiappori1992collective,lundberg1993separate}).

Such mechanisms can generate heterogeneous responses across male and female same-sex households. Female same-sex couples are substantially more likely to have children than male same-sex couples, and thus may be more sensitive to changes in expected job stability or income that facilitate childcare. A shift toward specialization in female households may therefore involve adjustments in fertility and household production, rather than intensive-margin labor supply. To test these mechanisms directly, we estimate a household-level difference-in-differences model comparing male and female same-sex partnerships:

\vspace{-1.5\baselineskip} 
\begin{equation}
\begin{aligned}
Y_{jct}
&=
\alpha
\;+\;
\beta_{1}
\bigl[ \mathrm{FemSSP}_{j} \times \mathrm{ENDA}_{ct} \bigr]
\;+\;
\beta_{2} \, \mathrm{FemSSP}_{j}
\;+\;
\beta_{3} \, \mathrm{ENDA}_{ct}
+ X_{jct}' \delta 
+ \mu_c 
+ \delta_t 
+ \epsilon_{jct}
\end{aligned}
\end{equation}

Here \(\beta_1\) captures how female same-sex households respond relative to male same-sex households. We examine three sets of outcomes: (i) whether the household is a one-earner family, (ii) the absolute difference in weekly hours worked between partners (capturing intra-household specialization), and (iii) fertility outcomes.

Table~\ref{tab:mvf_ssp} shows that ENDAs do not affect the probability of being a one-earner household, but they significantly increase intra-household differences in hours worked for both male and female same-sex households. The interaction term is negative because male households experience a larger increase in specialization; combining the ENDA main effect and the interaction yields a positive, though smaller, specialization response for female households, consistent with the specialization hypothesis.

\begin{table}[htbp]
\centering
\begin{threeparttable}
\scriptsize
\caption{Responses between Male and Female Same-Sex Partnerships}\label{tab:mvf_ssp}%
\begin{tabular}{l|cc|cc}
\toprule
      & (1)   & (2)   & (3)   & (4) \\
 & One Earner HH & Diff in Hours Worked & Any Children & Number of Children \\
\midrule
      &       &       &       &  \\
ENDA $\times$ Female &      0.0107         &     -1.0573\sym{*}  &      0.0338\sym{***}&      0.0562\sym{*}  \\
                    &    (0.0086)         &    (0.5452)         &    (0.0121)         &    (0.0303)         \\
ENDA   &     -0.0201         &      1.6692\sym{***}&     -0.0056         &     -0.0289         \\
                    &    (0.0133)         &    (0.4773)         &    (0.0148)         &    (0.0329)         \\
Female &      0.0029         &     -0.8066\sym{***}&      0.1428\sym{***}&      0.2273\sym{***}\\
                    &    (0.0061)         &    (0.2587)         &    (0.0068)         &    (0.0157)         \\ 
\midrule
Observations        &  41,382       &  41,382       &  41,382       &  41,382       \\
Mean (F)            &      0.2585         &     26.8599         &      0.3460         &      0.6036         \\
$R^2$               &      0.0534         &      0.1349         &      0.1306         &      0.1216         \\
\bottomrule
\end{tabular}%
\begin{tablenotes}
\scriptsize
\item Notes: Data is derived from the Annual Community Surveys (ACS) spanning 2005-2019 comparing women in same-sex partnerships to men. The first column examines if the household has only one earner. The second column tests the difference in absolute terms of hours worked between the two partners. The third and fourth columns examine how households differ in having children. The coefficients show the differential effect of anti-discrimination between female and male same-sex households. The ``Mean (F)" row reports pre-treatment averages of the outcome variables (including never-treated counties) for female same-sex couple households. Standard errors are clustered at the county level.  \sym{*} \(p<0.10\), \sym{**} \(p<0.05\), \sym{\sym{***}} \(p<0.01\).
\end{tablenotes}
\end{threeparttable}
\end{table}

Female same-sex households also exhibit significant increases in fertility, both in the likelihood of having any child and in the number of children, relative to male households. These results suggest that ENDA-induced improvements in job security may allow female same-sex households to realize fertility preferences and subsequently adjust labor supply to accommodate additional childcare needs.

Taken together, these findings show that male and female same-sex households respond differently to anti-discrimination laws not because the laws are ineffective for women, but because the relevant channels of adjustment differ. Male same-sex households respond primarily through increased labor-market specialization, while female same-sex households adjust fertility and household production. If female households value these additional children more than forgone labor market hours, these responses may reflect higher welfare even in the absence of average improvements in women's labor market outcomes.

\subsection{Sorting and Partnership Formation} \label{subsec:sorting}

A remaining concern is whether ENDA adoption changes who appears in the same-sex partnered sample. If ENDAs affect partnership formation, cohabitation decisions, or relationship stability, then the composition of same-sex partnerships could shift endogenously with treatment. Such changes would complicate interpretation of the triple-differences estimates, since the observed sample would no longer represent the same underlying population before and after ENDA adoption. To assess this possibility, we examine whether the probability of being observed in a same-sex partnership changes following ENDA adoption, and we additionally test for sorting on observable demographics. Specifically we estimate:

\vspace{-0.5\baselineskip} 
\begin{equation}
\begin{aligned}
\mathrm{Y}_{ict}
&=
\beta \, \mathrm{ENDA}_{ct}
+ \mu_{c}
+ \delta_{t}
+ \epsilon_{ict},
\end{aligned}
\end{equation}

where $\mathrm{Y}_{ict}$ is an indicator for living in a same-sex partnership, age, or years of education. $\mu_c$ are county fixed effects, and $\delta_t$ are year fixed effects. Our core specification tests whether the fraction of adults observed in same-sex partnerships shifts systematically in treated counties relative to untreated counties. Using age and years of education as outcomes provides a direct check for demographic sorting: if ENDAs systematically attract (or repel) younger or more educated individuals, we would observe corresponding shifts in treated counties. 

Table~\ref{tab:sort} reports the results. Across all specifications, and separately for men and women, we find no evidence that ENDAs affect the likelihood of living in a same-sex partnership. Coefficients are small, precisely estimated, and statistically insignificant. These findings indicate that ENDA adoption does not meaningfully alter partnership formation or cohabitation patterns.

\begin{table}[htbp]
\centering
\begin{threeparttable}
\scriptsize
\caption{Sorting and Reporting}\label{tab:sort}%
\begin{tabular}{l|ccccc}
\midrule
      & (1)   & (2)   & (3) & (4) & (5)\\
 & All SSP & Male SSP & Female SSP & Age & Years of Education \\
\midrule
      &       &       &  \\
ENDA    &     -0.0002         &      0.0004         &     -0.0008         &      0.0257         &     -0.0212         \\
                    &    (0.0005)         &    (0.0005)         &    (0.0005)         &    (0.0648)         &    (0.0182)         \\
\midrule                    
Observations        &16,071,420        &7,850,821        &8,220,599        &16,071,420        &16,071,420        \\
Mean          &      0.0093         &      0.0087         &      0.0099         &     45.5421         &     13.6441         \\
$R^2$               &      0.0043         &      0.0081         &      0.0024         &      0.0101         &      0.0467         \\
\bottomrule
\end{tabular}%
\begin{tablenotes}
\scriptsize
\item Notes: Data comes from the 2005-2019 yearly ACS seeing how the number of same-sex partnerships change in a county after the passage of an anti-discrimination law. The first column looks at both men and women with the next two columns separating the sexes. Column (4) investigates age and (5) looks at years of education. The ``Mean" row reports pre-treatment averages of the outcome variables (including never-treated counties) for same-sex couples. Standard errors are clustered at the county level.  \sym{*} \(p<0.10\), \sym{**} \(p<0.05\), \sym{\sym{***}} \(p<0.01\).
\end{tablenotes}
\end{threeparttable}
\end{table}

The absence of sorting responses strengthens the interpretation of the main results. Because ENDAs do not appear to change who enters or exits same-sex partnerships, the coupled-sample restriction does not generate treatment-induced compositional bias. Thus, the heterogeneous labor supply responses documented above are best understood as behavioral adjustments within stable households, rather than as artifacts of changes in relationship structure or partner selection.

\subsection{Mechanisms}

The results suggest that sexual-orientation anti-discrimination laws can meaningfully reduce labor-market gaps for men in male same-sex couples. One interpretation is that these laws are binding and help limit discrimination. At the same time, enforcing such protections is fundamentally different from enforcing protections based on race or sex, since sexual orientation is not always observable. These laws may become more binding when individuals are in publicly identifiable same-sex relationships, but less so otherwise.\footnote{This idea is consistent with research on the saliency of protected class membership, such as \cite{armour2018disability}, which shows that anti-discrimination protections are more effective when group membership is more readily perceived.} Thus, while the evidence points to important effects for male same-sex couples, the extent to which anti-discrimination laws eliminate broader forms of sexual-orientation discrimination remains uncertain.

One mechanism that the laws could be affecting labor market outcomes is through increased positive sentiment toward LGB workers. We show in Figure \ref{fig:event_support} that following state anti-discrimination laws there is a significant and persistent increase in the percentage of people in that state that support same-sex marriage. Our regression results already condition on the change in the county-level support for same-sex marriage via fixed effects. It seems plausible that local law changes could significantly change local sentiment and impact the labor supply and pay of LGB workers. Changing public sentiment in conjunction with greater protection for LGB workers seems more plausible as the mechanism than simply greater protection in the workplace. 

\subsection{Threats to External Validity}

A primary limitation of this study is that we cannot observe the full LGB population. The analysis compares individuals in same-sex partnerships to those in different-sex partnerships, but the ACS does not identify single LGB workers, whose labor market outcomes may differ systematically from partnered LGB individuals. Partnership status is also an imperfect proxy for sexual orientation: bisexual individuals in different-sex relationships are classified as heterosexual, and some gay or lesbian workers may similarly appear in the control group. Because lesbian women tend to earn more than heterosexual women, gay men tend to earn less than heterosexual men, and bisexual workers earn less than both \citep{mize2016sexual}, such misclassification does not shift group means in a uniform direction. While the bias in level pay gaps is generally toward zero, the sign of the bias in the estimated ENDA coefficients is theoretically ambiguous—especially if bisexual workers respond differently to anti-discrimination protections than gay or lesbian workers. Misclassification that places gay or lesbian individuals into the control group reduces the contrast between groups and thus produces the usual attenuation mechanism. These issues highlight the need for future work that can distinguish gay, lesbian, and bisexual workers more precisely to avoid masking heterogeneity across subpopulations.

A related concern is that individuals in same-sex partnerships may constitute the subset of LGB workers most vulnerable to discrimination. Single LGB workers may be more able to conceal their sexual orientation in the workplace, whereas partnered individuals face greater visibility. The effects documented here may therefore reflect treatment effects for the group most exposed to discrimination, potentially overstating impacts for the broader LGB population.

Crucially, these limitations pertain to external validity rather than internal validity. Misclassification of sexual orientation does not affect the within-county, within-sex comparisons that identify $\beta_{1g}$, but it does restrict the generalizability of the results to groups not directly observed in the data—such as single LGB workers or bisexual individuals in different-sex partnerships. How ENDA protections translate to these populations depends on their baseline outcomes, patterns of disclosure, and exposure to discrimination, none of which can be observed in the ACS.

Finally, the ACS does not allow us to identify transgender or nonbinary individuals, nor does it permit analysis of anti-discrimination laws that specifically protect gender minorities. This further limits generalizability to the broader LGBTQ+ community.

\section{Conclusion} \label{sec:conclusion}
The policy environment has evolved considerably in recent years. Prior to \emph{Bostock v. Clayton County} (2020), many states allowed employees to be fired solely based on sexual orientation. Although \emph{Bostock} extended federal Title VII protections to LGBTQI+ workers, new anti-discrimination statutes have stagnated at both the state and federal levels. The Equality Act, reintroduced in the 118th Congress on June 21, 2023, remains pending in the Senate, with critics voicing concerns related to religious liberties and women’s rights. Meanwhile, more states have chosen to preempt or restrict local anti-discrimination ordinances than those that have adopted new protections, reflecting persistent polarization.

Against this backdrop, our study offers quasi-experimental evidence on how local and state-level anti-discrimination laws shape labor market outcomes for individuals in same-sex couples. Exploiting a novel city-level panel dataset that captures the granular evolution of these laws, we document significant reductions in sexual orientation-based labor market disparities. In particular, for men in male same-sex couples, our estimates indicate a 3.8 percentage-point (78\%) decrease in the labor force participation gap and a 3.7 percentage-point (77\%) decline in the employment gap relative to different-sex counterparts, whereas we find no significant changes for women in female same-sex couples.

Beyond labor supply, our results reveal meaningful wage improvements for men in same-sex couples. Their average percentile rank in hourly wages and annual earnings increases by about 2.6 and 2.9 points, respectively, with a 3.7 percentage-point increase in the likelihood of surpassing the 25th wage percentile and a 3.3 percentage-point boost for exceeding the 50th percentile. These outcomes underscore the capacity of anti-discrimination laws to mitigate wage gaps rooted in sexual orientation, with no comparable impact observed for women in female same-sex couples.

We identify increased positive societal sentiments toward LGB Americans, reflected in growing support for same-sex marriage, as a potential mechanism by which these laws improve labor market outcomes. Taken together, our results suggest that local and state-level anti-discrimination measures can significantly mitigate labor market disadvantages faced by men in same-sex couples. By illustrating how these laws reduce labor supply and wage gaps, we offer timely insight into ongoing policy debates and judicial deliberations about extending employment protections to sexual minorities, paving the way for future research on the broader implications of such measures.

\singlespacing
\bibliographystyle{plainnat}
\bibliography{ref}
\nocite{*}

\doublespacing

\clearpage 

\appendix
\addcontentsline{toc}{section}{Appendices}
\counterwithin{figure}{section}
\counterwithin{table}{section}

\section{County Identification from PUMAs}

The ACS reports Public Use Microdata Areas (PUMAs) rather than counties. We assign respondents to counties using the IPUMS PUMA--county crosswalk. In practice, metro PUMAs map nearly one-to-one to counties: 475 counties in our sample are uniquely identified. Rural PUMAs aggregate multiple counties within a state; in these cases, we follow ACS public-use geography and treat each rural PUMA as a single county-group. This yields 48 county-groups, for a total of 523 county (or county-group) units. All ENDA codes are constructed at the county (or county-group)~$\times$~year level using this mapping.

\section{Using Only State Laws}

Local ordinances provide meaningful identifying variation. Among the 332 counties in our sample that ever receive ENDA-type protection, 133 (about 40 percent) adopt a county-level ordinance before their state enacts a statewide ENDA. A state-only measure would classify all residents in these county–years as untreated, which affects roughly 30 percent of post-treatment ACS person-year observations. Ignoring county ordinances therefore induces non-classical measurement error in the treatment variable and mechanically attenuates the estimated effects.

Table \ref{tab:ext_int_state} illustrates this attenuation by removing local-law variation. In Panel A, which replaces county fixed effects with state fixed effects, point estimates shrink relative to our baseline (Table \ref{tab:ext_int}), and, importantly, the male–female differences on the extensive margin become statistically insignificant. Panel B, which keeps the original fixed-effects structure but forces ENDA to vary only at the state level, shows a similar pattern: some effects shrink towards zero, and the male–female gap is less precisely estimated.

\begin{table}[htbp]
\scriptsize
\centering
\begin{threeparttable}
\caption{Effect of Anti-Discrimination Laws: Extensive \& Intensive Margin of Labor Supply (State Laws)}\label{tab:ext_int_state}
\begin{tabular}{l|cc|cccc}
\toprule
& (1) & (2) & (3) & (4) & (5) & (6) \\
& \multicolumn{2}{c}{\underline{Extensive Margin}} 
& \multicolumn{4}{c}{\underline{Intensive Margin}} \\
& Labor Force & Employed 
& Weekly Hours & Weeks Worked 
& Weekly Hours & Weeks Worked \\
\midrule
\multicolumn{7}{c}{\textbf{Panel A: State-Fixed Effects}} \\
\midrule

ENDA $\times$ SSP: Female                 
&      0.0040         &     -0.0089         &      1.5019\sym{**} &     -0.3407         &      1.0750\sym{***}&     -1.4526\sym{***}\\
                    &    (0.0136)         &    (0.0189)         &    (0.5864)         &    (0.7646)         &    (0.2932)         &    (0.3630)         \\[4pt]

ENDA $\times$ SSP: Male                  
   &      0.0270\sym{**} &      0.0243\sym{*}  &      0.2947         &     -0.1546         &     -0.2827         &     -1.0244\sym{***}\\
                    &    (0.0125)         &    (0.0125)         &    (0.3015)         &    (0.5411)         &    (0.3488)         &    (0.3585)         \\

\midrule
$p$-value: M $-$ F  
   &      0.2831         &      0.2062         &      0.0793         &      0.8527         &      0.0053         &      0.3204         \\

\midrule
\multicolumn{7}{c}{\textbf{Panel B: County-Fixed Effects}} \\
\midrule

ENDA $\times$ SSP: Female         
&     -0.0067         &     -0.0158         &      1.5380\sym{***}&     -0.3971         &      1.2360\sym{***}&     -1.3461\sym{***}\\
                    &    (0.0106)         &    (0.0140)         &    (0.5051)         &    (0.6383)         &    (0.3126)         &    (0.3535)         \\ [4pt]

ENDA $\times$ SSP: Male                 
  &      0.0326\sym{**} &      0.0321\sym{*}  &      0.6748\sym{**} &      0.2175         &     -0.2064         &     -0.7743\sym{**} \\
                    &    (0.0153)         &    (0.0169)         &    (0.3292)         &    (0.5922)         &    (0.5124)         &    (0.3419)         \\

\midrule
$p$-value: M $-$ F  
&      0.0294         &      0.0754         &      0.1967         &      0.5485         &      0.0117         &      0.1645         \\
\midrule 
Employed Only  &   &   &   &   & X & X \\
Observations   & 7,781,358 & 7,781,358 & 7,781,358 & 7,781,358 & 5,765,580 & 5,765,580 \\
Mean (F)              &       0.818         &       0.782         &       34.42         &       39.51         &       41.25         &       47.27         \\
Mean (M)              &       0.825         &       0.790         &       35.53         &       39.96         &       42.41         &       47.69         \\
\bottomrule
\end{tabular}

\begin{tablenotes}
\scriptsize
\item Notes: Data is derived from the American Community Survey (ACS) conducted annually between 2005 and 2019. We compare individuals in same-sex partnerships with those in different-sex partnerships within four years of ENDA enactment. For this table, we abstract from local laws and use only state-level ENDA variation. The regressions are estimated jointly, allowing the treatment effect to differ for men and women through interaction terms. Columns (1–2) focus on the extensive margin, examining changes in labor force participation and employment. Columns (3–6) explore the intensive margin using variation in weeks worked per year and usual weekly hours; Columns (5–6) restrict the sample to employed individuals. The ``Mean (F)" and ``Mean (M)" rows report pre-treatment averages of the outcome variables (including never-treated counties) for women and men in same-sex couples, respectively. Reported \(p\)-values test whether the ENDA effect differs significantly between men and women. Panel A replaces county fixed effects with state fixed effects in Equation (1), while Panel B reports the original specification. Standard errors are clustered at the state-level. \sym{*} \(p<0.10\), \sym{**} \(p<0.05\), \sym{***} \(p<0.01\).
\end{tablenotes}

\end{threeparttable}
\end{table}

These results confirm that omitting local ordinances discards meaningful within-state variation and leads to substantial attenuation of the estimated ENDA effects, particularly on the extensive margin where the gender differences weaken considerably.

\section{Additional Robustness Checks}\label{sec:app}

\subsection{Placebo Tests and Alternative Estimators}\label{sec:app_alternative}

\begin{figure}[!ht]
\footnotesize
\begin{center}
    \centering
    \begin{subfigure}{0.475\textwidth}
        \centering
        \includegraphics[width=\textwidth]{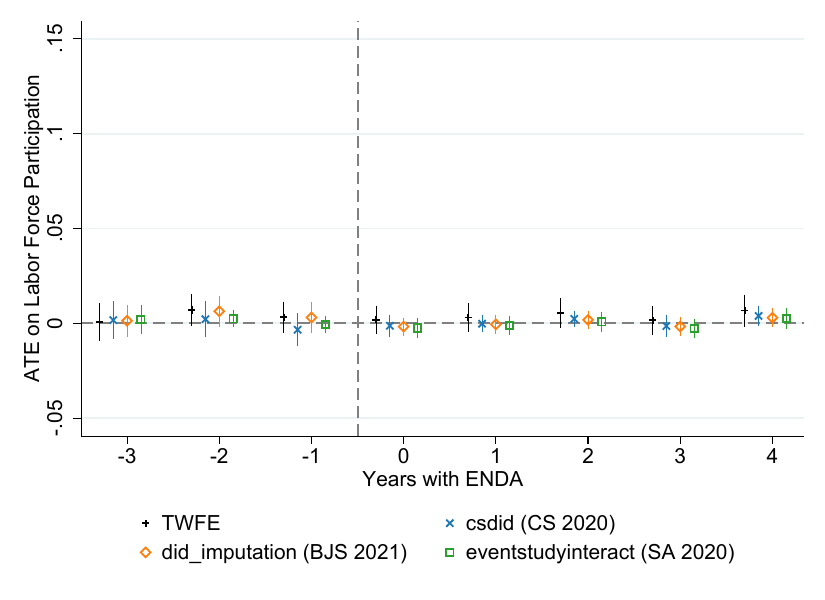}
        \caption{Labor Force Participation (Female Different-Sex Couples)}
    \end{subfigure}
    \hfill
    \begin{subfigure}{0.475\textwidth}
        \centering
        \includegraphics[width=\textwidth]{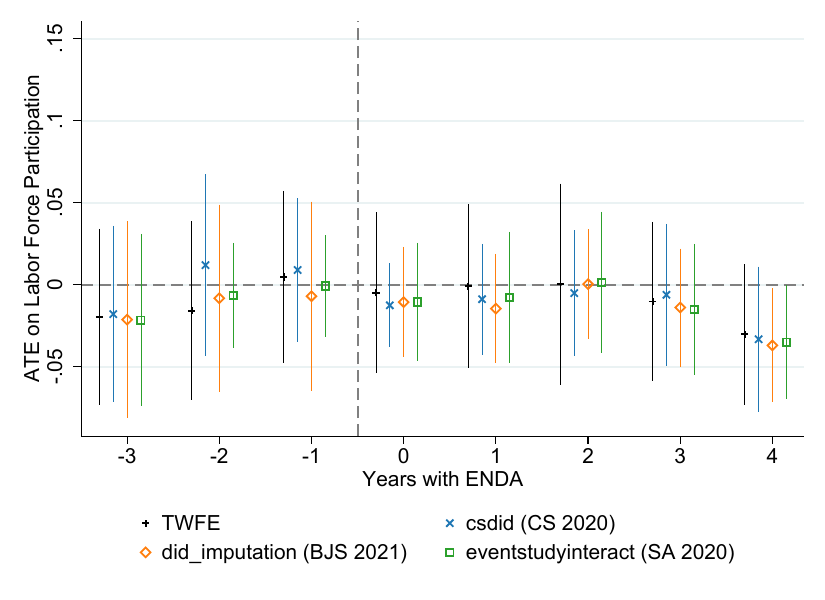}
        \caption{Labor Force Participation (Female Same-Sex Couples)}
    \end{subfigure} \\

    \begin{subfigure}{0.475\textwidth}
        \centering
        \includegraphics[width=\textwidth]{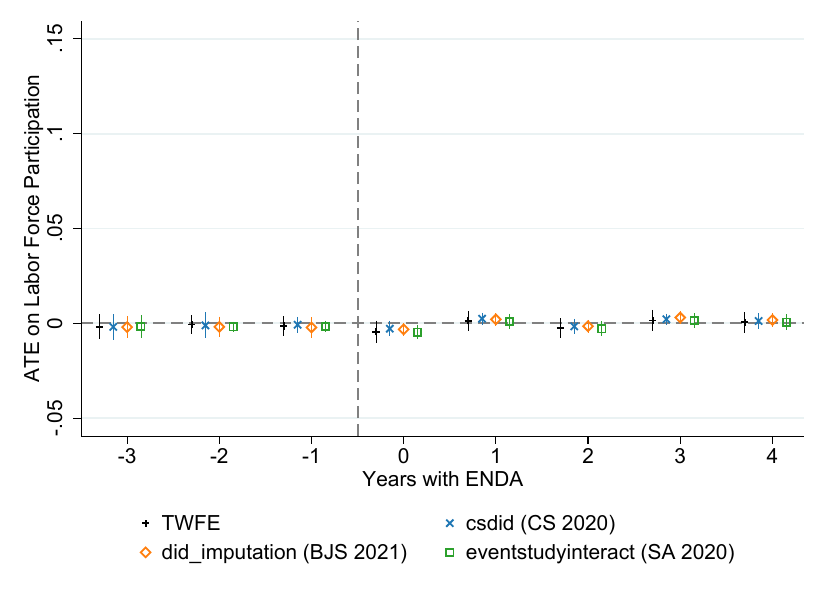}
        \caption{Labor Force Participation (Male Different-Sex Couples)}
    \end{subfigure}
    \hfill
    \begin{subfigure}{0.475\textwidth}
        \centering
        \includegraphics[width=\textwidth]{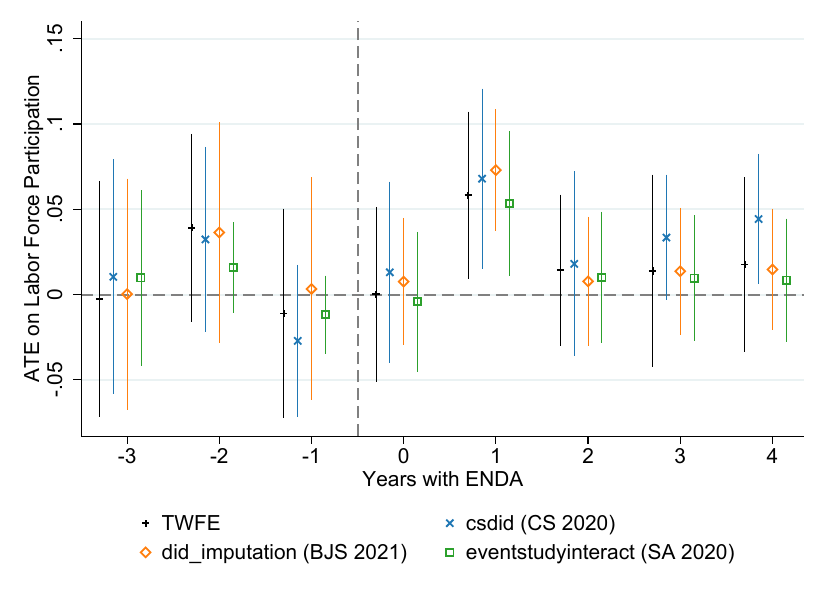}
        \caption{Labor Force Participation (Male Same-Sex Couples)}
    \end{subfigure}
\end{center}
    \caption{Impacts on Labor Force Participation by Partnership Type} \label{fig:lf_placebo}
\begin{flushleft}
\footnotesize{Notes: Event study plots on the effects of ENDAs on people in same-sex couples and different-sex couples broken down by sex following the county-level and state-level anti-discrimination laws. Coefficients with 95\% confidence intervals. Standard errors are clustered at the county level. Reference year is -4.}
\end{flushleft} 
\end{figure}

\begin{figure}[!ht]
\footnotesize
\begin{center}
    \centering
    \begin{subfigure}{0.475\textwidth}
        \centering
        \includegraphics[width=\textwidth]{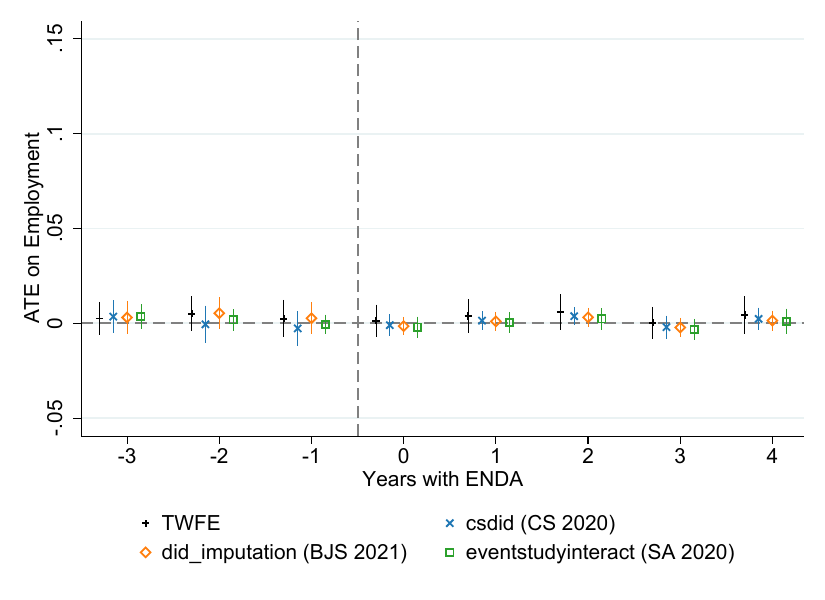}
        \caption{Employment (Female Different-Sex Couples)}
    \end{subfigure}
    \hfill
    \begin{subfigure}{0.475\textwidth}
        \centering
        \includegraphics[width=\textwidth]{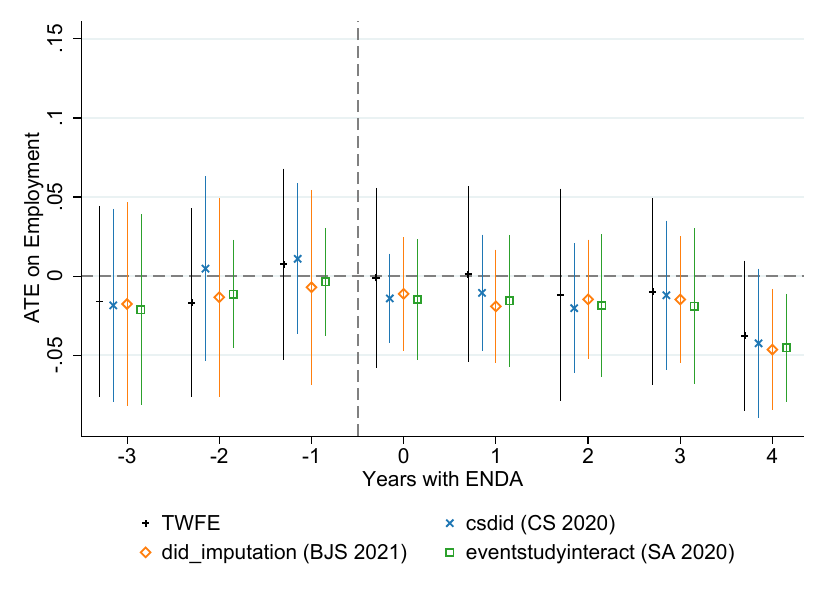}
        \caption{Employment (Female Same-Sex Couples)}
    \end{subfigure} \\

    \begin{subfigure}{0.475\textwidth}
        \centering
        \includegraphics[width=\textwidth]{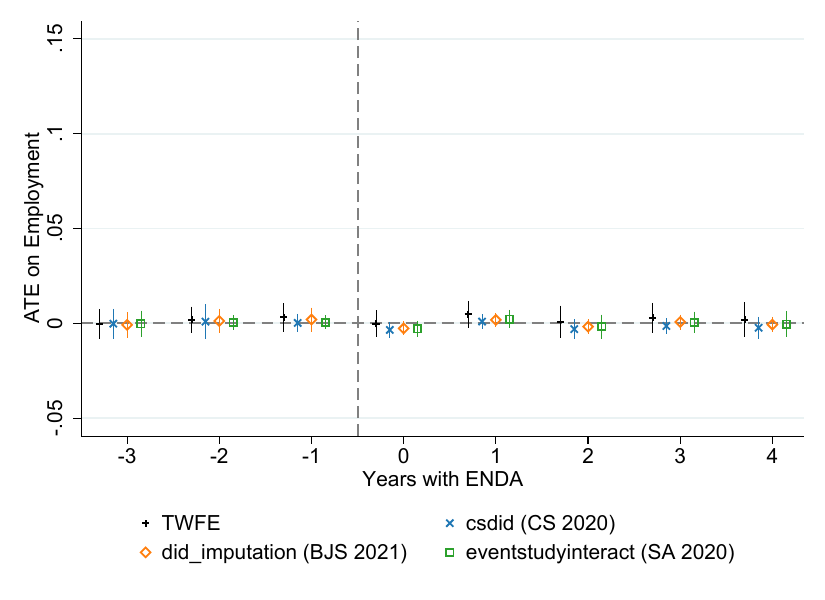}
        \caption{Employment (Male Different-Sex Couples)}
    \end{subfigure}
    \hfill
    \begin{subfigure}{0.475\textwidth}
        \centering
        \includegraphics[width=\textwidth]{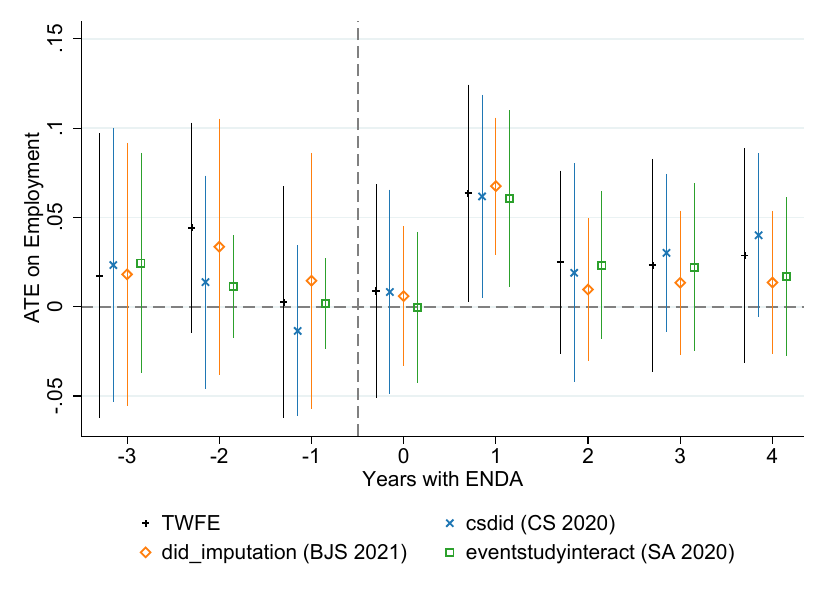}
        \caption{Employment (Male Same-Sex Couples)}
    \end{subfigure}
\end{center}
    \caption{Impacts on Employment by Partnership Type} \label{fig:employed_placebo}
\begin{flushleft}
\footnotesize{Notes: Event study plots on the effects of ENDAs on people in same-sex couples and different-sex couples broken down by sex following the county-level and state-level anti-discrimination laws. Coefficients with 95\% confidence intervals. Standard errors are clustered at the county level. Reference year is -4.}
\end{flushleft} 
\end{figure}

We next conduct a series of placebo tests and apply alternative estimators to assess the impact of anti‑discrimination laws on labor market outcomes among both same‑sex and different‑sex couples. Specifically, we utilize four distinct methods to assess the impact of anti‑discrimination laws on labor market outcomes among same‑sex and different‑sex couples: TWFE, \cite{callaway2021difference}, \cite{sun2021estimating}, and \cite{borusyak2021revisiting}. As a placebo test, we apply the DiD framework to different‑sex couples, who are not the target of these laws, and thus we expect no significant changes in their labor market outcomes. Panel (a) of Tables \ref{tab:ext_int_men} and \ref{tab:ext_int_women} report the DiD estimates for different‑sex couples, while Panels (a) and (c) of Figures \ref{fig:lf_placebo} and \ref{fig:employed_placebo} display the corresponding event study plots.

Our placebo tests show that the DiD estimates yield no statistically significant impacts on labor force participation or employment rates for different‑sex couples. The pre‑ENDA coefficients are insignificant, confirming the absence of differential trends prior to the law's enactment. Most post‑ENDA coefficients remain statistically insignificant, with only a few minor exceptions of small magnitude, largely consistent with a 5\% false positive rate. 

Next, Panel (b) of Tables \ref{tab:ext_int_men} and \ref{tab:ext_int_women} presents the impact of anti‑discrimination laws on same‑sex couples, while Panels (b) and (d) of Figures \ref{fig:lf_placebo} and \ref{fig:employed_placebo} display the corresponding event studies. In these analyses, we compare labor market outcomes for same‑sex couples in regions that have enacted the laws with those in regions that have not. The results reveal that male same‑sex couples in areas with anti‑discrimination laws experience significant improvements in labor market outcomes, particularly through increased labor force participation and higher employment rates. For female same‑sex couples, however, we observe no significant effects on the extensive margin; instead, the positive and significant impact is confined to the intensive margin, as evidenced by an increase in average weekly hours worked. Hence, our alternative estimators yield similar point estimates and comparable confidence intervals, reinforcing the robustness of our findings.

\begin{sidewaystable}[!h]
\footnotesize 
\centering
\begin{threeparttable}
\caption{Effect of Anti-Discrimination Laws on Men in Same-Sex and Different-Sex Partnerships}\label{tab:ext_int_men}%
\begin{tabular}{l|cc|cccc}
\toprule
      & (1)   & (2)   & (3)   & (4) & (5) & (6) \\
& \multicolumn{2}{c}{\underline{Extensive Margin}} & \multicolumn{4}{c}{\underline{Intensive Margin}}\\
ENDA $\times$ SSP Estimates & Labor Force & Employed & Weekly Hours & Weeks Worked & Weekly Hours & Weeks Worked \\
\midrule
\multicolumn{7}{c}{Panel A: Men in Different-Sex Partnerships} \\
\midrule
TWFE     &      0.0004         &     -0.0002         &      0.0629         &      0.0642         &     -0.0022         &     -0.0233         \\
                    &    (0.0015)         &    (0.0018)         &    (0.0834)         &    (0.0809)         &    (0.0602)         &    (0.0448)         \\
\cite{callaway2021difference} &      0.0008         &     -0.0012         &      0.1038         &      0.0596         &      0.0797         &      0.0016         \\
                    &    (0.0013)         &    (0.0017)         &    (0.0961)         &    (0.0817)         &    (0.0725)         &    (0.0595)         \\
\cite{sun2021estimating}   &      0.0008         &     -0.0005         &      0.1160         &      0.0827         &      0.0349         &     -0.0239         \\
            &      (0.0015)         &      (0.0019)         &      (0.0852)         &      (0.0911)       &      (0.0570)         &      (0.0496)         \\ 
\cite{borusyak2021revisiting}      &      0.0023         &      0.0006         &      0.1898\sym{**} &      0.1345         &      0.0554         &     -0.0415         \\
                    &    (0.0014)         &    (0.0020)         &    (0.0879)         &    (0.0958)         &    (0.0552)         &    (0.0465)         \\  
                    \hline 
      &       &       &       &   \\[\dimexpr-\normalbaselineskip+2pt]
Observations       &3,763,442      &3,763,442      &3,763,442      &3,763,442      &3,046,376        &3,046,376        \\
Employed  Only          &                     &                     &                     &                     &           X         &           X         \\
      &       &       &       &  \\[\dimexpr-\normalbaselineskip+2pt]
\midrule
\multicolumn{5}{c}{Panel B: Men in Same-Sex Partnerships} \\
\midrule
TWFE     &      0.0284\sym{**} &      0.0263\sym{*}  &      0.8832         &      1.1614\sym{*}  &     -0.1199         &     -0.2060         \\
                    &    (0.0137)         &    (0.0144)         &    (0.5618)         &    (0.6476)         &    (0.4351)         &    (0.3773)         \\
\cite{callaway2021difference}         &      0.0463\sym{**} &      0.0428\sym{**} &      1.2572\sym{**} &      1.3140         &      0.1736         &     -0.1863         \\
                    &    (0.0184)         &    (0.0213)         &    (0.6368)         &    (0.8613)         &    (0.5461)         &    (0.4090)         \\
\cite{sun2021estimating}  &      0.0337\sym{**}         &      0.0332\sym{**}          &      0.9484\sym{*}          &      1.3165\sym{**}          &     -0.1495         &     -0.1319         \\
          &      (0.0145)         &      (0.0152)         &      (0.5357)         &      (0.6657)         &      (0.3888)         &      (0.3290)         \\
\cite{borusyak2021revisiting}   &      0.0420\sym{***}&      0.0414\sym{***}&      1.3754\sym{***}&      1.7464\sym{***}&     -0.0756         &     -0.1261         \\
                    &    (0.0129)         &    (0.0137)         &    (0.5001)         &    (0.6246)         &    (0.3827)         &    (0.2779)         \\   
                    \hline 
      &       &       &       &   \\[\dimexpr-\normalbaselineskip+2pt]
Observations       &  35,194       &  35,194       &  35,194       &  35,194       &  27,504         &  27,504         \\
Employed  Only          &                     &                     &                     &                     &           X         &           X         \\
      &       &       &       &  \\[\dimexpr-\normalbaselineskip+2pt]
\bottomrule
\end{tabular}%
\begin{tablenotes}
\scriptsize
\item Notes:  Data is derived from the American Community Survey (ACS) conducted annually between 2005 and 2019. We compare men in counties with ENDAs with those without within four years of ENDA enactment. The regressions are estimated separately by partnership: Panel A for men in different-sex partnerships, while Panel B for men in same-sex partnerships. The coefficients indicate the impact of anti-discrimination laws. The analysis is divided into two parts: Columns (1-2) focus on the extensive margin, examining changes in labor force participation and employment status as a result of the laws. Columns (3-6), on the other hand, explore the intensive margin by leveraging variations in the number of weeks worked per year and average weekly working hours. Standard errors are clustered at the county level. Results are shown using six different estimators: Two-Way Fixed Effects (TWFE), \cite{callaway2021difference}, \cite{sun2021estimating}, and \cite{borusyak2021revisiting}. \sym{*} \(p<0.10\), \sym{**} \(p<0.05\), \sym{\sym{***}} \(p<0.01\).
\end{tablenotes}
\end{threeparttable}
\end{sidewaystable}

\begin{sidewaystable}[!h]
\footnotesize 
\centering
\begin{threeparttable}
\caption{Effect of Anti-Discrimination Laws on Women in Same-Sex and Different-Sex Partnerships}\label{tab:ext_int_women}%
\begin{tabular}{l|cc|cccc}
\toprule
      & (1)   & (2)   & (3)   & (4) & (5) & (6) \\
& \multicolumn{2}{c}{\underline{Extensive Margin}} & \multicolumn{4}{c}{\underline{Intensive Margin}}\\
ENDA $\times$ SSP Estimates & Labor Force & Employed & Weekly Hours & Weeks Worked & Weekly Hours & Weeks Worked \\
\midrule
\multicolumn{7}{c}{Panel A: Women in Different-Sex Partnerships} \\
\midrule
TWFE     &      0.0010         &      0.0008         &      0.0093         &      0.0045         &     -0.0882         &     -0.1083\sym{**} \\
                    &    (0.0019)         &    (0.0017)         &    (0.0781)         &    (0.0756)         &    (0.0566)         &    (0.0485)         \\
\cite{callaway2021difference} &      0.0017         &      0.0020         &     -0.0127         &     -0.0105         &     -0.0693         &     -0.0602         \\
                    &    (0.0019)         &    (0.0023)         &    (0.1021)         &    (0.1016)         &    (0.0733)         &    (0.0783)         \\
\cite{sun2021estimating} &      0.0010         &      0.0010         &     -0.0152         &     -0.0159         &     -0.0732         &     -0.0705         \\
           &      (0.0017)         &      (0.0018)         &      (0.0710)         &      (0.0830)         &      (0.0618)         &      (0.0602)         \\
\cite{borusyak2021revisiting}     &      0.0015         &      0.0013         &      0.0057         &     -0.0130         &     -0.0637         &     -0.1021         \\
                    &    (0.0016)         &    (0.0016)         &    (0.0733)         &    (0.0845)         &    (0.0620)         &    (0.0641)         \\
                    \hline 
      &       &       &       &   \\[\dimexpr-\normalbaselineskip+2pt]
Observations      &3,942,881        &3,942,881        &3,942,881        &3,942,881        &2,660,191        &2,660,191        \\
Employed  Only          &                     &                     &                     &                     &           X         &           X         \\
      &       &       &       &  \\[\dimexpr-\normalbaselineskip+2pt]
\midrule
\multicolumn{5}{c}{Panel B: Women in Same-Sex Partnerships} \\
\midrule
TWFE    &      0.0099         &      0.0090         &      1.3686\sym{***}&      0.4999         &      0.5733         &     -0.4873         \\
                    &    (0.0092)         &    (0.0105)         &    (0.4386)         &    (0.5327)         &    (0.3579)         &    (0.3743)         \\
\cite{callaway2021difference}      &     -0.0042         &     -0.0088         &      0.9758         &     -0.6803         &      0.8353\sym{*}  &     -1.2359\sym{**} \\
                    &    (0.0120)         &    (0.0125)         &    (0.7503)         &    (0.6246)         &    (0.4648)         &    (0.4891)         \\
\cite{sun2021estimating}   &      0.0022         &     -0.0009         &      1.1645\sym{**}         &     -0.0565         &      0.5538\sym{*}         &     -0.8987\sym{**}         \\
            &      (0.0103)         &      (0.0109)         &      (0.4920)         &      (0.5785)         &      (0.3148)         &      (0.3529)         \\
\cite{borusyak2021revisiting}     &      0.0000         &     -0.0013         &      1.0359\sym{**} &     -0.2617         &      0.5542\sym{*}  &     -0.9571\sym{***}\\
                    &    (0.0111)         &    (0.0129)         &    (0.4939)         &    (0.5924)         &    (0.2977)         &    (0.3228)         \\ 
                    \hline 
      &       &       &       &   \\[\dimexpr-\normalbaselineskip+2pt]
Observations &  40,071         &  40,071         &  40,071         &  40,071         &  31,707         &  31,707         \\
Employed  Only          &                     &                     &                     &                     &           X         &           X         \\
      &       &       &       &  \\[\dimexpr-\normalbaselineskip+2pt]
\bottomrule
\end{tabular}%
\begin{tablenotes}
\scriptsize
\item Notes:  Data is derived from the American Community Survey (ACS) conducted annually between 2005 and 2019. We compare women in counties with ENDAs with those without within four years of ENDA enactment. The regressions are estimated separately by partnership: Panel A for women in different-sex partnerships, while Panel B for women in same-sex partnerships. The coefficients indicate the impact of anti-discrimination laws. The analysis is divided into two parts: Columns (1-2) focus on the extensive margin, examining changes in labor force participation and employment status as a result of the laws. Columns (3-6), on the other hand, explore the intensive margin by leveraging variations in the number of weeks worked per year and average weekly working hours. Standard errors are clustered at the county level. Results are shown using six different estimators: Two-Way Fixed Effects (TWFE), \cite{callaway2021difference}, \cite{sun2021estimating}, and \cite{borusyak2021revisiting}. \sym{*} \(p<0.10\), \sym{**} \(p<0.05\), \sym{\sym{***}} \(p<0.01\).
\end{tablenotes}
\end{threeparttable}
\end{sidewaystable}

\clearpage 

\subsection{Same-Sex Marriage Fixed Effects}

The results in Table~\ref{tab:ext_int_no_ssm} reinforce the robustness of our primary findings for men. Although the extensive-margin effects for men in same-sex couples are slightly lower (0.0267 for labor force participation and 0.0270 for employment) than in our primary specification, they remain statistically significant, with generally smaller intensive-margin estimates. For women, the point estimates on the extensive margin are quite similar in Table~\ref{tab:ext_int_no_ssm} and Table~\ref{tab:ext_int}, but the employment effect turns negative and is marginally significant ($-0.0212$, $p<0.10$), which contrasts with the mostly null extensive-margin effects observed elsewhere, aside from the negative responses noted in metropolitan areas. Moreover, the joint model indicates that the ENDA effects on the extensive margin are significantly larger for men than for women, as reflected in the $p$-values reported in Columns~(1) and (2). This conflicting evidence for women suggests that further research is needed to better understand the mechanisms at play; our hypothesis remains that high child-care costs in metropolitan areas may be an important contributing factor.

\begin{table}[htbp]
\scriptsize
\centering
\begin{threeparttable}
\caption{Effects of Anti-Discrimination Laws on Labor Supply: Analysis Without Same-Sex Marriage Interactions}\label{tab:ext_int_no_ssm}
\begin{tabular}{l|cc|cccc}
\toprule
& (1) & (2) & (3) & (4) & (5) & (6) \\
& \multicolumn{2}{c}{\underline{Extensive Margin}} 
& \multicolumn{4}{c}{\underline{Intensive Margin}} \\
& Labor Force & Employed 
& Weekly Hours & Weeks Worked 
& Weekly Hours & Weeks Worked \\
\midrule

ENDA $\times$ SSP: Female  
& -0.0130 & -0.0212\sym{*} & 0.8477 & -0.7081 & 0.9893\sym{***} & -0.7657\sym{**} \\
& (0.0108) & (0.0115) & (0.5167) & (0.5762) & (0.3083) & (0.3165) \\[4pt]

ENDA $\times$ SSP: Male  
& 0.0267\sym{**} & 0.0270\sym{**} & 0.5377 & 0.8852 & -0.4480 & -0.3068 \\
& (0.0129) & (0.0133) & (0.4987) & (0.6178) & (0.3627) & (0.2820) \\

\midrule

$p$-value: M $-$ F  
& 0.0042 & 0.0034 & 0.6404 & 0.0259 & 0.0035 & 0.2688 \\
\midrule 

Employed Only  
&  &  &  &  & X & X \\

Observations  
& 7,781,493 & 7,781,493 & 7,781,493 & 7,781,493 & 5,765,693 & 5,765,693 \\

Mean (F)              &       0.818         &       0.782         &       34.42         &       39.51         &       41.25         &       47.27         \\
Mean (M)              &       0.825         &       0.790         &       35.53         &       39.96         &       42.41         &       47.69         \\

\bottomrule
\end{tabular}

\begin{tablenotes}
\scriptsize
\item Notes: Data are derived from the American Community Survey (ACS), 2005–2019. We compare individuals in same-sex partnerships with those in different-sex partnerships within four years of ENDA enactment. The regressions are jointly estimated, allowing ENDA effects to differ by sex. This table excludes the fully interacted same-sex-marriage term from Equation~(1). Columns (1–2) report extensive-margin effects; Columns (3–6) report intensive-margin effects, with Columns (5–6) restricted to employed individuals.  The ``Mean (F)" and ``Mean (M)" rows report pre-treatment averages of the outcome variables (including never-treated counties) for women and men in same-sex couples, respectively. Reported \(p\)-values test equality of male and female coefficients. Standard errors are clustered at the county level. \sym{*} \(p<0.10\), \sym{**} \(p<0.05\), \sym{***} \(p<0.01\).
\end{tablenotes}

\end{threeparttable}
\end{table}

\subsection{Occupation Fixed Effects}

LGB workers tend to sort to different occupations and industries based on sex and sexual orientation \citep{black2007economics}. They show that gay men tend to sort to more female-dominated fields and lesbian women sort to more male-dominated fields relative to their heterosexual counterparts. It is typical to not consider those factors because discrimination affects occupation and industry choice, which would make occupation and industry choice endogenous. 

\begin{table}[htbp]
\scriptsize 
\centering
\begin{threeparttable}
\caption{Anti-Discrimination Laws on Labor Supply with Occupation FEs}\label{tab:occ_ls}
\begin{tabular}{l|cc|cccc}
\toprule
      & (1)   & (2)   & (3)   & (4) & (5) & (6) \\
& \multicolumn{2}{c}{\underline{Extensive Margin}} 
& \multicolumn{4}{c}{\underline{Intensive Margin}} \\
 & Labor Force & Employed & Weekly Hours & Weeks Worked & Weekly Hours & Weeks Worked \\
\midrule

ENDA $\times$ SSP: Female  
& -0.0064 & -0.0132 & 0.5614 & -0.8344\sym{*} & 0.5854\sym{**} & -0.9524\sym{***} \\
& (0.0098) & (0.0111) & (0.4011) & (0.4630) & (0.2943) & (0.3218) \\[4pt]

ENDA $\times$ SSP: Male  
& 0.0228\sym{**} & 0.0246\sym{**} & 0.5170 & 0.8743\sym{**} & -0.1462 & -0.0550 \\
& (0.0093) & (0.0100) & (0.3722) & (0.4336) & (0.3306) & (0.2799) \\

\midrule
$p$-value: M $-$ F  
& 0.0105 & 0.0075 & 0.9335 & 0.0013 & 0.1021 & 0.0237 \\
\midrule 

Employed Only  
&  &  &  &  & X & X \\

Observations  
& 7,781,358 & 7,781,358 & 7,781,358 & 7,781,358 & 5,765,580 & 5,765,580 \\

Mean (F)              &       0.818         &       0.782         &       34.42         &       39.51         &       41.25         &       47.27         \\
Mean (M)              &       0.825         &       0.790         &       35.53         &       39.96         &       42.41         &       47.69         \\

\bottomrule
\end{tabular}

\begin{tablenotes}
\scriptsize
\item Notes: Data are from the American Community Survey (ACS), 2005–2019. We compare individuals in same-sex partnerships with those in different-sex partnerships within four years of ENDA enactment. Regressions are estimated jointly with four-digit occupation fixed effects, allowing ENDA effects to differ by sex through interaction terms. Columns (1–2) report extensive-margin outcomes (labor force participation and employment). Columns (3–6) report intensive-margin outcomes (usual weekly hours and weeks worked), with Columns (5–6) restricting to employed individuals.  The ``Mean (F)" and ``Mean (M)" rows report pre-treatment averages of the outcome variables (including never-treated counties) for women and men in same-sex couples, respectively. Reported \(p\)-values test equality of male and female coefficients. Standard errors are clustered at the county level. \sym{*} \(p<0.10\), \sym{**} \(p<0.05\), \sym{***} \(p<0.01\).
\end{tablenotes}

\end{threeparttable}
\end{table}

\begin{table}[htbp]
\scriptsize 
\centering
\begin{threeparttable}
\caption{Anti-Discrimination Laws on Pay with Occupation FEs}\label{tab:occ_wage}
\begin{tabular}{l|cccc|cccc}
\toprule 
\multicolumn{1}{r}{} & \multicolumn{4}{c}{Hourly Wage} & \multicolumn{2}{c}{Annual Earnings} \\
\midrule
      & (1)   & (2)   & (3)   & (4) & (5) & (6) & (7) & (8) \\
 & Percentile & $\ge$25th & $\ge$50th & $\ge$75th & Percentile & $\ge$25th & $\ge$50th & $\ge$75th \\
\midrule

ENDA $\times$ SSP: Female  
& -0.9026 & -0.0024 & -0.0182 & -0.0116 
& -1.1521\sym{*} & -0.0018 & -0.0274\sym{**} & -0.0059 \\
& (0.6991) & (0.0097) & (0.0121) & (0.0146)
& (0.6522) & (0.0097) & (0.0111) & (0.0155) \\[4pt]

ENDA $\times$ SSP: Male  
& 1.8922\sym{**} & 0.0243\sym{**} & 0.0244\sym{*} & 0.0200 
& 2.2207\sym{***} & 0.0242\sym{**} & 0.0391\sym{***} & 0.0308\sym{**} \\
& (0.9191) & (0.0113) & (0.0145) & (0.0141)
& (0.8332) & (0.0113) & (0.0152) & (0.0128) \\

\midrule
$p$-value: M $-$ F  
& 0.0204 & 0.0699 & 0.0508 & 0.0983
& 0.0017 & 0.0778 & 0.0015 & 0.0375 \\
\midrule

Employed Only  
&  &  &  &  &  &  &  &  \\

Observations  
& 7,781,358 & 7,781,358 & 7,781,358 & 7,781,358
& 7,781,358 & 7,781,358 & 7,781,358 & 7,781,358 \\

Mean (F)              &       48.73         &       0.799         &       0.513         &       0.247         &       48.68         &       0.798         &       0.515         &       0.241         \\
Mean (M)              &       51.51         &       0.791         &       0.557         &       0.305         &       51.69         &       0.791         &       0.561         &       0.305         \\

\bottomrule

\end{tabular}

\begin{tablenotes}
\scriptsize
\item Notes: Data are from the 2005–2019 ACS. We compare individuals in same-sex partnerships with those in different-sex partnerships within four years of ENDA enactment. Regressions are estimated separately by sex and include four-digit occupation fixed effects. Columns (1–4) report effects on hourly real wage percentiles; Columns (5–8) report effects on annual real earnings percentiles. Columns (2–4) and (6–8) estimate the probability of being above the 25th, 50th, and 75th percentiles. All wage and earnings variables are in 2019 dollars.  The ``Mean (F)" and ``Mean (M)" rows report pre-treatment averages of the outcome variables (including never-treated counties) for women and men in same-sex couples, respectively. Reported \(p\)-values test equality of male and female coefficients. Standard errors are clustered at the county level. \sym{*} \(p<0.10\), \sym{**} \(p<0.05\), \sym{***} \(p<0.01\).
\end{tablenotes}

\end{threeparttable}
\end{table}

Although controlling for occupation can pose selection concerns, because individuals not in the labor force typically lack occupation codes, we nonetheless examine within-occupation effects of anti-discrimination laws on wages and labor supply. Specifically, we employ regressions with four-digit occupation fixed effects to assess how wage and labor supply disparities shift following the enactment of these policies.\footnote{In order to estimate within-occupation effects while still capturing individuals who are not in the labor force, we include a separate ``no occupation" category for those without valid occupational codes. This approach allows us to retain non-employed respondents in the analysis, rather than excluding them outright due to missing occupation information. We acknowledge that categorizing all non-employed individuals into a single group can mask heterogeneity in reasons for not working (e.g., unemployment, home production, or retirement). Nonetheless, by creating a distinct ``no occupation" category, we preserve a measure of occupational affiliation for those who are employed while ensuring that non-employed individuals remain part of the sample.} One potential mechanism is that workers in male same-sex couples may transition into more in-demand occupations, thereby mitigating relative labor supply and wage gaps. However, as shown in Tables \ref{tab:occ_ls} and \ref{tab:occ_wage}, even after accounting for occupation, men in same-sex couples exhibit significant changes in both labor supply and wages. These findings suggest that the impact of anti-discrimination laws is not merely a by-product of occupational sorting. Nonetheless, caution is warranted when interpreting these results, given the inherent limitations of conditioning on occupation in a broader labor force context.

\subsection{Clustering}
Our baseline results report standard errors clustered at the county level, which is the level of treatment variation for local ENDA ordinances. As \cite{CameronMiller2015Practitioners} emphasize, however, serial correlation at higher levels of aggregation can bias standard errors downward when treatment varies at both local and state levels. To address this concern, we re-estimate all specifications using (i) state-level clustering, and (ii) two-way clustering by state and year.

Table \ref{tab:ext_int_clustering} shows that clustering at the state level produces narrower confidence intervals than the original county-clustered results. Two-way clustering by state $\times$ year yields standard errors nearly identical to state clustering. Across all outcomes, the qualitative patterns remain unchanged: men in same-sex partnerships exhibit positive and statistically significant extensive-margin responses to ENDA, while women’s responses are smaller and generally not statistically distinguishable from zero. Overall, the results are robust to alternative clustering choices, and inference is, if anything, more conservative under our original county-level specification.

\begin{table}[htbp]
\scriptsize
\centering
\begin{threeparttable}
\caption{Effect of Anti-Discrimination Laws: Extensive \& Intensive Margin of Labor Supply (Clustering)}\label{tab:ext_int_clustering}
\begin{tabular}{l|cc|cccc}
\toprule
& (1) & (2) & (3) & (4) & (5) & (6) \\
& \multicolumn{2}{c}{\underline{Extensive Margin}} 
& \multicolumn{4}{c}{\underline{Intensive Margin}} \\
& Labor Force & Employed 
& Weekly Hours & Weeks Worked 
& Weekly Hours & Weeks Worked \\
\midrule
\multicolumn{7}{c}{\textbf{Panel A: Clustered by State}} \\
\midrule

ENDA $\times$ SSP: Female         
& -0.0037 & -0.0050 & 0.9519\sym{**} & -0.4143 & 0.7017\sym{***} & -0.8915\sym{***} \\
& (0.0100) & (0.0141) & (0.4116) & (0.5690) & (0.2426) & (0.3293) \\[4pt]

ENDA $\times$ SSP: Male           
& 0.0379\sym{***} & 0.0370\sym{***} & 1.1289\sym{***} & 1.5089\sym{***} & -0.0920 & -0.0719 \\
& (0.0106) & (0.0096) & (0.3741) & (0.3729) & (0.3834) & (0.2740) \\

\midrule
$p$-value: M $-$ F  
& 0.0011 & 0.0191 & 0.7778 & 0.0086 & 0.1114 & 0.0365 \\

\midrule
\multicolumn{7}{c}{\textbf{Panel B: Clustered by State-Year}} \\
\midrule

ENDA $\times$ SSP: Female          
& -0.0037 & -0.0050 & 0.9519\sym{*}  & -0.4143 & 0.7017\sym{**} & -0.8915\sym{***} \\
& (0.0103) & (0.0120) & (0.4907) & (0.5992) & (0.3487) & (0.3277) \\[4pt]

ENDA $\times$ SSP: Male                 
& 0.0379\sym{***} & 0.0370\sym{***} & 1.1289\sym{***} & 1.5089\sym{***} & -0.0920 & -0.0719 \\
& (0.0098) & (0.0104) & (0.4205) & (0.5173) & (0.4045) & (0.2842) \\

\midrule
$p$-value: M $-$ F  
& 0.0059 & 0.0128 & 0.8063 & 0.0191 & 0.1434 & 0.0632 \\
\midrule 
Employed Only  &   &   &   &   & X & X \\
Observations   & 7,781,358 & 7,781,358 & 7,781,358 & 7,781,358 & 5,765,580 & 5,765,580 \\
Mean (F)              &       0.818         &       0.782         &       34.42         &       39.51         &       41.25         &       47.27         \\
Mean (M)              &       0.825         &       0.790         &       35.53         &       39.96         &       42.41         &       47.69         \\
\bottomrule
\end{tabular}

\begin{tablenotes}
\scriptsize
\item Notes: Data is derived from the American Community Survey (ACS) conducted annually between 2005 and 2019. We compare individuals in same-sex partnerships with those in different-sex partnerships within four years of ENDA enactment. The regressions are estimated jointly, allowing the treatment effect to differ for men and women through interaction terms. Columns (1–2) focus on the extensive margin, examining changes in labor force participation and employment. Columns (3–6) explore the intensive margin using variation in weeks worked per year and usual weekly hours; Columns (5–6) restrict the sample to employed individuals. The ``Mean (F)" and ``Mean (M)" rows report pre-treatment averages of the outcome variables (including never-treated counties) for women and men in same-sex couples, respectively. Reported \(p\)-values test whether the ENDA effect differs significantly between men and women. Panel A clusters at the state level, Panel B at the state-year level. \sym{*} \(p<0.10\), \sym{**} \(p<0.05\), \sym{***} \(p<0.01\).
\end{tablenotes}

\end{threeparttable}
\end{table}

\subsection{Weighting}\label{subsec:weighting}

Although the ACS recommends applying sample weights to ensure representativeness, we conduct a robustness check by re‐estimating our models without these weights, following \citet{solon2015we}. This approach allows us to verify that our main results are not driven solely by the weighting scheme.

\begin{table}[htbp]
\scriptsize
\centering
\begin{threeparttable}
\caption{Effect of Anti-Discrimination Laws: Extensive \& Intensive Margin of Labor Supply (Unweighted)}\label{tab:ext_int_unweighted}
\begin{tabular}{l|cc|cccc}
\toprule
& (1) & (2) & (3) & (4) & (5) & (6) \\
& \multicolumn{2}{c}{\underline{Extensive Margin}} 
& \multicolumn{4}{c}{\underline{Intensive Margin}} \\
& Labor Force & Employed 
& Weekly Hours & Weeks Worked 
& Weekly Hours & Weeks Worked \\
\midrule

ENDA $\times$ SSP: Female  
& -0.0085 & -0.0112 & 0.7029 & -0.5207 & 0.6289\sym{**} & -0.7879\sym{**} \\
& (0.0102) & (0.0118) & (0.4868) & (0.5804) & (0.3008) & (0.3315) \\[4pt]

ENDA $\times$ SSP: Male  
& 0.0265\sym{*} & 0.0251\sym{*} & 0.8506\sym{*} & 1.0802\sym{*} & -0.0666 & 0.0419 \\
& (0.0144) & (0.0149) & (0.4873) & (0.6126) & (0.3540) & (0.2871) \\

\midrule

$p$-value: M $-$ F  
& 0.0091 & 0.0345 & 0.7768 & 0.0102 & 0.0909 & 0.0344 \\
\midrule 

Employed Only  
&  &  &  &  & X & X \\

Observations  
& 7,781,358 & 7,781,358 & 7,781,358 & 7,781,358 & 5,765,580 & 5,765,580 \\

Mean (F)  
  &       0.818         &       0.782         &       34.42         &       39.51         &       41.25         &       47.27         \\

Mean (M)  
    &       0.825         &       0.790         &       35.53         &       39.96         &       42.41         &       47.69         \\

\bottomrule
\end{tabular}

\begin{tablenotes}
\scriptsize
\item Notes: Data is derived from the American Community Survey (ACS) conducted annually between 2005 and 2019. We compare individuals in same-sex partnerships with those in different-sex partnerships within four years of ENDA enactment. The regressions are estimated jointly, allowing the treatment effect to differ for men and women through interaction terms. Columns (1–2) focus on the extensive margin, examining changes in labor force participation and employment. Columns (3–6) explore the intensive margin using variation in weeks worked per year and usual weekly hours; Columns (5–6) restrict the sample to employed individuals. The reported \(p\)-values test whether the ENDA effect differs significantly between men and women. Standard errors are clustered at the county level. \sym{*} \(p<0.10\), \sym{**} \(p<0.05\), \sym{***} \(p<0.01\).
\end{tablenotes}

\end{threeparttable}
\end{table}

The results in Table \ref{tab:ext_int_unweighted} indicate that the overall impacts of anti‑discrimination laws on labor supply remain robust even when ACS weights are removed. For men in same‑sex couples, the magnitude of the estimated effects on both labor force participation and employment is slightly smaller compared to the weighted estimates, but they continue to be statistically significant. This suggests that the positive impact of these laws for men is not driven solely by the weighting procedure. For women, the pattern of the results is similar to the weighted sample, with some variations in magnitude across different labor outcomes, though most effects remain small and statistically insignificant.

\subsection{Event Studies}

\begin{figure}[!ht]
\footnotesize
\begin{center}
	\begin{subfigure}{0.8\textwidth}  
		\centering
		\includegraphics[width=0.8\textwidth]{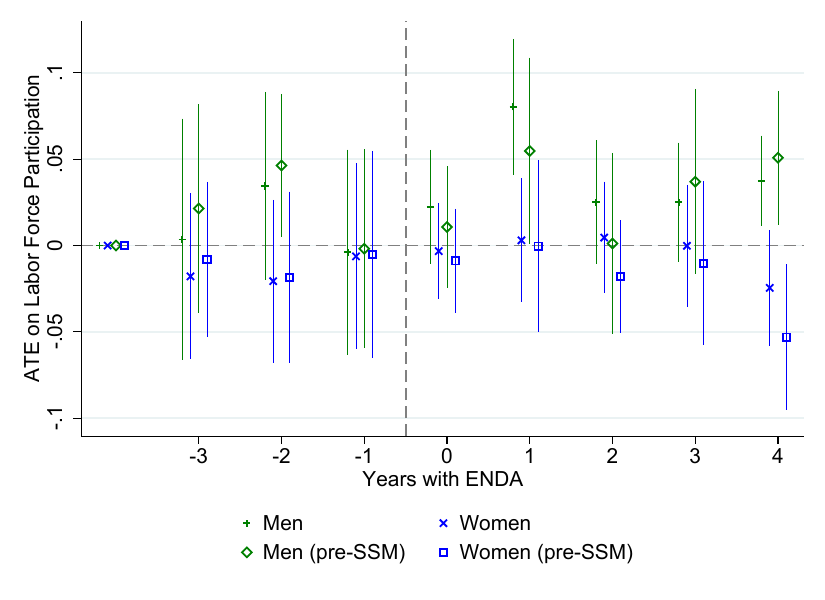}
		\caption{Labor Force Participation}
	\end{subfigure}
 
    \hfill  
    
	\begin{subfigure}{0.8\textwidth}  
		\centering
		\includegraphics[width=0.8\textwidth]{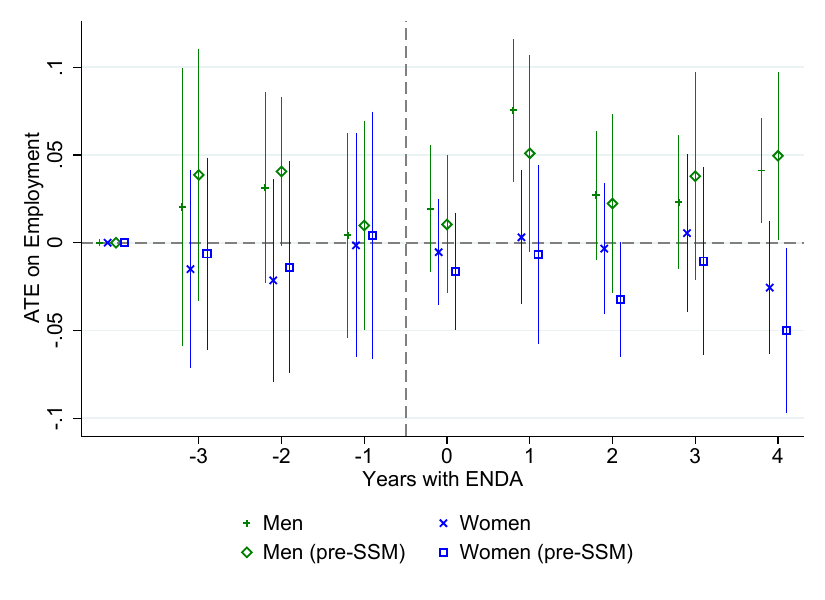}
		\caption{Employment}
	\end{subfigure} \\
\end{center}
	\caption{Labor Supply Impacts on the Extensive Margin: Same-Sex Marriage Legalization}\label{fig:event_ls_ssm}   
\begin{flushleft}
\footnotesize{Notes: Event study plots on the difference in labor supply between people in same-sex couples and different-sex couples broken down by sex following the county-level and state-level anti-discrimination laws. Sample is limited to observations before same-sex marriage legalization. Coefficients with 95\% confidence intervals. Standard errors are clustered at the county level. Reference year is -4.}
\end{flushleft} 
\end{figure}

\begin{figure}[!ht]
\footnotesize
\begin{center}
	\begin{subfigure}{0.8\textwidth}  
		\centering
		\includegraphics[width=0.8\textwidth]{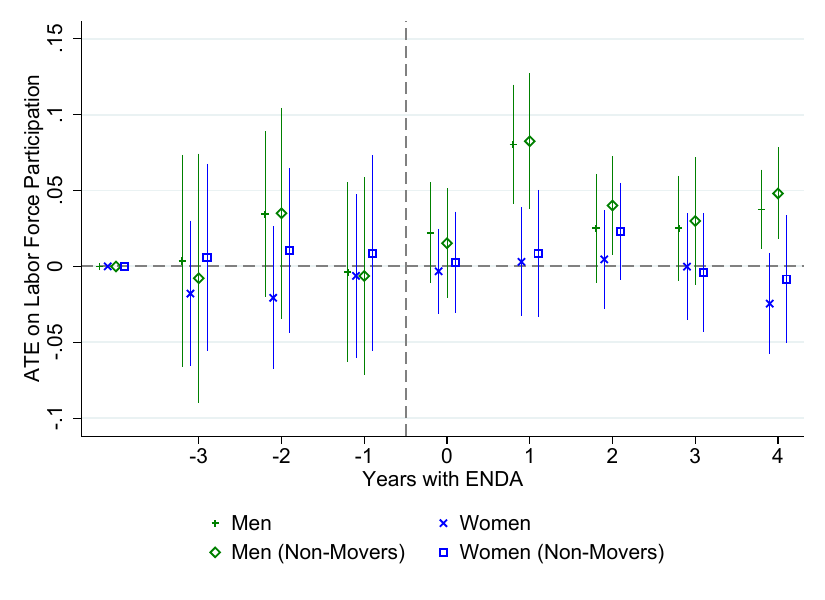}
		\caption{Labor Force Participation}
	\end{subfigure}
 
    \hfill  
    
	\begin{subfigure}{0.8\textwidth}  
		\centering
		\includegraphics[width=0.8\textwidth]{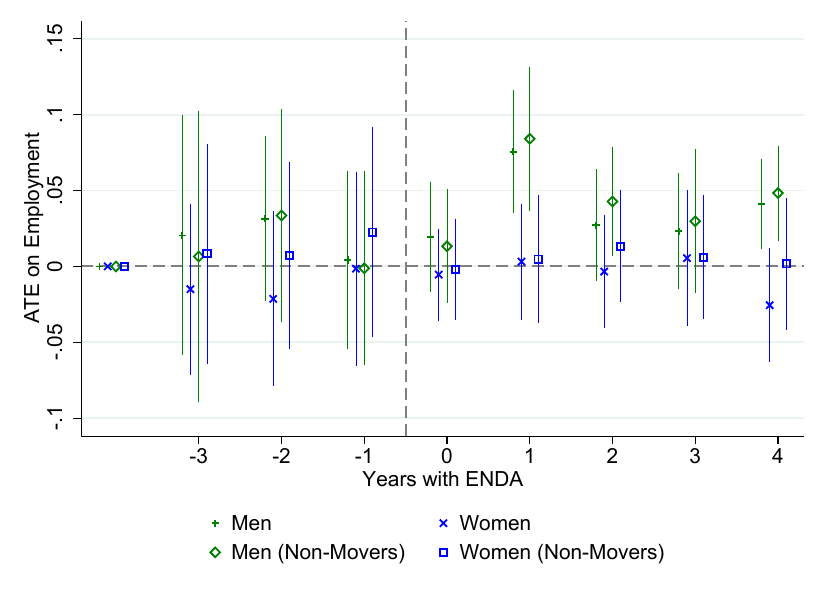}
		\caption{Employment}
	\end{subfigure} \\
\end{center}
	\caption{Labor Supply Impacts on the Extensive Margin: Non-Movers}\label{fig:event_ls_non_movers}   
\begin{flushleft}
\footnotesize{Notes: Event study plots on the difference in labor supply between people in same-sex couples and different-sex couples broken down by sex following the county-level and state-level anti-discrimination laws. Sample is limited to individuals who have not moved in the past year. Coefficients with 95\% confidence intervals. Standard errors are clustered at the county level. Reference year is -4.}
\end{flushleft} 
\end{figure}

\begin{figure}[!ht]
\footnotesize
\begin{center}
	\begin{subfigure}{0.8\textwidth}  
		\centering
		\includegraphics[width=0.8\textwidth]{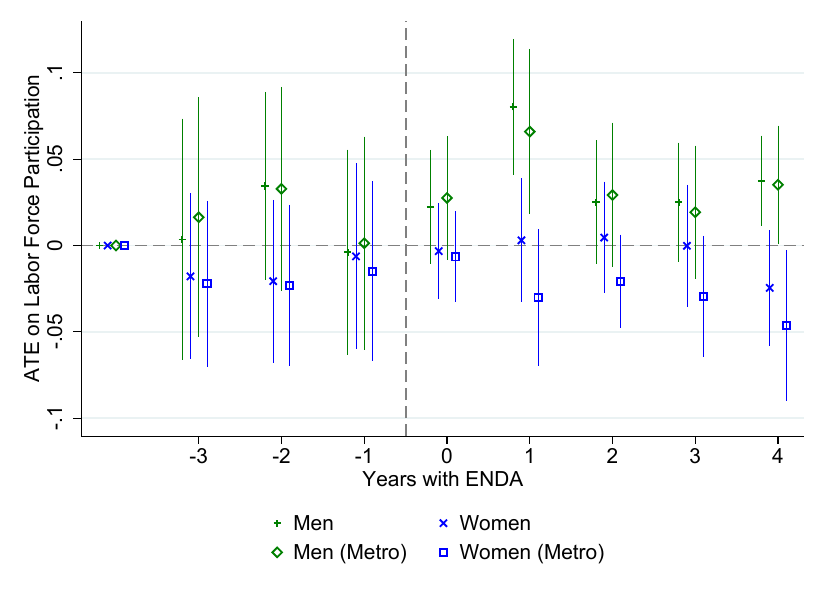}
		\caption{Labor Force Participation}
	\end{subfigure}
 
    \hfill  
    
	\begin{subfigure}{0.8\textwidth}  
		\centering
		\includegraphics[width=0.8\textwidth]{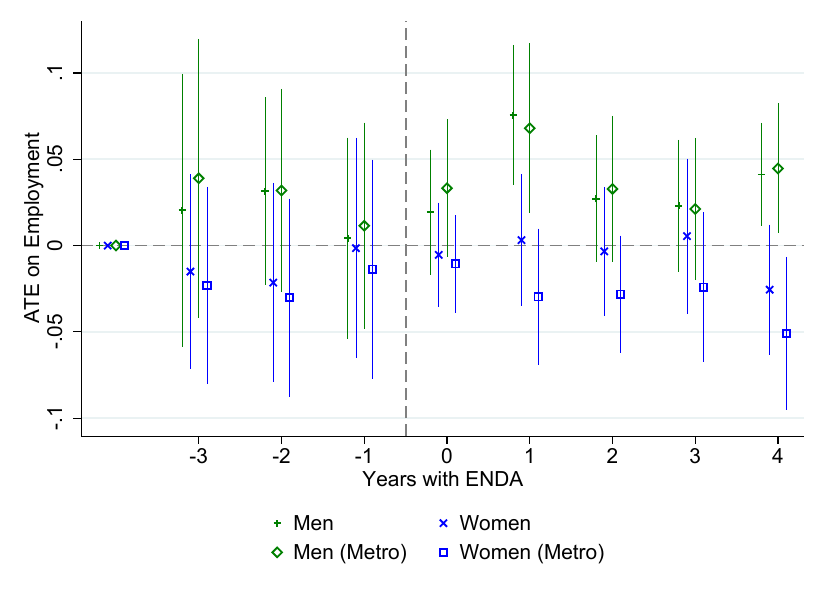}
		\caption{Employment}
	\end{subfigure} \\
\end{center}
	\caption{Labor Supply Impacts on the Extensive Margin: Metropolitan Areas}\label{fig:event_ls_metro}   
\begin{flushleft}
\footnotesize{Notes: Event study plots on the difference in labor supply between people in same-sex couples and different-sex couples broken down by sex following the county-level and state-level anti-discrimination laws. Sample is limited to individuals who live in metropolitan areas. Coefficients with 95\% confidence intervals. Standard errors are clustered at the county level. Reference year is -4.}
\end{flushleft} 
\end{figure}

\begin{figure}[!ht]
\footnotesize
\begin{center}
	\begin{subfigure}{0.8\textwidth}  
		\centering
		\includegraphics[width=0.8\textwidth]{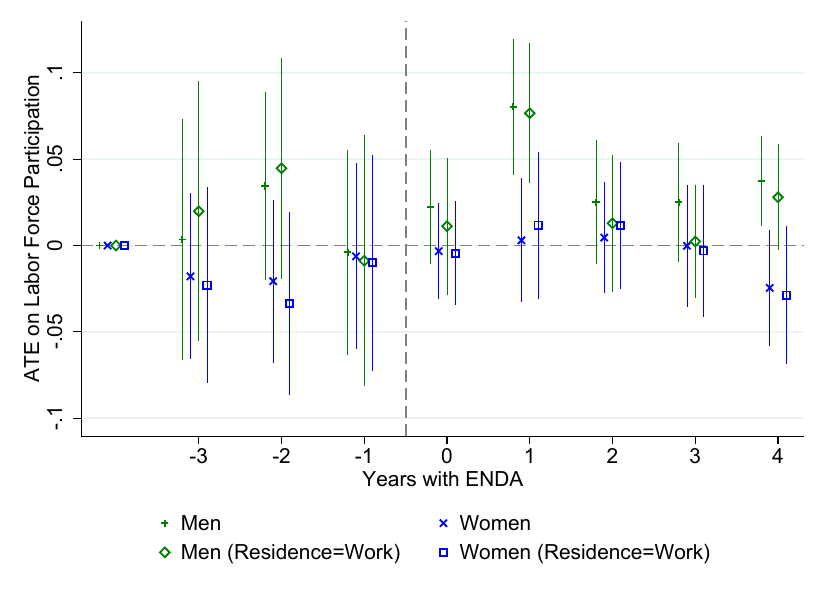}
		\caption{Labor Force Participation}
	\end{subfigure}
 
    \hfill  
    
	\begin{subfigure}{0.8\textwidth}  
		\centering
		\includegraphics[width=0.8\textwidth]{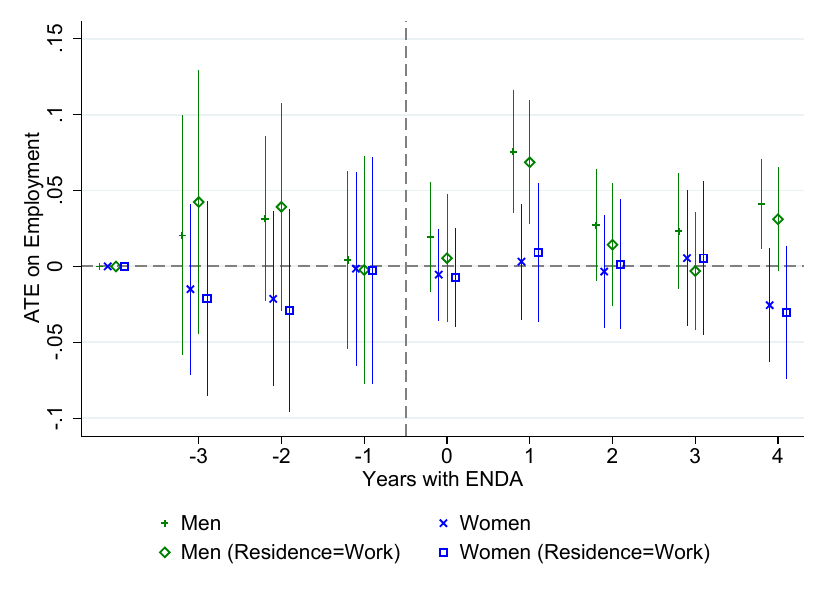}
		\caption{Employment}
	\end{subfigure} \\
\end{center}
	\caption{Labor Supply Impacts on the Extensive Margin: Work Residence same as Home Residence}\label{fig:event_ls_noncommute}   
\begin{flushleft}
\footnotesize{Notes: Event study plots on the difference in labor supply between people in same-sex couples and different-sex couples broken down by sex following the county-level and state-level anti-discrimination laws. Sample is limited to individuals who live in the same county as they work in. Coefficients with 95\% confidence intervals. Standard errors are clustered at the county level. Reference year is -4.}
\end{flushleft} 
\end{figure}

\begin{figure}[!ht]
\footnotesize
\begin{center}
	\begin{subfigure}{0.8\textwidth}  
		\centering
		\includegraphics[width=0.8\textwidth]{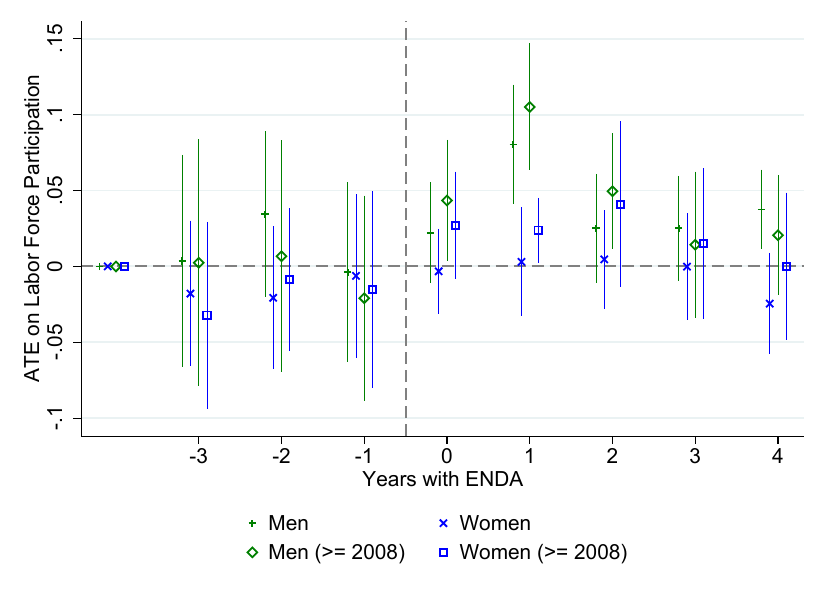}
		\caption{Labor Force Participation}
	\end{subfigure}
 
    \hfill  
    
	\begin{subfigure}{0.8\textwidth}  
		\centering
		\includegraphics[width=0.8\textwidth]{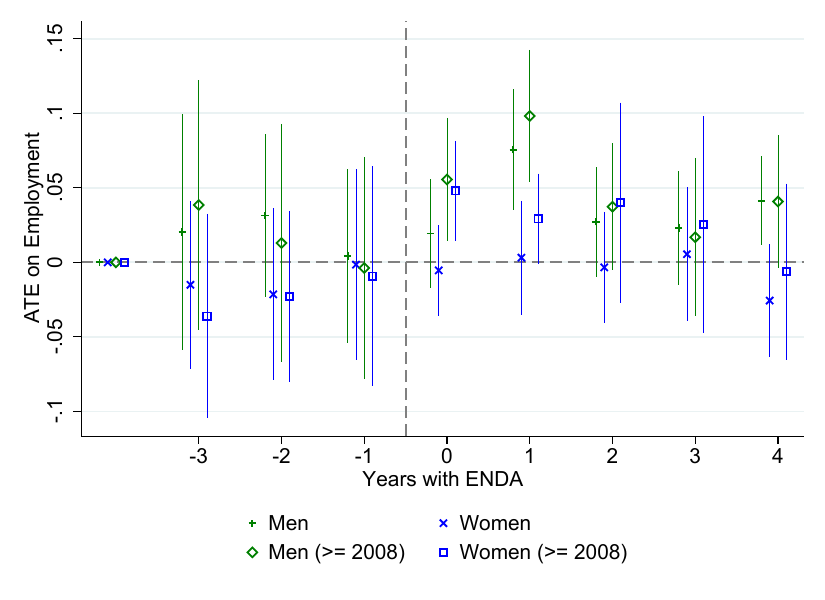}
		\caption{Employment}
	\end{subfigure} \\
\end{center}
	\caption{Labor Supply Impacts on the Extensive Margin: 2008-2019}\label{fig:event_ls_2008}  
\begin{flushleft}
\footnotesize{Notes: Event study plots on the difference in labor supply between people in same-sex couples and different-sex couples broken down by sex following the county-level and state-level anti-discrimination laws. Sample is limited to cover 2008 through 2019. Coefficients with 95\% confidence intervals. Standard errors are clustered at the county level. Reference year is -4.}
\end{flushleft} 
\end{figure}

\begin{figure}[!ht]
\footnotesize
\begin{center}
	\begin{subfigure}{0.8\textwidth}  
		\centering
		\includegraphics[width=0.8\textwidth]{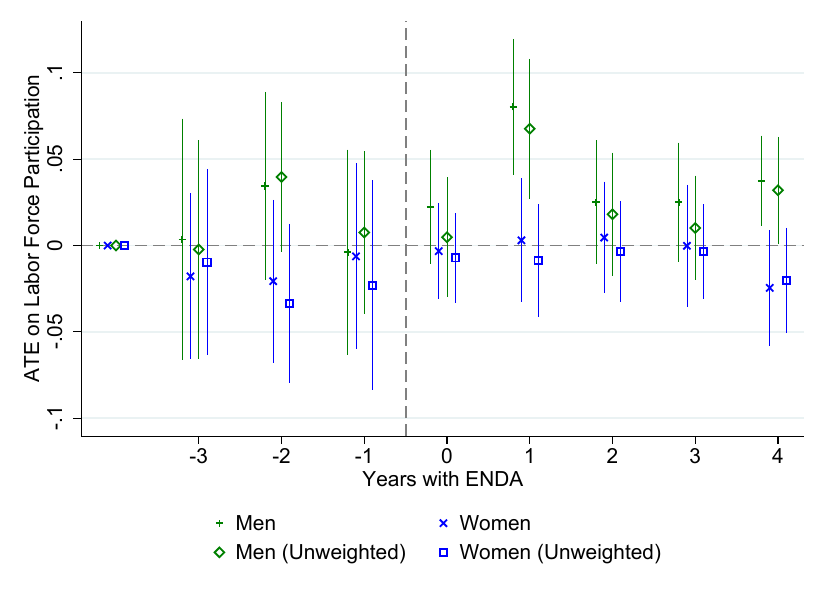}
		\caption{Labor Force Participation}
	\end{subfigure}
 
    \hfill  
    
	\begin{subfigure}{0.8\textwidth}  
		\centering
		\includegraphics[width=0.8\textwidth]{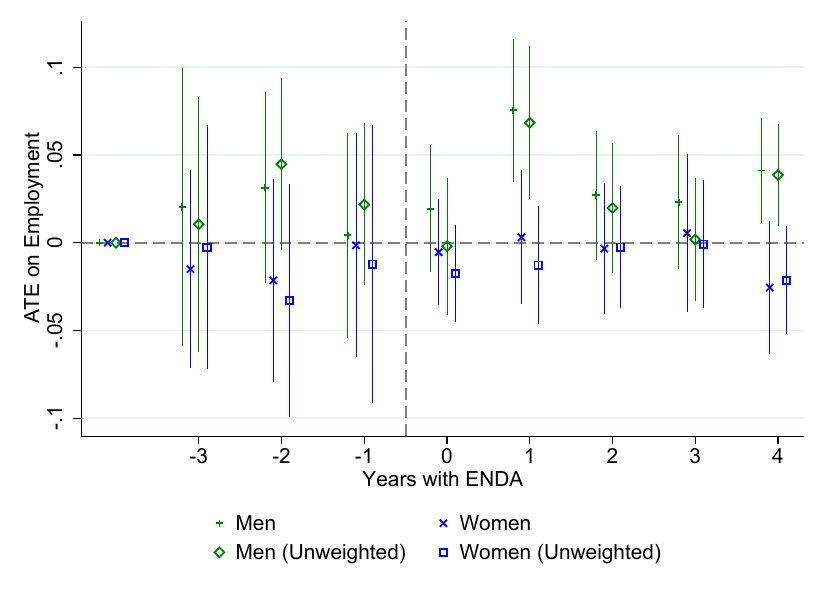}
		\caption{Employment}
	\end{subfigure} \\
\end{center}
	\caption{Labor Supply Impacts on the Extensive Margin: Weighting Schemes}\label{fig:event_ls_weights}  
\begin{flushleft}
\footnotesize{Notes: Event study plots on the difference in labor supply between people in same-sex couples and different-sex couples broken down by sex following the county-level and state-level anti-discrimination laws. We compare estimates weighted by ACS weights and without. Coefficients with 95\% confidence intervals. Standard errors are clustered at the county level. Reference year is -4.}
\end{flushleft} 
\end{figure}

Figures \ref{fig:event_ls_ssm}--\ref{fig:event_ls_weights} present the event study regressions corresponding to the robustness checks discussed in Sections \ref{subsec:marriage}--\ref{subsec:reporting} and Appendix \ref{subsec:weighting}. In each figure, we trace the evolution of labor supply impacts on the external margin relative to the enactment of anti-discrimination laws. Across all specifications, the pre-treatment coefficients are statistically insignificant, and the post-treatment dynamics closely mirror those observed in our main analyses. Specifically, estimates for women in metropolitan areas are smaller on average, unweighted models and those estimated before the legalization of same-sex marriage typically produce slightly lower coefficients, and both the post-2008 sample and non-movers yield somewhat larger effects.

\subsection{Sensitivity Analysis for Parallel Trends}\label{app:sensitivity}
To address potential violations of the parallel trends assumption, we conduct a sensitivity analysis following \cite{RambachanRoth2023}. Instead of relying on potentially underpowered pre-trend tests (\cite{roth2022pretest}), we construct robust confidence intervals using the ``Relative Magnitude'' restriction. This approach bounds the potential post-treatment bias to be no larger than $\bar{M}$ times the maximum observed pre-treatment trend deviation.

We estimate 90\% robust confidence intervals ($\alpha = 0.1$) for the main effects reported in Columns (1-4) of Table \ref{tab:ext_int} across sensitivity parameters $\bar{M} \in [0.1, 2.0]$. As shown in Table \ref{tab:sensitivity}, male extensive margin estimates (Panel A) remain statistically significant up to $\bar{M}=0.4$, indicating robustness to moderate trend violations, while intensive margin effects are slightly more sensitive (robust up to $\bar{M}=0.3$). Conversely, estimates for women (Panel B) are statistically indistinguishable from zero across all sensitivity parameters, confirming the main null results hold even under relaxed assumptions.

\begin{table}[htbp]
\scriptsize
\centering
\begin{threeparttable}
\caption{Sensitivity Analysis for Parallel Trends (HonestDiD)}\label{tab:sensitivity}
\begin{tabular}{lcccc}
\toprule
& (1) & (2) & (3) & (4) \\
& \multicolumn{2}{c}{\underline{Extensive Margin}} 
& \multicolumn{2}{c}{\underline{Intensive Margin}} \\
Sensitivity ($M$) & Labor Force & Employed & Weekly Hours & Weeks Worked \\
\midrule

\multicolumn{5}{l}{\underline{Panel A: Men}} \\
Original ($M=0$) & [0.017, 0.059] & [0.015, 0.059] & [0.310, 1.948] & [0.478, 2.540] \\
$M=0.1$ & [0.013, 0.062] & [0.013, 0.061] & [0.233, 2.044] & [0.325, 2.693] \\
$M=0.2$ & [0.010, 0.065] & [0.010, 0.064] & [0.144, 2.155] & [0.128, 2.916] \\
$M=0.3$ & [0.006, 0.069] & [0.007, 0.067] & [0.012, 2.308] & \textbf{[-0.205, 3.249]} \\
$M=0.4$ & [0.001, 0.073] & [0.001, 0.073] & \textbf{[-0.185, 2.526]} & [-0.600, 3.644] \\
$M=0.5$ & \textbf{[-0.004, 0.078]} & \textbf{[-0.005, 0.079]} & [-0.406, 2.768] & [-1.005, 4.049] \\
$M=0.6$ & [-0.010, 0.084] & [-0.012, 0.086] & [-0.652, 3.015] & [-1.447, 4.492] \\
$M=0.7$ & [-0.016, 0.090] & [-0.019, 0.093] & [-0.902, 3.266] & [-1.872, 4.917] \\
$M=0.8$ & [-0.022, 0.096] & [-0.026, 0.100] & [-1.156, 3.521] & [-2.335, 5.381] \\
$M=0.9$ & [-0.028, 0.102] & [-0.033, 0.107] & [-1.414, 3.780] & [-2.781, 5.827] \\
$M=1.0$ & [-0.035, 0.108] & [-0.039, 0.114] & [-1.676, 4.043] & [-3.208, 6.283] \\
$M=1.5$ & [-0.067, 0.140] & [-0.076, 0.150] & [-3.006, 5.378] & [-5.512, 8.560] \\
$M=2.0$ & [-0.100, 0.172] & [-0.113, 0.187] & [-4.365, 6.741] & [-7.794, 10.844] \\

\midrule
\multicolumn{5}{l}{\underline{Panel B: Women}} \\
Original ($M=0$) & \textbf{[-0.023, 0.015]} & \textbf{[-0.023, 0.015]} & \textbf{[-0.023, 0.015]} & \textbf{[-0.023, 0.015]} \\
$M=0.1$ & [-0.025, 0.017] & [-0.025, 0.017] & [-0.025, 0.017] & [-0.025, 0.017] \\
$M=0.2$ & [-0.027, 0.019] & [-0.027, 0.019] & [-0.027, 0.019] & [-0.027, 0.019] \\
$M=0.3$ & [-0.029, 0.021] & [-0.029, 0.021] & [-0.029, 0.021] & [-0.029, 0.021] \\
$M=0.4$ & [-0.032, 0.023] & [-0.032, 0.023] & [-0.032, 0.023] & [-0.032, 0.023] \\
$M=0.5$ & [-0.036, 0.027] & [-0.036, 0.027] & [-0.036, 0.027] & [-0.036, 0.027] \\
$M=1.0$ & [-0.059, 0.049] & [-0.059, 0.049] & [-0.059, 0.049] & [-0.059, 0.049] \\
$M=1.5$ & [-0.083, 0.073] & [-0.083, 0.073] & [-0.083, 0.073] & [-0.083, 0.073] \\
$M=2.0$ & [-0.108, 0.098] & [-0.108, 0.098] & [-0.108, 0.098] & [-0.108, 0.098] \\

\bottomrule
\end{tabular}

\begin{tablenotes}
\scriptsize
\item Notes: This table presents 90\% robust confidence intervals calculated using the methods proposed by \cite{RambachanRoth2023}. We assume limited violations of parallel trends based on the relative magnitude of pre-treatment trends. We revisit estimates from Table \ref{tab:ext_int} Columns (1-4). The parameter $M$ governs the amount of slope violation allowed between consecutive periods, relative to the maximum observed violation in the pre-treatment period. $M=0$ corresponds to the standard DDD estimate (assuming perfect parallel trends). A bolded entry indicates the value of $M$ at which the confidence interval first includes zero (i.e., the result becomes statistically insignificant). Panel A reports estimates for men, while Panel B for women.
\end{tablenotes}

\end{threeparttable}
\end{table}

\clearpage

\appendix
\addcontentsline{toc}{section}{Appendices}
\counterwithin{figure}{section}
\counterwithin{table}{section}
\renewcommand{\thefigure}{O\arabic{figure}}
\renewcommand{\thetable}{O\arabic{table}}
\renewcommand{\thesection}{O} 
\setcounter{figure}{0}
\setcounter{table}{0}

\section*{Online Appendix*: Effects of State and Local Sexual Orientation Anti-Discrimination Laws on Labor Market Differentials}\label{app:online_app}

In an earlier version of the paper, we estimated all regressions separately by gender. The point estimates and standard errors are identical to those from the fully interacted specification, but the separate-gender approach does not allow us to test for differences efficiently. For transparency, we report those results here.

\begin{table}[htbp]
\footnotesize 
\centering
\begin{threeparttable}
\caption{Effect of Anti-Discrimination Laws: Extensive \& Intensive Margin of Labor Supply}\label{tab:ext_int_app}%
\begin{tabular}{l|cc|cccc}
\toprule
      & (1)   & (2)   & (3)   & (4) & (5) & (6) \\
& \multicolumn{2}{c}{\underline{Extensive Margin}} & \multicolumn{4}{c}{\underline{Intensive Margin}}\\
 & Labor Force & Employed & Weekly Hours & Weeks Worked & Weekly Hours & Weeks Worked \\
\midrule
\multicolumn{7}{c}{Panel A: Men} \\
\midrule
ENDA $\times$ SSP        &      0.0379\sym{***}&      0.0370\sym{***}&      1.1289\sym{**} &      1.5089\sym{**} &     -0.0920         &     -0.0719         \\
                    &    (0.0130)         &    (0.0135)         &    (0.4978)         &    (0.6268)         &    (0.3727)         &    (0.2820)         \\ \hline 
      &       &       &       &  & & \\[\dimexpr-\normalbaselineskip+2pt]
Observations     &3,798,542       &3,798,542       &3,798,542       &3,798,542       &3,073,797       &3,073,797       \\
Mean (M)  
    &       0.825         &       0.790         &       35.53         &       39.96         &       42.41         &       47.69         \\
Employed  Only          &                     &                     &                     &                     &           X         &           X         \\
      &       &       &       &  & & \\[\dimexpr-\normalbaselineskip+2pt]
\midrule
\multicolumn{7}{c}{Panel B: Women} \\
\midrule
ENDA $\times$ SSP      &     -0.0037         &     -0.0050         &      0.9519\sym{*}  &     -0.4143         &      0.7017\sym{**} &     -0.8915\sym{***}\\
                    &    (0.0116)         &    (0.0135)         &    (0.5324)         &    (0.6230)         &    (0.3074)         &    (0.3350)         \\ \hline 
      &       &       &       &  & & \\[\dimexpr-\normalbaselineskip+2pt]
Observations       &3,982,816       &3,982,816       &3,982,816       &3,982,816       &2,691,783      &2,691,783      \\
Mean (F)  
  &       0.818         &       0.782         &       34.42         &       39.51         &       41.25         &       47.27         \\

Employed  Only          &                     &                     &                     &                     &           X         &           X         \\ \hline 
      &       &       &       &  & & \\[\dimexpr-\normalbaselineskip+2pt]
\bottomrule
\end{tabular}%
\begin{tablenotes}
\scriptsize
\item Notes:  Data is derived from the American Community Survey (ACS) conducted annually between 2005 and 2019. We compare individuals in same-sex partnerships with those in different-sex partnerships within four years of ENDA enactment. The regressions are estimated separately by sex: Panel A for men, while Panel B for women. The coefficients indicate the impact of anti-discrimination laws. The analysis is divided into two parts: Columns (1-2) focus on the extensive margin, examining changes in labor force participation and employment status as a result of the laws. Columns (3-6), on the other hand, explore the intensive margin by leveraging variations in the number of weeks worked per year and average weekly working hours.  The ``Mean (F)" and ``Mean (M)" rows report pre-treatment averages of the outcome variables (including never-treated counties) for women and men in same-sex couples, respectively. Standard errors are clustered at the county level.  \sym{*} \(p<0.10\), \sym{**} \(p<0.05\), \sym{\sym{***}} \(p<0.01\).
\end{tablenotes}
\end{threeparttable}
\end{table}

\begin{table}[htbp]
\footnotesize 
\centering
\begin{threeparttable}
\caption{Effect of Anti-Discrimination Laws: Wages and Earnings}\label{tab:wage_app}%
\begin{tabular}{l|cccc|cccc}
\toprule 
\multicolumn{1}{r}{} & \multicolumn{4}{c}{Hourly Wage} & \multicolumn{2}{c}{Annual Earnings} \\
\midrule
      & (1)   & (2)   & (3)   & (4) & (5) & (6) & (7) & (8) \\
 & Percentile & $\ge$25th & $\ge$50th & $\ge$75th & Percentile & $\ge$25th & $\ge$50th & $\ge$75th \\
\midrule
\multicolumn{9}{c}{Panel A: Men} \\
\midrule
ENDA $\times$ SSP     &      2.6310\sym{**} &      0.0366\sym{***}&      0.0326\sym{*}  &      0.0232         &      2.9315\sym{***}&      0.0365\sym{***}&      0.0467\sym{**} &      0.0334\sym{**} \\
                    &    (1.1766)         &    (0.0138)         &    (0.0182)         &    (0.0178)         &    (1.1014)         &    (0.0139)         &    (0.0185)         &    (0.0154)         \\ \hline 
      &       &       &       &  \\[\dimexpr-\normalbaselineskip+2pt]
Observations       &3,798,542         &3,798,542         &3,798,542         &3,798,542         &3,798,542         &3,798,542         &3,798,542         &3,798,542         \\
Mean (M)              &       51.51         &       0.791         &       0.557         &       0.305         &       51.69         &       0.791         &       0.561         &       0.305         \\
&       &       &       &  & & \\[\dimexpr-\normalbaselineskip+2pt]
\midrule
\multicolumn{9}{c}{Panel B: Women} \\
\midrule
ENDA $\times$ SSP    &     -0.3333         &      0.0045         &     -0.0117         &     -0.0061         &     -0.5994         &      0.0051         &     -0.0204         &     -0.0016         \\
                    &    (0.8442)         &    (0.0118)         &    (0.0143)         &    (0.0154)         &    (0.8307)         &    (0.0118)         &    (0.0134)         &    (0.0161)         \\ \hline 
      &       &       &       &  \\[\dimexpr-\normalbaselineskip+2pt]
Observations         &3,982,816       &3,982,816       &3,982,816       &3,982,816       &3,982,816       &3,982,816       &3,982,816       &3,982,816       \\
Mean (F)              &       48.73         &       0.799         &       0.513         &       0.247         &       48.68         &       0.798         &       0.515         &       0.241         \\
\bottomrule
\end{tabular}%
\begin{tablenotes}
\scriptsize
\item Notes: Data is derived from the American Community Survey (ACS) conducted annually between 2005 and 2019. We compare individuals in same-sex partnerships with those in different-sex partnerships within four years of ENDA enactment. The regressions are estimated separately by sex: Panel A for men, while Panel B for women. The coefficients in the first row indicate the impact of anti-discrimination laws. Columns (1–4) report estimates for hourly real wage ranks, while Columns (5–8) report estimates for annual real wage ranks. In Columns 2–4 and 6–8, the estimates are indicator variables for being above the 25th, 50th, and 75th percentiles, respectively. All income variables are measured in 2019 dollars and expressed as percentile ranks.  The ``Mean (F)" and ``Mean (M)" rows report pre-treatment averages of the outcome variables (including never-treated counties) for women and men in same-sex couples, respectively. Standard errors are clustered at the county level. \sym{*} \(p<0.10\), \sym{**} \(p<0.05\), \sym{\sym{***}} \(p<0.01\).
\end{tablenotes}
\end{threeparttable}
\end{table}

\begin{table}[htbp]
\footnotesize 
\centering
\begin{threeparttable}
\caption{Effects of Anti-Discrimination Laws on Labor Supply: Analysis of Same-Sex Marriage Legalization}\label{tab:ext_int_ssm_app}%
\begin{tabular}{l|cc|cccc}
\toprule
      & (1)   & (2)   & (3)   & (4) & (5) & (6) \\
& \multicolumn{2}{c}{\underline{Extensive Margin}} & \multicolumn{4}{c}{\underline{Intensive Margin}}\\
 & Labor Force & Employed & Weekly Hours & Weeks Worked & Weekly Hours & Weeks Worked \\
\midrule
\multicolumn{7}{c}{Panel A: Men} \\
\midrule
ENDA $\times$ SSP    &      0.0283         &      0.0311         &      0.8892         &      0.4945         &     -0.1655         &     -0.7909\sym{**} \\
                    &    (0.0186)         &    (0.0193)         &    (0.6504)         &    (0.8716)         &    (0.4310)         &    (0.3950)         \\ \hline 
      &       &       &       &  & & \\[\dimexpr-\normalbaselineskip+2pt]
Observations       &2,553,515        &2,553,515        &2,553,515        &2,553,515        &2,058,535        &2,058,535        \\
Mean (M)  
&       0.818         &       0.778         &       35.39         &       39.44         &       42.40         &       47.27         \\

Employed  Only          &                     &                     &                     &                     &           X         &           X         \\
      &       &       &       &  & & \\[\dimexpr-\normalbaselineskip+2pt]
\midrule
\multicolumn{7}{c}{Panel B: Women} \\
\midrule
ENDA $\times$ SSP         &     -0.0157         &     -0.0215         &      1.1714\sym{*}  &     -0.6654         &      1.0376\sym{***}&     -1.2330\sym{***}\\
                    &    (0.0153)         &    (0.0158)         &    (0.6724)         &    (0.7928)         &    (0.3679)         &    (0.3998)         \\ \hline 
      &       &       &       &  & & \\[\dimexpr-\normalbaselineskip+2pt]
Observations     &2,665,702      &2,665,702      &2,665,702      &2,665,702      &1,802,056       &1,802,056       \\
Mean (F)  
&       0.817         &       0.774         &       34.44         &       39.13         &       41.28         &       46.85         \\

Employed  Only          &                     &                     &                     &                     &           X         &           X         \\ \hline 
      &       &       &       &  & & \\[\dimexpr-\normalbaselineskip+2pt]
\bottomrule
\end{tabular}%
\begin{tablenotes}
\scriptsize
\item Notes:  Data is derived from the American Community Survey (ACS) conducted annually between 2005 and 2019. We compare individuals in same-sex partnerships with those in different-sex partnerships within four years of ENDA enactment. The regressions are estimated separately by sex: Panel A for men, while Panel B for women. Sample is limited to observations before same-sex marriage legalization. The coefficients indicate the impact of anti-discrimination laws. The analysis is divided into two parts: Columns (1-2) focus on the extensive margin, examining changes in labor force participation and employment status as a result of the laws. Columns (3-6), on the other hand, explore the intensive margin by leveraging variations in the number of weeks worked per year and average weekly working hours. The ``Mean (F)" and ``Mean (M)" rows report pre-treatment averages of the outcome variables (including never-treated counties) for women and men in same-sex couples, respectively. Standard errors are clustered at the county level.  \sym{*} \(p<0.10\), \sym{**} \(p<0.05\), \sym{\sym{***}} \(p<0.01\).
\end{tablenotes}
\end{threeparttable}
\end{table}

\begin{table}[htbp]
\footnotesize 
\centering
\begin{threeparttable}
\caption{Effects of Anti-Discrimination Laws on Labor Supply: Analysis of Non-Movers}\label{tab:ext_int_migration_app}%
\begin{tabular}{l|cc|cccc}
\toprule
      & (1)   & (2)   & (3)   & (4) & (5) & (6) \\
& \multicolumn{2}{c}{\underline{Extensive Margin}} & \multicolumn{4}{c}{\underline{Intensive Margin}}\\
 & Labor Force & Employed & Weekly Hours & Weeks Worked & Weekly Hours & Weeks Worked \\
\midrule
\multicolumn{7}{c}{Panel A: Men} \\
\midrule
ENDA $\times$ SSP          &      0.0421\sym{***}&      0.0424\sym{***}&      1.2112\sym{**} &      1.7705\sym{**} &     -0.0583         &      0.2447         \\
                    &    (0.0145)         &    (0.0153)         &    (0.5686)         &    (0.7283)         &    (0.4027)         &    (0.2938)         \\ \hline 
      &       &       &       &  & & \\[\dimexpr-\normalbaselineskip+2pt]
Observations         &3,414,649         &3,414,649         &3,414,649         &3,414,649         &2,748,449        &2,748,449        \\
Mean (M)         &       0.823         &       0.791         &       35.44         &       39.99         &       42.45         &       47.89         \\
Employed  Only          &                     &                     &                     &                     &           X         &           X         \\
      &       &       &       &  & & \\[\dimexpr-\normalbaselineskip+2pt]
\midrule
\multicolumn{7}{c}{Panel B: Women} \\
\midrule
ENDA $\times$ SSP              &      0.0050         &      0.0048         &      1.1290\sym{*}  &      0.1140         &      0.7276\sym{**} &     -0.5031         \\
                    &    (0.0142)         &    (0.0147)         &    (0.6188)         &    (0.7469)         &    (0.3082)         &    (0.3944)         \\ \hline 
      &       &       &       &  & & \\[\dimexpr-\normalbaselineskip+2pt]
Observations      &3,611,058       &3,611,058       &3,611,058       &3,611,058       &2,437,454         &2,437,454         \\
Mean (F)          &       0.815         &       0.783         &       34.34         &       39.57         &       41.32         &       47.55         \\
Employed  Only          &                     &                     &                     &                     &           X         &           X         \\ \hline 
      &       &       &       &  & & \\[\dimexpr-\normalbaselineskip+2pt]
\bottomrule
\end{tabular}%
\begin{tablenotes}
\scriptsize
\item Notes:  Data is derived from the American Community Survey (ACS) conducted annually between 2005 and 2019. We compare individuals in same-sex partnerships with those in different-sex partnerships within four years of ENDA enactment. The regressions are estimated separately by sex: Panel A for men, while Panel B for women. Sample is limited to individuals who have not moved in the past year. The coefficients indicate the impact of anti-discrimination laws. The analysis is divided into two parts: Columns (1-2) focus on the extensive margin, examining changes in labor force participation and employment status as a result of the laws. Columns (3-6), on the other hand, explore the intensive margin by leveraging variations in the number of weeks worked per year and average weekly working hours. The ``Mean (F)" and ``Mean (M)" rows report pre-treatment averages of the outcome variables (including never-treated counties) for women and men in same-sex couples, respectively. Standard errors are clustered at the county level.  \sym{*} \(p<0.10\), \sym{**} \(p<0.05\), \sym{\sym{***}} \(p<0.01\).
\end{tablenotes}
\end{threeparttable}
\end{table}

\begin{table}[htbp]
\footnotesize 
\centering
\begin{threeparttable}
\caption{Effects of Anti-Discrimination Laws on Labor Supply: Analysis of Metro Areas}\label{tab:ext_int_metro_app}%
\begin{tabular}{l|cc|cccc}
\toprule
      & (1)   & (2)   & (3)   & (4) & (5) & (6) \\
& \multicolumn{2}{c}{\underline{Extensive Margin}} & \multicolumn{4}{c}{\underline{Intensive Margin}}\\
 & Labor Force & Employed & Weekly Hours & Weeks Worked & Weekly Hours & Weeks Worked \\
\midrule
\multicolumn{7}{c}{Panel A: Men} \\
\midrule
ENDA $\times$ SSP               &      0.0352\sym{**} &      0.0398\sym{**} &      0.8139         &      1.4503\sym{*}  &     -0.4661         &     -0.2430         \\
                    &    (0.0159)         &    (0.0162)         &    (0.5417)         &    (0.7402)         &    (0.3735)         &    (0.3454)         \\ \hline 
      &       &       &       &  & & \\[\dimexpr-\normalbaselineskip+2pt]
Observations   &2,278,474        &2,278,474        &2,278,474        &2,278,474        &1,904,970        &1,904,970        \\
Mean (M)           &       0.847         &       0.812         &       36.50         &       41.09         &       42.62         &       47.96         \\
Employed  Only          &                     &                     &                     &                     &           X         &           X         \\
      &       &       &       &  & & \\[\dimexpr-\normalbaselineskip+2pt]
\midrule
\multicolumn{7}{c}{Panel B: Women} \\
\midrule
ENDA $\times$ SSP            &     -0.0262\sym{**} &     -0.0284\sym{**} &     -0.4601         &     -1.3804\sym{**} &     -0.4479         &     -1.3355\sym{***}\\
                    &    (0.0115)         &    (0.0129)         &    (0.7241)         &    (0.6505)         &    (0.6670)         &    (0.3471)         \\ \hline 
      &       &       &       &  & & \\[\dimexpr-\normalbaselineskip+2pt]
Observations    &2,376,723         &2,376,723         &2,376,723         &2,376,723         &1,625,459         &1,625,459         \\
Mean (F)           &       0.834         &       0.801         &       35.20         &       40.42         &       41.43         &       47.52         \\
Employed  Only          &                     &                     &                     &                     &           X         &           X         \\ \hline 
      &       &       &       &  & & \\[\dimexpr-\normalbaselineskip+2pt]
\bottomrule
\end{tabular}%
\begin{tablenotes}
\scriptsize
\item Notes:  Data is derived from the American Community Survey (ACS) conducted annually between 2005 and 2019. We compare individuals in same-sex partnerships with those in different-sex partnerships within four years of ENDA enactment. The regressions are estimated separately by sex: Panel A for men, while Panel B for women. Sample is limited to individuals who live in metropolitan areas. The coefficients indicate the impact of anti-discrimination laws. The analysis is divided into two parts: Columns (1-2) focus on the extensive margin, examining changes in labor force participation and employment status as a result of the laws. Columns (3-6), on the other hand, explore the intensive margin by leveraging variations in the number of weeks worked per year and average weekly working hours. The ``Mean (F)" and ``Mean (M)" rows report pre-treatment averages of the outcome variables (including never-treated counties) for women and men in same-sex couples, respectively. Standard errors are clustered at the county level.  \sym{*} \(p<0.10\), \sym{**} \(p<0.05\), \sym{\sym{***}} \(p<0.01\).
\end{tablenotes}
\end{threeparttable}
\end{table}

\begin{table}[htbp]
\footnotesize 
\centering
\begin{threeparttable}
\caption{Effects of Anti-Discrimination Laws on Labor Supply: Analysis of ACS Redesign}\label{tab:ext_int_2008_app}%
\begin{tabular}{l|cc|cccc}
\toprule
      & (1)   & (2)   & (3)   & (4) & (5) & (6) \\
& \multicolumn{2}{c}{\underline{Extensive Margin}} & \multicolumn{4}{c}{\underline{Intensive Margin}}\\
 & Labor Force & Employed & Weekly Hours & Weeks Worked & Weekly Hours & Weeks Worked \\
\midrule
\multicolumn{7}{c}{Panel A: Men} \\
\midrule
ENDA $\times$ SSP  &      0.0459\sym{*}  &      0.0492\sym{*}  &      1.5759         &      3.1075\sym{**} &     -0.2548         &      0.8834\sym{***}\\
                    &    (0.0255)         &    (0.0267)         &    (1.2059)         &    (1.2909)         &    (0.7122)         &    (0.3361)         \\ \hline 
      &       &       &       &  & & \\[\dimexpr-\normalbaselineskip+2pt]
Observations      &2,966,542       &2,966,542       &2,966,542       &2,966,542       &2,396,977         &2,396,977         \\
Mean (M)        &       0.837         &       0.802         &       35.66         &       40.48         &       42.41         &       48.11         \\
Employed  Only          &                     &                     &                     &                     &           X         &           X         \\
      &       &       &       &  & & \\[\dimexpr-\normalbaselineskip+2pt]
\midrule
\multicolumn{7}{c}{Panel B: Women} \\
\midrule
ENDA $\times$ SSP       &      0.0217         &      0.0272         &      1.3482         &      0.6313         &      0.5888         &     -0.1525         \\
                    &    (0.0230)         &    (0.0245)         &    (1.1363)         &    (1.2235)         &    (0.4428)         &    (0.4731)         \\ \hline 
      &       &       &       &  & & \\[\dimexpr-\normalbaselineskip+2pt]
Observations        &3,119,738      &3,119,738      &3,119,738      &3,119,738      &2,100,379       &2,100,379       \\
Mean (F)        &       0.824         &       0.787         &       34.27         &       39.72         &       41.09         &       47.51         \\
Employed  Only          &                     &                     &                     &                     &           X         &           X         \\ \hline 
      &       &       &       &  & & \\[\dimexpr-\normalbaselineskip+2pt]
\bottomrule
\end{tabular}%
\begin{tablenotes}
\scriptsize
\item Notes:  Data is derived from the American Community Survey (ACS) conducted annually between 2005 and 2019. We compare individuals in same-sex partnerships with those in different-sex partnerships within four years of ENDA enactment. The regressions are estimated separately by sex: Panel A for men, while Panel B for women. Sample is limited to cover 2008 through 2019. The coefficients indicate the impact of anti-discrimination laws. The analysis is divided into two parts: Columns (1-2) focus on the extensive margin, examining changes in labor force participation and employment status as a result of the laws. Columns (3-6), on the other hand, explore the intensive margin by leveraging variations in the number of weeks worked per year and average weekly working hours. The ``Mean (F)" and ``Mean (M)" rows report pre-treatment averages of the outcome variables (including never-treated counties) for women and men in same-sex couples, respectively. Standard errors are clustered at the county level.  \sym{*} \(p<0.10\), \sym{**} \(p<0.05\), \sym{\sym{***}} \(p<0.01\).
\end{tablenotes}
\end{threeparttable}
\end{table}
\begin{table}[htbp]
\footnotesize 
\centering
\begin{threeparttable}
\caption{Effects of Anti-Discrimination Laws on Labor Supply: Analysis of Unweighted Estimates}\label{tab:ext_int_unweighted_app}%
\begin{tabular}{l|cc|cccc}
\toprule
      & (1)   & (2)   & (3)   & (4) & (5) & (6) \\
& \multicolumn{2}{c}{\underline{Extensive Margin}} & \multicolumn{4}{c}{\underline{Intensive Margin}}\\
 & Labor Force & Employed & Weekly Hours & Weeks Worked & Weekly Hours & Weeks Worked \\
\midrule
\multicolumn{7}{c}{Panel A: Men} \\
\midrule
ENDA $\times$ SSP      &      0.0265\sym{*}  &      0.0251\sym{*}  &      0.8506\sym{*}  &      1.0802\sym{*}  &     -0.0666         &      0.0419         \\
                    &    (0.0144)         &    (0.0149)         &    (0.4860)         &    (0.6116)         &    (0.3540)         &    (0.2870)         \\ \hline 
      &       &       &       &  & & \\[\dimexpr-\normalbaselineskip+2pt]
Observations     &3,798,542       &3,798,542       &3,798,542       &3,798,542       &3,073,797       &3,073,797       \\
Mean (M)  
    &       0.825         &       0.790         &       35.53         &       39.96         &       42.41         &       47.69         \\
Employed  Only          &                     &                     &                     &                     &           X         &           X         \\
      &       &       &       &  & & \\[\dimexpr-\normalbaselineskip+2pt]
\midrule
\multicolumn{7}{c}{Panel B: Women} \\
\midrule
ENDA $\times$ SSP       &     -0.0085         &     -0.0112         &      0.7029         &     -0.5207         &      0.6289\sym{**} &     -0.7879\sym{**} \\
                    &    (0.0102)         &    (0.0118)         &    (0.4860)         &    (0.5798)         &    (0.3004)         &    (0.3316)         \\ \hline 
      &       &       &       &  & & \\[\dimexpr-\normalbaselineskip+2pt]
Observations       &3,982,816       &3,982,816       &3,982,816       &3,982,816       &2,691,783      &2,691,783      \\
Mean (F)  
  &       0.818         &       0.782         &       34.42         &       39.51         &       41.25         &       47.27         \\
Employed  Only          &                     &                     &                     &                     &           X         &           X         \\ \hline 
      &       &       &       &  & & \\[\dimexpr-\normalbaselineskip+2pt]
\bottomrule
\end{tabular}%
\begin{tablenotes}
\scriptsize
\item Notes:  Data is derived from the American Community Survey (ACS) conducted annually between 2005 and 2019. We compare individuals in same-sex partnerships with those in different-sex partnerships within four years of ENDA enactment. The regressions are estimated separately by sex: Panel A for men, while Panel B for women. The coefficients indicate the impact of anti-discrimination laws. The analysis is divided into two parts: Columns (1-2) focus on the extensive margin, examining changes in labor force participation and employment status as a result of the laws. Columns (3-6), on the other hand, explore the intensive margin by leveraging variations in the number of weeks worked per year and average weekly working hours. Standard errors are clustered at the county level.  \sym{*} \(p<0.10\), \sym{**} \(p<0.05\), \sym{\sym{***}} \(p<0.01\).
\end{tablenotes}
\end{threeparttable}
\end{table}

\begin{table}[htbp]
\footnotesize 
\centering
\begin{threeparttable}
\caption{Effects of Anti-Discrimination Laws on Labor Supply: Analysis of Same-Sex Marriage Fixed Effects}\label{tab:ext_int_no_ssm_app}%
\begin{tabular}{l|cc|cccc}
\toprule
      & (1)   & (2)   & (3)   & (4) & (5) & (6) \\
& \multicolumn{2}{c}{\underline{Extensive Margin}} & \multicolumn{4}{c}{\underline{Intensive Margin}}\\
 & Labor Force & Employed & Weekly Hours & Weeks Worked & Weekly Hours & Weeks Worked \\
\midrule
\multicolumn{7}{c}{Panel A: Men} \\
\midrule
ENDA $\times$ SSP        &      0.0267\sym{**} &      0.0270\sym{**} &      0.5377         &      0.8852         &     -0.4480         &     -0.3068         \\
                    &    (0.0129)         &    (0.0133)         &    (0.4982)         &    (0.6173)         &    (0.3629)         &    (0.2818)         \\ \hline 
      &       &       &       &  & & \\[\dimexpr-\normalbaselineskip+2pt]
Observations       &3,798,599        &3,798,599        &3,798,599        &3,798,599        &3,073,844        &3,073,844        \\
Mean (M)              &       0.825         &       0.790         &       35.53         &       39.96         &       42.41         &       47.69         \\
Employed  Only          &                     &                     &                     &                     &           X         &           X         \\
      &       &       &       &  & & \\[\dimexpr-\normalbaselineskip+2pt]
\midrule
\multicolumn{7}{c}{Panel B: Women} \\
\midrule
ENDA $\times$ SSP       &     -0.0130         &     -0.0212\sym{*}  &      0.8477         &     -0.7081         &      0.9893\sym{***}&     -0.7657\sym{**} \\
                    &    (0.0108)         &    (0.0115)         &    (0.5163)         &    (0.5760)         &    (0.3081)         &    (0.3166)         \\ \hline 
      &       &       &       &  & & \\[\dimexpr-\normalbaselineskip+2pt]
Observations        &3,982,894       &3,982,894       &3,982,894       &3,982,894       &2,691,849        &2,691,849        \\
Mean (F)              &       0.818         &       0.782         &       34.42         &       39.51         &       41.25         &       47.27         \\

Employed  Only          &                     &                     &                     &                     &           X         &           X         \\ \hline 
      &       &       &       &  & & \\[\dimexpr-\normalbaselineskip+2pt]
\bottomrule
\end{tabular}%
\begin{tablenotes}
\scriptsize
\item Notes:  Data is derived from the American Community Survey (ACS) conducted annually between 2005 and 2019. We compare individuals in same-sex partnerships with those in different-sex partnerships within four years of ENDA enactment. The regressions are estimated separately by sex: Panel A for men, while Panel B for women. The coefficients indicate the impact of anti-discrimination laws. The analysis is divided into two parts: Columns (1-2) focus on the extensive margin, examining changes in labor force participation and employment status as a result of the laws. Columns (3-6), on the other hand, explore the intensive margin by leveraging variations in the number of weeks worked per year and average weekly working hours. This table excludes the fully interacted same-sex-marriage term from Equation~(1). The ``Mean (F)" and ``Mean (M)" rows report pre-treatment averages of the outcome variables (including never-treated counties) for women and men in same-sex couples, respectively. Standard errors are clustered at the county level.  \sym{*} \(p<0.10\), \sym{**} \(p<0.05\), \sym{\sym{***}} \(p<0.01\).
\end{tablenotes}
\end{threeparttable}
\end{table}

\begin{table}[htbp]
\footnotesize 
\centering
\begin{threeparttable}
\caption{Anti-Discrimination Laws on Labor Supply with Occupation FEs}\label{tab:occ_ls_app}%
\begin{tabular}{l|cc|cccc}
\toprule
      & (1)   & (2)   & (3)   & (4) & (5) & (6) \\
& \multicolumn{2}{c}{\underline{Extensive Margin}} & \multicolumn{4}{c}{\underline{Intensive Margin}}\\
 & Labor Force & Employed & Weekly Hours & Weeks Worked  & Weekly Hours & Weeks Worked \\
\midrule
\multicolumn{7}{c}{Panel A: Men} \\
\midrule
ENDA $\times$ SSP           &      0.0228\sym{**} &      0.0246\sym{**} &      0.5170         &      0.8743\sym{**} &     -0.1462         &     -0.0550         \\
                    &    (0.0093)         &    (0.0100)         &    (0.3720)         &    (0.4334)         &    (0.3307)         &    (0.2797)         \\ \hline 
      &       &       &       &  \\[\dimexpr-\normalbaselineskip+2pt]
Observations       &3,798,542        &3,798,542        &3,798,542        &3,798,542        &3,073,797        &3,073,797        \\
Mean (M)              &       0.825         &       0.790         &       35.53         &       39.96         &       42.41         &       47.69         \\
Employed  Only          &                     &                     &                     &                     &           X         &           X         \\
      &       &       &       &  & & \\[\dimexpr-\normalbaselineskip+2pt]
\midrule
\multicolumn{7}{c}{Panel B: Women} \\
\midrule
ENDA $\times$ SSP      &     -0.0064         &     -0.0132         &      0.5614         &     -0.8344\sym{*}  &      0.5854\sym{**} &     -0.9524\sym{***}\\
                    &    (0.0098)         &    (0.0111)         &    (0.4008)         &    (0.4630)         &    (0.2939)         &    (0.3219)         \\ \hline 
      &       &       &       & & &  \\[\dimexpr-\normalbaselineskip+2pt]
Observations         &3,982,816       &3,982,816       &3,982,816       &3,982,816       &2,691,783        &2,691,783        \\
Mean (F)              &       0.818         &       0.782         &       34.42         &       39.51         &       41.25         &       47.27         \\
Employed  Only          &                     &                     &                     &                     &           X         &           X         \\ \hline 
\bottomrule
\end{tabular}%
\begin{tablenotes}
\scriptsize
\item Notes:  Data is derived from the American Community Survey (ACS) conducted annually between 2005 and 2019. We compare individuals in same-sex partnerships with those in different-sex partnerships within four years of ENDA enactment. The regressions are estimated separately by sex: Panel A for men, while Panel B for women, and include four-digit occupation fixed effects. The coefficients indicate the impact of anti-discrimination laws. The analysis is divided into two parts: Columns (1-2) focus on the extensive margin, examining changes in labor force participation and employment status as a result of the laws. Columns (3-6), on the other hand, explore the intensive margin by leveraging variations in the number of weeks worked per year and average weekly working hours.  The ``Mean (F)" and ``Mean (M)" rows report pre-treatment averages of the outcome variables (including never-treated counties) for women and men in same-sex couples, respectively. Standard errors are clustered at the county level.  \sym{*} \(p<0.10\), \sym{**} \(p<0.05\), \sym{\sym{***}} \(p<0.01\).
\end{tablenotes}
\end{threeparttable}
\end{table}

\begin{table}[htbp]
\footnotesize 
\centering
\begin{threeparttable}
\caption{Anti-Discrimination Laws on Pay with Occupation FEs}\label{tab:occ_wage_app}%
\begin{tabular}{l|cccc|cccc}
\toprule 
\multicolumn{1}{r}{} & \multicolumn{4}{c}{Hourly Wage} & \multicolumn{2}{c}{Annual Earnings} \\
\midrule
      & (1)   & (2)   & (3)   & (4) & (5) & (6) & (7) & (8) \\
 & Percentile & $\ge$25th & $\ge$50th & $\ge$75th & Percentile & $\ge$25th & $\ge$50th & $\ge$75th \\
\midrule
\multicolumn{9}{c}{Panel A: Men} \\
\midrule
ENDA $\times$ SSP            &      1.8922\sym{**} &      0.0243\sym{**} &      0.0244\sym{*}  &      0.0200         &      2.2207\sym{***}&      0.0242\sym{**} &      0.0391\sym{***}&      0.0308\sym{**} \\
                    &    (0.9189)         &    (0.0113)         &    (0.0146)         &    (0.0141)         &    (0.8331)         &    (0.0113)         &    (0.0152)         &    (0.0128)         \\ \hline 
  & & & &    &       &       &       &  \\[\dimexpr-\normalbaselineskip+2pt]
Observations       &3,798,542         &3,798,542         &3,798,542         &3,798,542         &3,798,542         &3,798,542         &3,798,542         &3,798,542         \\
Mean (M)              &       51.51         &       0.791         &       0.557         &       0.305         &       51.69         &       0.791         &       0.561         &       0.305         \\
 & & & &    &       &       &       &  \\[\dimexpr-\normalbaselineskip+2pt]
\midrule
\multicolumn{9}{c}{Panel B: Women} \\
\midrule
ENDA $\times$ SSP        &     -0.9026         &     -0.0024         &     -0.0182         &     -0.0116         &     -1.1521\sym{*}  &     -0.0018         &     -0.0274\sym{**} &     -0.0059         \\
                    &    (0.6989)         &    (0.0097)         &    (0.0121)         &    (0.0146)         &    (0.6519)         &    (0.0097)         &    (0.0111)         &    (0.0155)         \\ \hline 
 & & & &    &       &       &       &  \\[\dimexpr-\normalbaselineskip+2pt]
Observations         &3,982,816       &3,982,816       &3,982,816       &3,982,816       &3,982,816       &3,982,816       &3,982,816       &3,982,816       \\
Mean (F)              &       48.73         &       0.799         &       0.513         &       0.247         &       48.68         &       0.798         &       0.515         &       0.241         \\
\bottomrule
\end{tabular}%
\begin{tablenotes}
\scriptsize
\item Notes: Data is derived from the American Community Survey (ACS) conducted annually between 2005 and 2019. We compare individuals in same-sex partnerships with those in different-sex partnerships within four years of ENDA enactment. The regressions are estimated separately by sex and include four-digit occupation fixed effects. The coefficients in the first row indicate the impact of anti-discrimination laws. Columns (1–4) report estimates for hourly real wage ranks, while Columns (5–8) report estimates for annual real wage ranks. In Columns 2–4 and 6–8, the estimates are indicator variables for being above the 25th, 50th, and 75th percentiles, respectively. All income variables are measured in 2019 dollars and expressed as percentile ranks.   The ``Mean (F)" and ``Mean (M)" rows report pre-treatment averages of the outcome variables (including never-treated counties) for women and men in same-sex couples, respectively. Standard errors are clustered at the county level. \sym{*} \(p<0.10\), \sym{**} \(p<0.05\), \sym{\sym{***}} \(p<0.01\).
\end{tablenotes}
\end{threeparttable}
\end{table}

\end{document}